\documentclass[11pt]{article}

\usepackage{amsmath}
\usepackage{amssymb}
\usepackage{amsthm}
\usepackage{authblk}
\usepackage{booktabs}
\usepackage{comment}
\usepackage{enumitem}
\usepackage[margin=1in]{geometry}
\usepackage{graphicx}
\usepackage{hyperref}
\usepackage{mathrsfs}
\usepackage{mathtools}
\usepackage{physics}
\usepackage[labelformat=simple]{subcaption}
\usepackage[normalem]{ulem}
\usepackage{xcolor}

\hypersetup{
  colorlinks,
  linkcolor={red!50!black},
  citecolor={blue!50!black},
  urlcolor={blue!80!black}
}

\theoremstyle{plain}
\newtheorem{theorem}{Theorem}[section]
\newtheorem{lemma}[theorem]{Lemma}
\newtheorem{proposition}[theorem]{Proposition}
\newtheorem{corollary}[theorem]{Corollary}

\theoremstyle{definition}

\theoremstyle{remark}

\newtheorem{claim}[theorem]{Claim}

\numberwithin{equation}{section}

\usepackage{dsfont}
\newcommand{\eps}{\epsilon}
\newcommand{\indicator}[1]{\mathds{1}_{#1}}
\DeclarePairedDelimiter{\set}{\{}{\}}
\DeclarePairedDelimiter{\bars}{\lvert}{\rvert}
\DeclarePairedDelimiter{\dbars}{\lVert}{\rVert}
\DeclarePairedDelimiter{\brackets}{[}{]}
\DeclarePairedDelimiter{\ceil}{\lceil}{\rceil}
\DeclarePairedDelimiter{\floor}{\lfloor}{\rfloor}
\DeclarePairedDelimiter{\parentheses}{(}{)}

\newcommand{\E}{\mathbb{E}}
\newcommand{\F}{\mathbb{F}}
\renewcommand{\P}{\mathbb{P}}
\newcommand{\R}{\mathbb{R}}
\newcommand{\Z}{\mathbb{Z}}
\newcommand{\bA}{\mathbb{A}}
\newcommand{\cA}{\mathcal{A}}
\newcommand{\cB}{\mathcal{B}}

\newcommand{\cE}{\mathcal{E}}

\newcommand{\cL}{\mathcal{L}}
\newcommand{\cO}{\mathcal{O}}
\newcommand{\cP}{\mathcal{P}}
\newcommand{\cQ}{\mathcal{Q}}

\newcommand{\fE}{\mathfrak{E}}

\newcommand{\fI}{\mathfrak{I}}
\newcommand{\fm}{\mathfrak{m}}

\newcommand{\fR}{\mathfrak{R}}

\newcommand{\fT}{\mathfrak{T}}

\newcommand{\sC}{\mathscr{C}}
\newcommand{\sE}{\mathscr{E}}
\newcommand{\sF}{\mathscr{F}}
\newcommand{\sG}{\mathscr{G}}
\newcommand{\sP}{\mathscr{P}}
\newcommand{\sR}{\mathscr{R}}
\newcommand{\sV}{\mathscr{V}}
\newcommand{\Cov}{\mathrm{Cov}}
\newcommand{\diam}{\operatorname{diam}}

\newcommand{\dom}{\mathrm{dom}}
\newcommand{\maj}{\mathrm{maj}}
\newcommand{\obs}{\mathrm{obs}}
\newcommand{\per}{\mathrm{per}}
\newcommand{\reg}{\mathrm{reg}}
\newcommand{\rmd}{\mathrm{d}}
\newcommand{\syn}{\mathrm{syn}}
\newcommand\rect[2]{\mathrm{R}_{#1,#2}}
\newcommand\peierls[1]{\mathfrak{p}_{#1}}
\newcommand\es[1]{\overline{#1}}
\newcommand\xre[2]{X_{#1}#2}
\newcommand{\Phase}[1]{\mathrm{Phase}(#1)}

\newcommand{\Side}{\mathrm{Side}}
\newcommand{\Ori}[1]{\mathrm{Ori}(#1)}
\newcommand{\Par}[1]{\mathrm{Par}(#1)}

\let\emptyset\varnothing
\let\vec\mathbf

\allowdisplaybreaks

\title{Discotic columnar phase in a three-dimensional monomer-tetramer model of Heilmann and Lieb with side-to-side attraction}
\author{Qidong He\thanks{Email address: \url{q.he@northeastern.edu}}}
\date{}
\affil{Department of Mathematics, Northeastern University}

\begin{document}

\maketitle

\begin{abstract}
    We study Model~III of Heilmann--Lieb (1979), a lattice model for liquid crystals consisting of tetramers (plaquettes) on the cubic lattice at chemical potential $\mu$, interacting via a hard-core repulsion and a lateral attraction of strength $a>0$ between coplanar, aligned tetramers.
    The model was conjectured to exhibit liquid-crystalline order at low temperature provided that $\mu+2a>0$.
    We establish discotic columnar order throughout this regime, fully resolving the conjecture.
    Specifically, we prove the existence of exactly twelve extremal and periodic Gibbs measures, each breaking rotational symmetry, with one of three tetramer orientations dominant, and translational symmetry in exactly two directions, with one of four tetramer parities dominant within that orientation.
    Thus, the tetramers form one-dimensional columns arranged in a square lattice.
    This arrangement is not a priori evident from the zero-temperature picture, where energy only enforces a layered structure: each layer is fully packed and crystalline, but different layers are free to shift independently within their planes.
    The main novelty here lies in showing how thermal fluctuations lift this degeneracy in favor of the twelve ground states in which all tetramers share a common parity.
\end{abstract}


\section{Introduction}

Liquid crystals are distinguished from ordinary liquids by the presence of long-range orientational order and from crystalline solids by the absence of complete positional order~\cite{degennes1993physics}.
Thermotropic liquid crystals, so called because their formation is driven primarily by temperature, are commonly classified according to the extent of the positional order of their constituent molecules.
In the \textit{nematic} phase, molecules align along a preferred direction but exhibit no long-range positional order.
In the \textit{cholesteric} phase, the direction of alignment rotates helically about an orthogonal axis.
In the \textit{smectic} phase, molecules exhibit positional order in one direction, organizing into layers within which they remain disordered.
In the \textit{columnar} phase, molecules exhibit positional order in two directions, organizing into columns arranged in a two-dimensional lattice.
The columnar phase, in particular, is characteristic of disk-shaped molecules whose flat faces tend to stack~\cite{chandrasekhar1977liquid}.
For instance, in the columnar phase of the extensively studied compound HAT6, the aromatic cores of the molecules, held together by attractive ($\pi$--$\pi$) interactions acting perpendicular to their flat faces~\cite{zbiri2010density}, form columns arranged in a hexagonal lattice, with charge transport occurring predominantly along the direction of the columns~\cite{boden1995mechanism}.
This anisotropy, together with the tendency of discotic materials to self-organize, has led to their extensive applications as organic semiconductors~\cite{schmidt2001self} and tunable optical materials~\cite{sentker2019self}.

Lattice gases of hard particles provide a simple, natural setting in which to isolate the roles of particle geometry and molecular interactions in determining the nature of orientational and positional order in a macroscopic system.
In the simplest cases, the particles interact only by geometric exclusion, so that any order that arises is purely entropic in origin.
Monte Carlo simulations of $2\times2\times1$ hard plates on the (simple) cubic lattice suggest that, with increasing density, the system transitions from a disordered phase first to a layered phase and then to a sublattice phase, in which translational symmetry is broken in all three lattice directions~\cite{mandal2023phases}.
For $2\times 2\times 2$ hard cubes on the cubic lattice, simulations instead suggest transitions from a disordered phase to a layered phase, a sublattice phase, and finally a columnar phase, in which translational symmetry is broken in only two directions~\cite{vigneshwar2019phase}.
Thus, at least on the level of numerics, the number of directions in which translational symmetry is broken need not increase monotonically with the density.
From a mathematical point of view, a particularly stimulating problem is therefore to derive such behaviors rigorously and directly from microscopic interactions and, in particular, to determine precisely the extent of positional order in such systems.

\begin{figure}
    \centering
    \includegraphics[width=0.55\columnwidth]{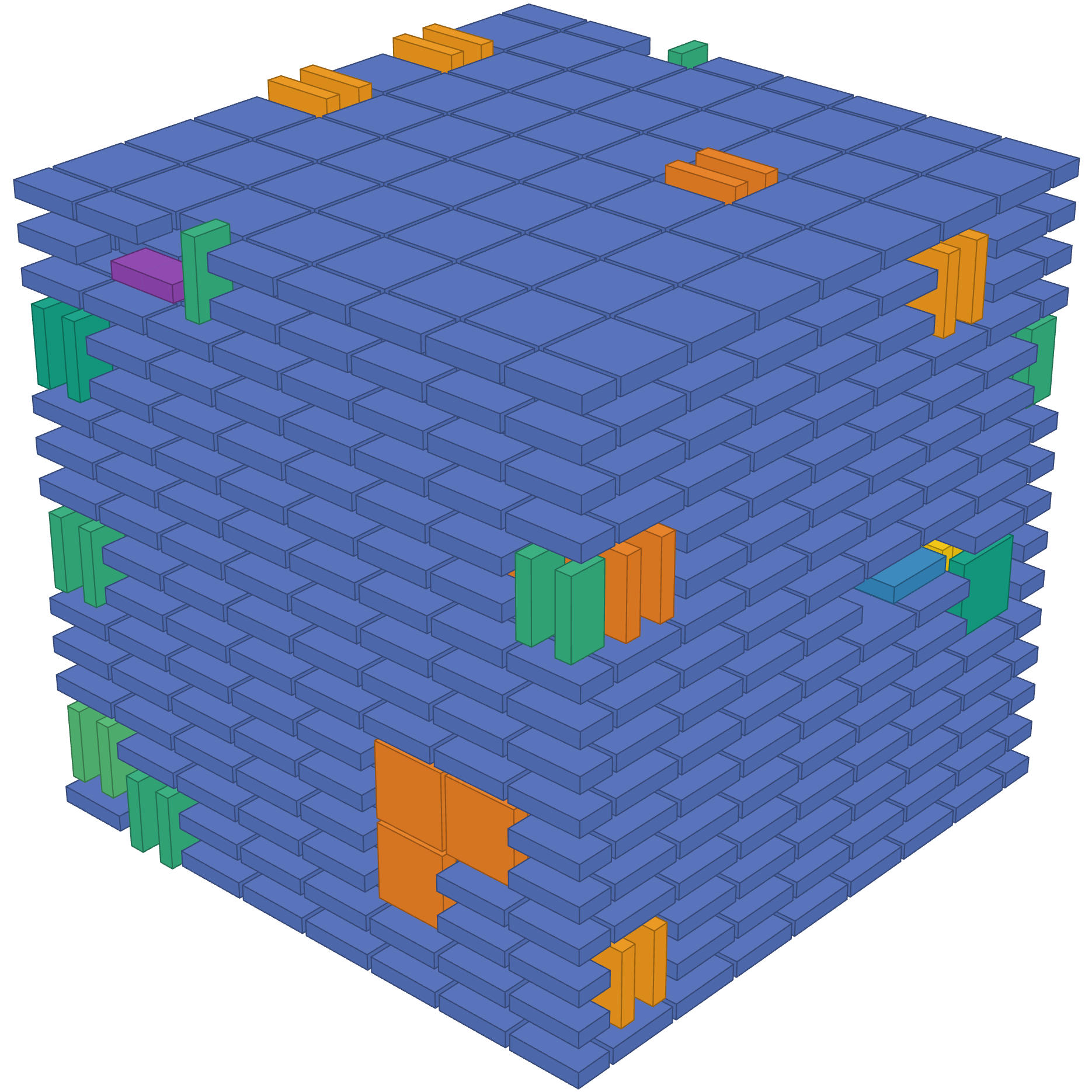}
    \caption{A configuration of Model~III on a $16\times 16\times 16$ torus at $\beta=5.65$, $\mu=0.8$, and $a=0.1$.
    Most tetramers stack into vertical columns arranged in a square lattice.
    The colors follow the convention of Figure~\ref{fig: ground states}.}
    \label{fig: columnar}
\end{figure}

\subsection{Mathematical background}

The mathematical study of liquid-crystalline order in three dimensions has a history stretching at least back to the 1970s.
In 1979, Heilmann--Lieb~\cite{heilmann1979lattice} proposed five lattice models of hard-core particles with attractive interactions, two on $\Z^2$ and three on $\Z^3$, and used the chessboard estimate to prove low-temperature orientational order in all five models, while leaving the absence of complete positional order unproved.
Abraham--Heilmann~\cite{abraham1980interacting} subsequently extended this approach to two other interacting dimer models on $\Z^3$ but left the precise orientational structure unresolved.
In 1982, Angelescu--Zagrebnov~\cite{angelescu1982lattice} considered a three-dimensional lattice gas with continuous molecular orientations and used a combination of reflection positivity methods to establish long-range orientational order at low temperature.
This approach was further developed by Angelescu--Romano--Zagrebnov~\cite{angelescu1995long} and Zagrebnov~\cite{zagrebnov1996long} in lower dimensions, although the status of long-range positional order remained unresolved.
In 1993, Koteck{\'y}--Laanait--Messager--Miracle-Sol{\'e}~\cite{kotecky1993spin} used an extension of Pirogov--Sinai theory~\cite{pirogov1975phase,pirogov1976phase,zahradnik1984alternate} due to Bricmont--Slawny~\cite{bricmont1989phase} to rigorously analyze the low-temperature phase diagram of a spin-$1$ lattice model of microemulsions in dimensions $d\ge3$.
In 2015, Giuliani--Lieb~\cite{giuliani2015columnar} used reflection positivity methods to prove columnar long-range order in a two-dimensional \textit{quantum} dimer model with an attractive parallel dimer-dimer interaction and a flipping term, which they noted could be easily extended to dimers and plaquettes in three dimensions.
In 2020, building on the Pirogov--Sinai techniques developed by Disertori--Giuliani~\cite{disertori2013nematic}, Disertori--Giuliani--Jauslin~\cite{disertori2020plate} proved the existence of a uniaxial nematic phase in a system of anisotropic hard plates in $\R^3$ with six allowed orientations.

In this paper, we continue the study of the Heilmann--Lieb models~\cite{heilmann1979lattice}, focusing specifically on Model~III.
Following their introduction, the question of whether low-temperature orientational order is accompanied by long-range positional order in these models remained open for several decades despite early attempts by Heilmann--Kj{\ae}r (unpublished) and Letawe~\cite{letawe1994module}. 
Renewed progress came in 2016, when Alberici~\cite{alberici2016cluster} used the cluster expansion method to rule out positional order in a variant of Model~I, a monomer-dimer model on $\Z^2$ where horizontal and vertical dimers are assigned different chemical potentials.
Shortly after, in 2018, Jauslin--Lieb~\cite{jauslin2018nematic} used Pirogov--Sinai theory to prove nematic order in the original Model~I, although their argument restricted the dimer chemical potential $\mu$ and the strength of the attraction $a$ in an ``unnatural'' way.
In 2025, we proved nematic order in Model~I in a substantially improved parameter regime~\cite{he2025extended}.
Though it still fell short of the Heilmann--Lieb prediction ($\mu+a>0$), the extra condition we required ($3a>\mu$) has a clear physical meaning: it ensures that misaligned dimers remain rare on scales comparable to the correlation length of the one-dimensional reference system.
This separation of scales allowed us to build on the Heilmann--Lieb approach, combining the chessboard estimate with the abstract framework of strongly percolating sets introduced by Hadas--Peled~\cite{hadas2026columnar} and the disagreement percolation method of van den Berg~\cite{van1993uniqueness} to rule out positional order.
More recently~\cite{he2026nematic}, we obtained analogous results for Models~II and~V, consisting of dimers on $\Z^2$ and $\Z^3$ with a different attractive interaction, where we relied on the cluster swapping method of Sheffield~\cite{sheffield2005random} rather than disagreement percolation to exploit a critical simplification special to these models.

\subsection{Model and main results}

Let $\F$ be the set of plaquettes of the cubic lattice $\Z^3$, each identified with its geometric center in $\R^3$.
We refer to the elements of $\F$ as \textbf{tetramers}, since each of them covers four vertices of $\Z^3$.\footnote{Heilmann--Lieb~\cite{heilmann1979lattice} referred to the plaquettes as ``fourmers''; we prefer the more standard term ``tetramer'', in line with ``monomer'' and ``dimer''.}
A \textbf{configuration} is a function $\sigma:\F\to\set{0,1}$ such that $\sigma(t)\sigma(t')=0$ whenever $t,t'\in\F$ are distinct and intersecting; we identify each configuration $\sigma$ with the set $\sigma^{-1}(\set{1})$ of pairwise disjoint tetramers.
Thus, each vertex of $\Z^3$ is covered by at most one tetramer in any configuration; we refer to the uncovered ones as \textbf{vacancies} (or \textbf{monomers}).
Let $\Omega$ denote the set of all configurations.

Whereas the particles in Models~I, II, and~V are dimers and rod-like~\cite{alberici2016cluster,jauslin2018nematic,he2025extended,he2026nematic}, the tetramers featured in Model~III (as well as Model~IV) are disk-like.
This is perhaps best seen by enlarging each tetramer $t$, viewed as a closed unit square in $\R^3$, to the closed set $t+[-1/2,1/2]^3$, which converts it into a $2\times2\times 1$ \textit{plate} with two square \textit{faces} and four rectangular \textit{sides}.
In this picture, the hard-core constraint simply requires the plates in a configuration to have pairwise disjoint interiors.
Hence, each configuration corresponds to a packing of plates, and a configuration without any vacancies corresponds to a tiling of $\R^3$ by such plates.

Let $\beta>0$ be the inverse temperature, $\mu\in\R$ be the chemical potential, and $a>0$ be the strength of the attraction.
Denote by $\vec{e}_1,\vec{e}_2,\vec{e}_3$ the standard basis vectors of $\R^3$.
We say that two tetramers $t,t'\in\F$ are \textbf{linked}, and write $t\sim t'$, if they are coplanar and their centers differ by $\pm 2\vec{e}_i$ for some $i\in\set{1,2,3}$.
Thus, each tetramer can be linked to at most four other tetramers, one across each of its sides.
In the picture of plates, this means that two plates are linked precisely when they share a full side; in particular, those stacked face to face or in contact along only part of their sides are not linked.
The formal Hamiltonian of the monomer-tetramer model with side-to-side attraction, introduced by Heilmann--Lieb~\cite{heilmann1979lattice} as Model~III, is
\begin{equation}
    \label{eqn:formal Hamiltonian}
    H(\sigma)= -\mu\abs{\sigma}-a\,\sum_{\set{t,t'}\subseteq\sigma}\indicator{t\sim t'},
\end{equation}
where $\abs{\cdot}$ denotes the cardinality.
The formal Hamiltonian~\eqref{eqn:formal Hamiltonian} is adapted to finite volumes in the usual way and gives rise to finite-volume Gibbs measures as well as infinite-volume ones via the usual Dobrushin--Lanford--Ruelle formalism; a detailed review of these definitions and constructions is given in Section~\ref{sec:the Hamiltonian}.

\begin{figure}[tbp]
    \centering

    \begin{subfigure}[t]{0.32\textwidth}
        \centering
        \includegraphics[width=\linewidth]{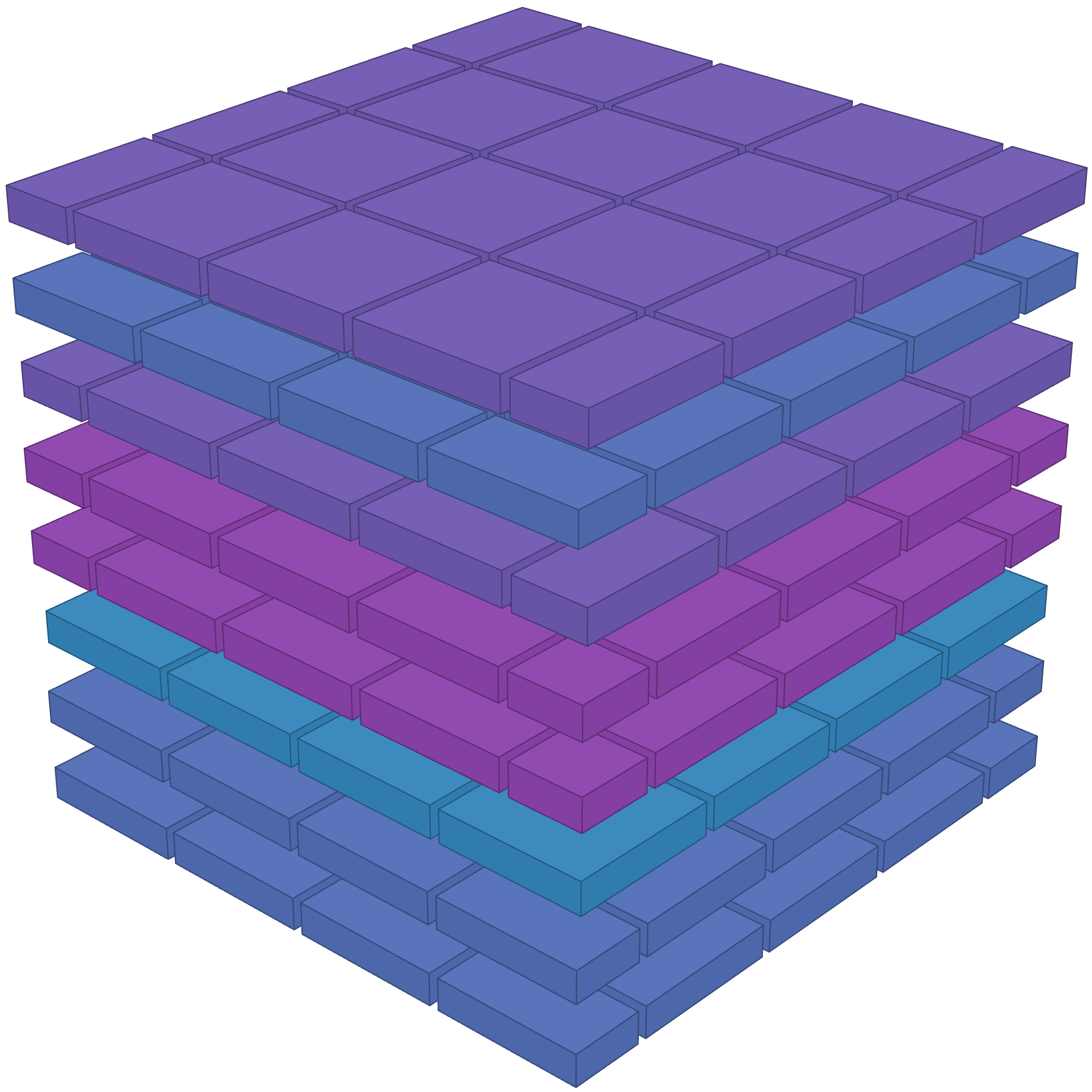}
        \caption{}
        \label{fig: xy}
    \end{subfigure}~
    \begin{subfigure}[t]{0.32\textwidth}
        \centering
        \includegraphics[width=\linewidth]{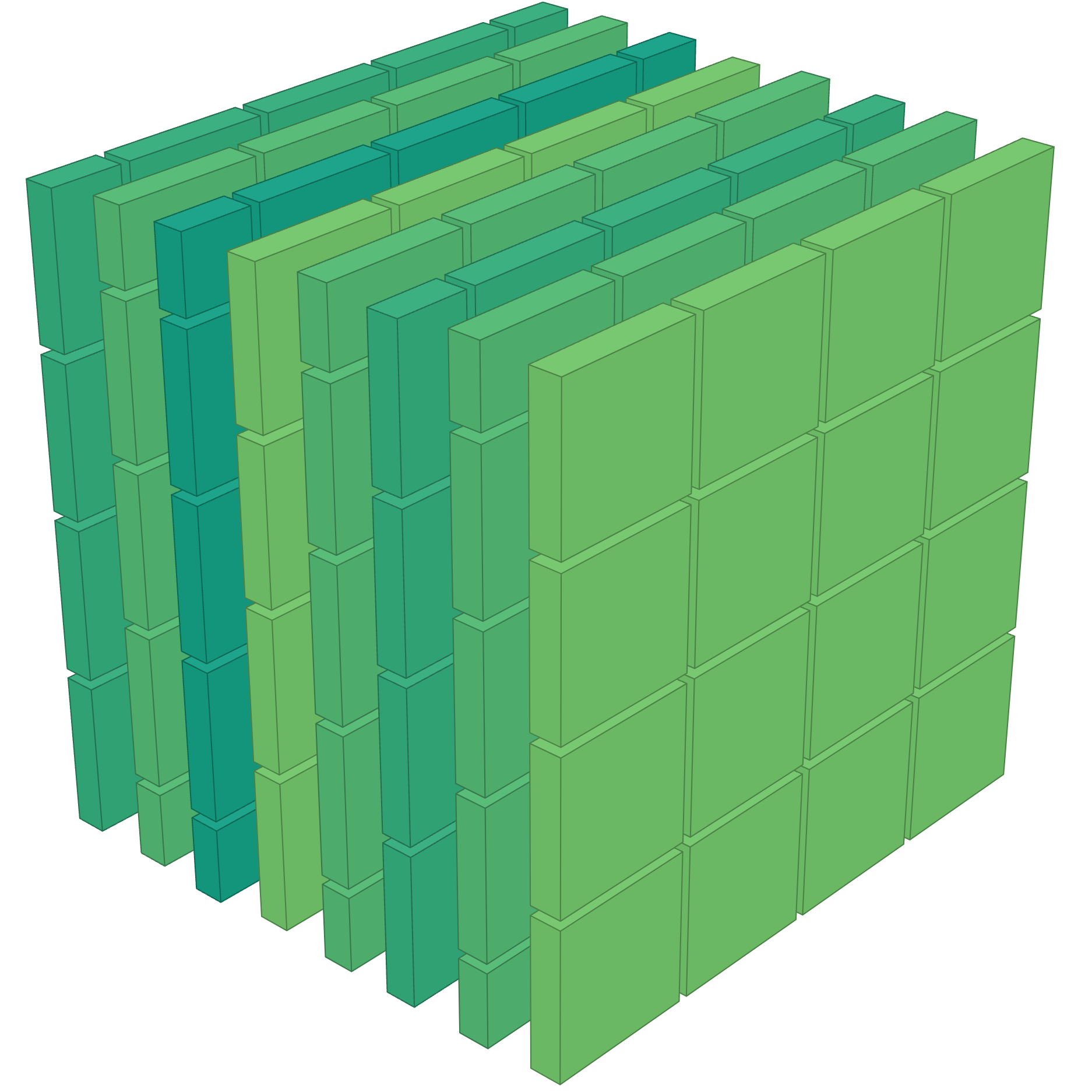}
        \caption{}
        \label{fig: yz}
    \end{subfigure}~
    \begin{subfigure}[t]{0.32\textwidth}
        \centering
        \includegraphics[width=\linewidth]{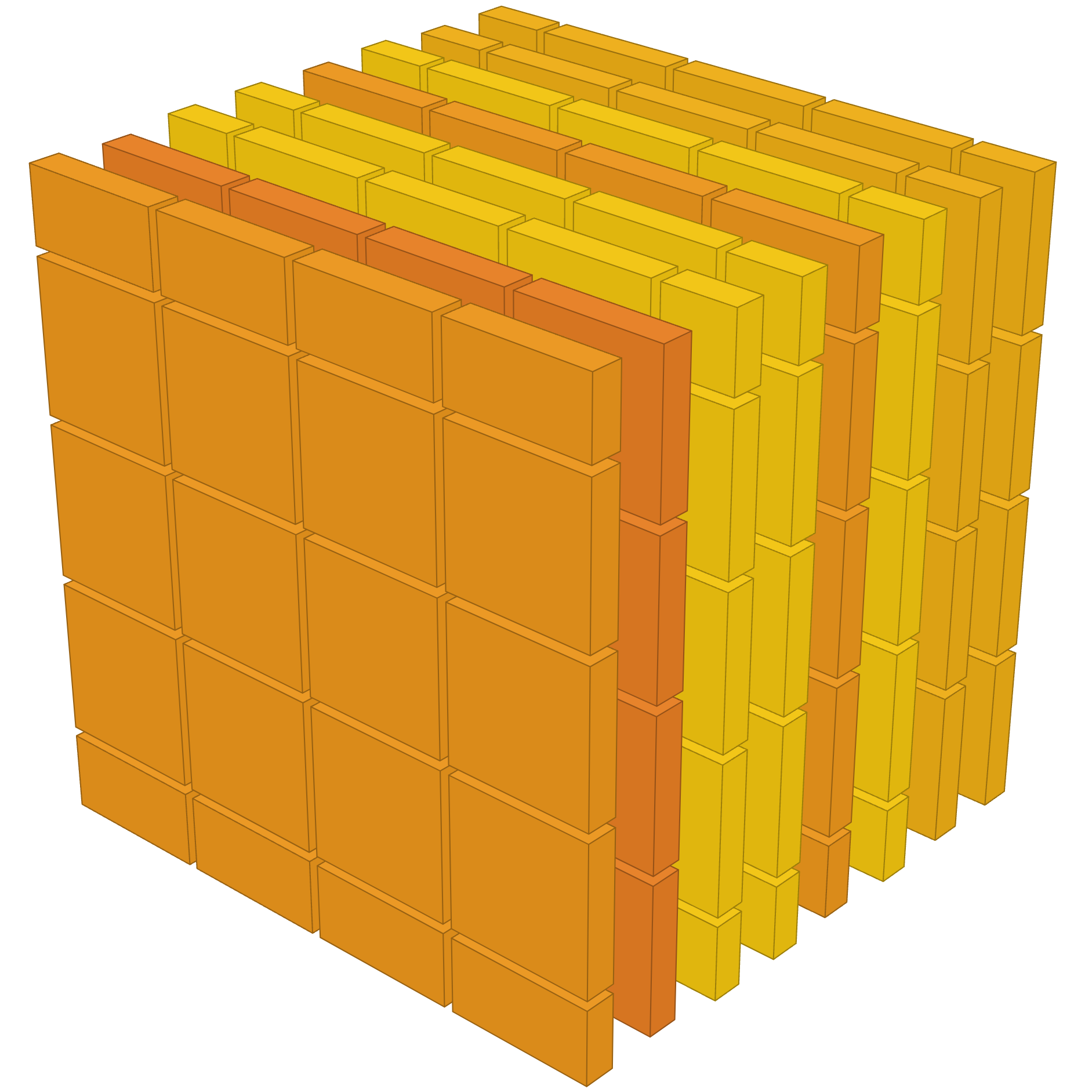}
        \caption{}
        \label{fig: xz}
    \end{subfigure}

    \caption{Layered ground states of the monomer-tetramer model, with all tetramers parallel to the \subref{fig: xy}~$x_1x_2$-, \subref{fig: yz}~$x_2x_3$-, and \subref{fig: xz}~$x_3x_1$-planes.
    Each layer is fully packed and crystalline, but different layers can be shifted independently within their planes.
    In these examples, the parities in different layers are not synchronized, so the tetramers do not form columns.
    Throughout the paper, tetramers of each orientation are drawn in one family of colors, with a different color for each parity, as in \subref{fig: xy}--\subref{fig: xz}.}
    \label{fig: ground states}
\end{figure}

In~\cite{heilmann1979lattice}, Heilmann--Lieb proved that the model exhibits orientational order at low temperature throughout the regime $\mu+2a>0$.
As a shifting of the Hamiltonian makes clear (see~\eqref{eqn:shifted Hamiltonian monomer-tetramer}), this condition is precisely what penalizes vacancies and drives the system toward configurations densely packed with tetramers at low temperature.
In fact, the shifting shows that, in this regime, the \textit{ground states} of the model are precisely those configurations that have no vacancies and in which every tetramer is linked to four other tetramers.\footnote{In this paper, by a \textit{ground state} we always mean a configuration with zero energy under the non-negative shifted Hamiltonian~\eqref{eqn:shifted Hamiltonian monomer-tetramer}. Every such configuration is also a ground state in the sense of Georgii~\cite[Definition~(6.18)]{georgii2011gibbs}.\label{fn: ground state}}
A simple argument then shows that every ground state is \textit{layered} (see Proposition~\ref{prop: every ground state is layered}), meaning that all tetramers have the same orientation and are organized in parallel layers, with those in each layer arranged periodically in a square lattice; see Figure~\ref{fig: ground states}.
An important feature of this ground state structure is that different layers do not interact and are, therefore, free to slide relative to one another, giving rise to uncountably many ground states and hence a highly degenerate landscape at zero temperature.
Thus, a priori, it is not obvious how this picture should change at low positive temperature.

In view of the picture at zero temperature, it may be somewhat surprising that our main result, stated below, shows that thermal fluctuations at low temperature lift the degeneracy in favor of exactly \textit{twelve} ground states in which the tetramers line up across all layers to form \textit{columns}, as is characteristic of a discotic columnar liquid crystal; see Figure~\ref{fig: columnar} for an illustration.

To state our main result precisely, we fix some notation and conventions, which we use throughout the paper.
We interpret indices of coordinate directions as elements of $\Z/3\Z$, and write $\cO\coloneqq\set{x_ix_{i+1}\mid i\in\Z/3\Z}=\set{x_1x_2,x_2x_3,x_3x_1}$.
We say that a tetramer $t$ is \textbf{$x_ix_{i+1}$-oriented} if it is parallel to the $x_ix_{i+1}$-plane, in which case we define its \textbf{parity} as $(t_i-1/2\bmod 2,t_{i+1}-1/2\bmod 2)\in\cP\coloneqq(\Z/2\Z)^2$, where $(t_1,t_2,t_3)$ is the geometric center of $t$.
Whenever the context calls for an integer representative, we use $1,2,3$ for the elements of $\Z/3\Z$ and $0,1$ for those of $\Z/2\Z$.

\begin{theorem}[Discotic columnar order]
    \label{thm:main}
    Fix $\mu\in\R$ and $a>0$ such that $\mu+2a>0$.
    There exist constants $\beta_0,C,c>0$ such that, for all $\beta\ge\beta_0$, $o=x_ix_{i+1}\in\cO$, and $p\in\cP$, the monomer-tetramer model admits a unique Gibbs measure $\P_{o,p}$ such that
    \begin{enumerate}[label=(\alph*)]
        \item \label{itm:main - invariance and extremality} $\P_{o,p}$ is $2\Z\vec{e}_i\oplus 2\Z\vec{e}_{i+1}\oplus\Z\vec{e}_{i+2}$-invariant and extremal.
        \item \label{itm:main - tetramer probabilities} For all $t\in\F$,
        \begin{equation}
            \P_{o,p}(\sigma(t)=1)=
            \begin{cases}
                1-\order{e^{-\beta a/2}} & \text{if }t\text{ is $o$-oriented of parity $p$} \\ 
                \order{e^{-\beta a/2}} & \text{otherwise}
            \end{cases}.
        \end{equation}
        \item \label{itm:main - decay of correlations} Let $A,B\subseteq\F$ and $f,g:\Omega\to[-1,1]$ be measurable functions such that $f$ depends only on the restriction of $\sigma$ to $A$ and $g$ only on the restriction of $\sigma$ to $B$.
        Then,
        \begin{equation}
            \abs{\Cov_{\P_{o,p}}(f,g)}\le \sum_{s\in A}\sup_{t\in B}\alpha_{o}(s,t),
        \end{equation}
        where $\ell\gg e^{\frac{1}{2}\beta\min\set{\mu+8a,8a}}$ is a mesoscopic length scale (see~\eqref{eqn: mesoscopic length scale} for an exact expression), and
        \begin{equation}
            \alpha_{o}(s,t)\coloneqq C\exp\set[\bigg]{-c\brackets[\bigg]{\frac{\abs{s_i-t_i}+\abs{s_{i+1}-t_{i+1}}}{\ell}+\abs{s_{i+2}-t_{i+2}}}}.
        \end{equation}
    \end{enumerate}
    Furthermore, let $\varphi:\R^3\to\R^3$ be the coordinate permutation $(x_1,x_2,x_3)\mapsto(x_3,x_1,x_2)$ and, for $z\in\Z^3$, let $\eta_z:\R^3\to\R^3$ be the translation $v\mapsto v+z$.
    Then,
    \begin{enumerate}[label=(\alph*)]
        \setcounter{enumi}{3}
        \item \label{itm:main - transformation} For all $z\in\Z^3$ and $k\in\Z$,
        \begin{equation}
            \varphi^k\eta_z\P_{o,p}=\P_{x_{i+k}x_{i+k+1},\,p+(z_i,z_{i+1})}.
        \end{equation}
        \item \label{itm:main - classification of periodic Gibbs measures} Every periodic Gibbs measure of the model is a convex combination of the measures $\P_{o,p}$, where $(o,p)\in\cO\times\cP$.
    \end{enumerate}
\end{theorem}

Theorem~\ref{thm:main} establishes that the monomer-tetramer model exhibits a discotic columnar phase at low temperature.
In the $(o,p)$-phase, as described by $\P_{o,p}$, the tetramers are predominantly $o$-oriented and occupy mainly positions of parity $p$ in every layer.
Thus, they organize into columns perpendicular to the $o$-plane, which are themselves arranged in a square lattice.
Theorem~\ref{thm:main} also answers a question raised by Heilmann--Lieb specifically for Model~III~\cite[Section~7]{heilmann1979lattice}:
\begin{quote}
    ``The attractive interaction in model III clearly favors a complete ordering within a plane and it is easy to prove by Peierls' argument that the two-dimensional version (where one only has fourmers of one orientation) does exist in a completely ordered phase at low enough temperatures. This effect would be expected to carry over to the three-dimensional model, in which case it raises the intriguing question of whether the model can exist in six different phases.''
\end{quote}
Indeed, it confirms that the intraplane ordering does carry over to three dimensions and that the model, in fact, exists in \textit{twelve} different phases.

From a physical point of view, perhaps the most surprising aspect of Theorem~\ref{thm:main} is that columns should form at all.
As discussed in the introduction, the discotic columnar phase is characteristic of disk-shaped molecules whose flat faces tend to stack.
Unlike the attractive interactions that act perpendicular to these flat faces and therefore directly favor stacking, the attraction in this model acts within the planes of the tetramers, and the energy is oblivious to the relative positions of tetramers lying in different layers.
Furthermore, the columns cannot be attributed to the hard-core constraint alone since, as mentioned earlier, simulations~\cite{mandal2023phases} suggest sublattice order at high density for the lattice gas of $2\times2\times1$ hard plates, which corresponds to the monomer-tetramer model with $a=0$.
Thus, even though the side-to-side attraction never acts between tetramers in different layers, it qualitatively changes the nature of the high-density phase in a way that allows the system to break translational symmetry in exactly two directions.
As we will discuss in Section~\ref{sec: proof strategy}, this is accomplished via an order-by-disorder mechanism mediated by the minimum-energy excitations involving tetramers transverse to the layers.

We mention in passing that tilings of $\R^3$ by $2\times2\times1$ plates (also called \textit{slabs} in this context) are of independent interest in combinatorics and discrete geometry as a natural but much less understood counterpart of three-dimensional domino tilings~\cite{pak2013complexity,alencar2026slab,klivans2026domino}.
Theorem~\ref{thm:main} offers a complementary, probabilistic perspective on these objects, demonstrating how their dense random packings can be proven to form nontrivial structures under a simple, physically inspired interaction.

\subsection{Proof strategy}
\label{sec: proof strategy}

The proof of Theorem~\ref{thm:main} proceeds in three main steps: establishing the existence of three orientational phases, splitting each into four according to the dominant parity of the tetramers, and finally ruling out any periodic phase beyond the resulting twelve.

\subsubsection{Three orientations}

\paragraph{The Heilmann--Lieb approach}

Consider a $2\times2\times2$ cube.
Following Heilmann--Lieb~\cite[Section~5]{heilmann1979lattice}, we call the cube \textbf{good} if all eight vertices of $\Z^3$ it contains are covered by tetramers of a common orientation, and \textbf{bad} otherwise.
A bad cube contains a vacancy or a broken link local to it, while overlapping good cubes intersect tetramers of the same orientation.
By the chessboard estimate, we immediately deduce the existence of three orientational phases, one for each $o\in\cO$, in which good cubes intersecting $o$-oriented tetramers percolate on the grid $\Z^3$ and the others form only finite components (Proposition~\ref{prop: three orientations}).
This defines a unique orientation label $\Ori{\P}$ for every periodic-ergodic Gibbs measure $\P$.
In the sequel, we work with measures labeled $x_1x_2$.

\paragraph{Orientation-regularity in mesoscopic slabs}

The next step is to extend the Heilmann--Lieb characterization of orientational order to mesoscopic scales.
In our work on Models~I, II, and~V, we undertook similar extensions by deriving near-optimal estimates of the rarity of misaligned dimers in an orientationally ordered phase (see for instance~\cite[Theorem~4.2]{he2025extended}).
These estimates allowed us to exclude misaligned dimers, which couple adjacent one-dimensional columns (or rows), on scales long enough for vacancies to appear: vacancies decorrelate the columns, leading to the absence of positional order.
Here, we need the extension for a fundamentally different reason.
As noted earlier, the ground states of Model~III consist of fully decoupled layers.
Adjacent layers can only couple through \textit{transverse} tetramers which straddle them, and the minimum-energy defects containing such tetramers appear only on scales $\ell\gg e^{\frac{1}{2}\beta\min\set{\mu+8a,8a}}$ (Proposition~\ref{prop: minimum-energy defect components}).
Hence, we do not want to exclude such ``misalignments''; on the contrary, we \textit{need} them to appear in order to see their coupling effect on the layers.
We would therefore like to quantify the prevalence of good cubes on the scale $\ell$ (chosen in~\eqref{eqn: mesoscopic length scale}).

Consider a $2\ell\times2\ell\times 2$ slab and the $2\times2\times2$ cubes it contains.
We say that the slab is \textbf{orientation-regular} if there exists a high-density component of good cubes crossing the slab in both horizontal directions; the \textbf{dominant orientation} in the slab is then defined as the common orientation of the cubes in this component.
This definition is in line with similar ones employed in the context of high-density percolation~\cite{deuschel1996surface,barlow2004random} and helps avoid two pitfalls: requiring all cubes to be good would be too strong, since transverse tetramers \textit{are} expected to appear in the slab, whereas requiring only that a majority of cubes be good would be too weak, since the good cubes could then exist in components of different orientations.
The density threshold is chosen so that neighboring orientation-regular slabs on the grid $\ell\Z\times\ell\Z\times\Z$ contain intersecting good cubes, which then forces the slabs to have the same dominant orientation; consequently, slabs in an $x_1x_2$-oriented phase typically have dominant orientation $x_1x_2$ (Corollary~\ref{cor: mesoscopic slabs have a dominant orientation}).
Moreover, in an orientation-regular slab with dominant orientation $x_1x_2$, both of its $2\ell\times2\ell\times 1$ slices are mostly covered by $x_1x_2$-oriented tetramers.
The proof that orientation-regular slabs percolate on the grid $\ell\Z\times\ell\Z\times\Z$ is fairly straightforward: since the model permits no intra-layer sliding, every defect is penalized right where it occurs, and crude bounds based on applying the chessboard estimate to joint defect events already suffice (Proposition~\ref{prop: mesoscopic slabs have a dominant orientation}).

\subsubsection{Four parities}

\paragraph{Parity dominance in mesoscopic slices}

Recall that the \textbf{parity} of an $x_1x_2$-oriented tetramer $t$ centered at $(t_1,t_2,t_3)\in\R^3$ is defined as the pair $(t_1-1/2\bmod 2,\,t_2-1/2\bmod 2)\in\cP$.
Consider a $2\ell\times2\ell\times 1$ slice.
We say that the slice has \textbf{majority parity} $p\in\cP$ if most of the $x_1x_2$-oriented tetramers of parity $p$ intersecting the slice are present, and \textbf{dominant parity} $p$ if, in addition, the set of vertices in the slice covered by such tetramers has a component crossing the slice in both horizontal directions.
Since coplanar $x_1x_2$-oriented tetramers of different parities must be separated by broken links, an argument similar to the one for orientation-regularity shows that a slice of an orientation-regular slab with dominant orientation $x_1x_2$ typically has a dominant parity (Proposition~\ref{prop: mesoscopic slices have a dominant parity}).
As before, the density threshold used to define dominant parities is chosen so that horizontally neighboring slices on the grid $\ell\Z\times\ell\Z\times\Z$ cannot have different dominant parities.
It remains to prove the same for vertically neighboring slices, which is the heart of the proof.

\paragraph{A motivating counting problem}

If parallel tetramers in different layers do not interact, what could possibly synchronize the dominant parities of vertically adjacent slices?
The answer is entropic effects, mediated by the minimum-energy excitations involving transverse tetramers.
We illustrate the main idea here with a simple counting problem, as follows.

It is convenient to work under the shifted Hamiltonian alluded to earlier, where the energy of a configuration is given by $\frac{1}{4}(\mu+2a)$ times the number of vacancies plus $\frac{1}{2}a$ times the number of broken links, that is, sides of tetramers not linked to another tetramer.
Thus, a layered ground state has energy zero.
Moreover, any configuration that differs from such a ground state in finitely many tetramers and contains at least one transverse tetramer has energy at least $E_0=\min\set{\mu+8a,8a}$ (Proposition~\ref{prop: minimum-energy defect components}).
We can now state the problem:
\textit{Consider two adjacent layers in an $x_1x_2$-oriented ground state.
Fix a horizontal lattice site and remove the tetramer covering that site in each layer.
In how many ways can transverse tetramers be inserted into the resulting vacancies to produce a configuration of energy $E_0$?}

The answer depends on the parity difference between the two layers and the sign of $\mu$ (see Figure~\ref{fig: minimum energy defects} for illustrations of the respective configurations):
\begin{center}
    \begin{tabular}{lccc}
        \toprule
        & $\mu>0$ & $\mu=0$ & $-2a<\mu<0$ \\
        \midrule
        Parities differ in both coordinates & $0$ & $0$ & $0$ \\
        Parities differ in one coordinate & $0$ & $1$ & $1$ \\
        Parities agree & $2$ & $6$ & $4$ \\
        \bottomrule
    \end{tabular}
\end{center}
Observe that the last row is the largest in each column, which shows that, throughout the regime $\mu+2a>0$, layers with the same parity admit strictly more transverse defects of minimum energy than layers with differing parities.
Since each such defect is suppressed by a factor of $e^{-\beta E_0}$, this entropic advantage is weak, but it is sufficient to synchronize the parities in different layers.

\paragraph{The synchronization estimate} 

To turn this heuristic into a proof, we consider a $2\ell\times2\ell\times2$ slab $S$ and the event $M_{p,p'}$ that its bottom and top slices have majority parities $p$ and $p'$, respectively, where $p\ne p'$.
The main result here is the \textit{synchronization estimate} (Theorem~\ref{thm: synchronization estimate}), which bounds the chessboard seminorm\footnote{Here, we assume familiarity with the chessboard estimate; see Section~\ref{sec:reflection positivity} for a review.} of this event by
\begin{equation}
    \label{eqn: introduction - synchronization estimate}
    \norm{M_{p,p'}}_S
    \le \exp\set[\Big]{-e^{-\beta E_0}\ell^2/3}
    +8\ell(2\ell-1) 6^{2r} \parentheses[\big]{Ce^{-\beta a/2}}^{r/8},
\end{equation}
where $C>0$ is a universal constant and $r$ is a large integer independent of $\beta$ (see~\eqref{eqn: microscopic length scale}); with $\ell$ chosen as in~\eqref{eqn: mesoscopic length scale}, the right-hand side vanishes as $\beta\to\infty$.

To prove this estimate, we need to compare the partition function on a large torus constrained by the disseminated version of $M_{p,p'}$ with the unconstrained one.
Up to an inconsequential factor, we may replace the boundary condition of the former with that of the reference ground state with parity pattern $\dots,p,p',p',p,p,p',p',p,\dots$
The idea then is to construct a \textit{repair map} that synchronizes the parities of all layers and to show that the synchronized configuration admits more minimum-energy transverse defects than the original one.
The construction of the repair map is inspired by related ideas in~\cite{duminil2017exponential,galvin2004phase,peled2014odd} and consists of three steps:
\begin{enumerate}
    \item \label{itm: overview - repair step 1} Erasing every transverse defect containing either (a) a transverse tetramer straddling two layers of different reference parities or (b) two vertically stacked transverse tetramers, and restoring the reference tetramers in their place.
    \item \label{itm: overview - repair step 2} Translating every tetramer that intersects a layer of reference parity $p'$ by $(p-p',0)$.
    Thanks to Step~1, these translations are well-defined and do not create any new broken links.
    \item \label{itm: overview - repair step 3} Optionally and independently inserting minimum-energy transverse defects along the interfaces between formerly mismatched layers.
\end{enumerate}
By construction, every original configuration can be uniquely recovered from any of its repaired versions given the set of defects erased in Step~\ref{itm: overview - repair step 1} (Lemma~\ref{lem: repair map disjointness}).
Therefore, it suffices to show that the gain from Step~\ref{itm: overview - repair step 3} outweighs the information lost in Step~\ref{itm: overview - repair step 1}.
This requires a slight modification to the program, as we now explain.

To keep track of the amount of information lost in the erasure in Step~\ref{itm: overview - repair step 1}, we would like the transverse defects to satisfy a \textit{Peierls condition}: each defect should cost energy proportional to its size, with a constant of proportionality $c$ satisfying $\beta c\gg 1$ so as to beat the entropy of defects of a given size.
However, no such condition holds in general. 
Indeed, a simple counterexample is obtained by drawing a large cubic box in a ground state, and replacing everything inside it by the restriction of a ground state with a different orientation.
The size of this defect scales with the volume of the box, whereas its energy scales only with the surface area, since broken links and vacancies appear only near the surface.
We encountered a similar situation in~\cite{he2025extended,he2026nematic}, where a large cluster of misaligned dimers likewise costs nothing in its interior.
There, it sufficed to work with the boundaries of such clusters (see for instance~\cite[Proposition~4.6]{he2025extended}).
Unfortunately, no such shortcut is available here, since the repair map can only be inverted if every erased defect is recorded in full.
Nor does the disseminated majority condition remove the difficulty, since it merely excludes defects from occupying more than a fixed fraction of any $2\ell\times 2\ell\times 1$ slice.
That yields a Peierls constant of order $\ell^{-1}$, which is far too small since $\ell$ is exponentially large in $\beta$. 
Hence, the problematic defects are precisely those whose cross-sections in the slices are too small to violate the majority condition, yet too large for a Peierls condition with $\beta c\gg 1$.
It turns out that we can rule out all such defects by simply excluding, in each slice, \textit{chains} of broken links of length exceeding some $\beta$-independent threshold $r>0$.
This accounts for the second term in~\eqref{eqn: introduction - synchronization estimate} (Proposition~\ref{prop: long chains of broken links are rare}) and delivers a Peierls constant of order $r^{-1}$ (Lemma~\ref{lem: obstructive defect components}), so the rest of the program goes through unchanged, except with $M_{p,p'}$ replaced by the subevent in which neither slice of $S$ contains a long chain of broken links.

Thanks to the Peierls condition, comparing the gain from Step~\ref{itm: overview - repair step 3} with the information lost in Step~\ref{itm: overview - repair step 1} comes down almost exactly to the counting problem above (see~\eqref{eqn: motivating problem manifestation}) and leads to the first term in~\eqref{eqn: introduction - synchronization estimate}.

\subsubsection{Twelve phases}

The preceding steps provide all the ingredients needed to split each orientational phase into four according to the dominant parity of the tetramers (Proposition~\ref{prop: four parities}).
Together with the three orientations, this yields the twelve phase labels of Theorem~\ref{thm:main}.
The main remaining tasks, then, are to show that each phase label corresponds to exactly one (extremal) Gibbs measure, and that the model admits no other extremal and periodic Gibbs measure.
These are accomplished using the disagreement percolation method of van den Berg~\cite{van1993uniqueness}, which consists in showing that, almost surely, there is no infinite path on which independent samples from two Gibbs measures with the same phase label disagree.
We prove the stronger statement that the probability of a disagreement path connecting two distant plaquettes decays exponentially in their distance, with length scale $\ell$ in directions parallel to the layers and of order one perpendicular to them (Proposition~\ref{prop:disagreement percolation decay}).

By symmetry, we may specialize to the $x_1x_2$-oriented phase with dominant parity $(0,0)$.
The reference ground state here is the configuration in which every tetramer is $x_1x_2$-oriented and has parity $(0,0)$.
Every plaquette in the disagreement set of independent samples from two Gibbs measures in this phase forces the reference tetramers covering its vertices to be missing in at least one sample.
Hence, every disagreement component gives rise to a connected cluster of missing reference tetramers, and it suffices to show that such clusters rarely extend far.
Since the reference ground state is crystalline, each cluster is lined with broken links along its entire boundary within each layer, making a Peierls condition possible as in the proof of the synchronization estimate.
The crux lies in the cross-sections of clusters within slices, where the only danger is a cross-section stretching across a slice since its boundary may then lie far away.
We rule out this scenario using the crossings supplied by parity dominance, which force the cross-section to be surrounded by a number of broken links proportional to its diameter.
This leads to a Peierls-type condition (Lemma~\ref{lem: escape leads to broken links}) from which the required decay follows.
We note that this argument does not apply to Models~I, II, and~V, where sliding in fact allows disagreements to propagate for free along one-dimensional columns~\cite{he2025extended,he2026nematic}.

Finally, the microscopic characterizations (Theorem~\ref{thm:main}\ref{itm:main - tetramer probabilities}) rest on showing that each reference tetramer is missing with probability $\order{e^{-\beta a/2}}$.
By translation invariance, this probability equals the expected fraction of vertices in a $2\ell\times2\ell\times1$ slice that are not covered by reference tetramers.
On the typical event that the slice has the correct dominant parity and does not contain a chain of broken links of length $\ge r$, the number of such vertices is bounded by $r/2$ times the number of broken links in the slice.
Since each link is broken with probability $\order{e^{-\beta a/2}}$ by the chessboard estimate, this yields the desired bound.

\subsection{Organization of the paper}

Section~\ref{sec:preliminaries} introduces the basic definitions and tools used in the rest of the paper.
Section~\ref{sec: ground states and defects} classifies the ground states and minimum-energy transverse defects of the model.
Section~\ref{sec: orientational symmetry breaking} establishes orientational symmetry breaking, first following the Heilmann--Lieb approach in Section~\ref{sec: Heilmann-Lieb approach} and then extending it to mesoscopic scales in Section~\ref{sec: mesoscopic slabs have a dominant orientation}.
Section~\ref{sec: partial translational symmetry breaking} establishes partial translational symmetry breaking, with the repair map constructed and the synchronization estimate proven in Section~\ref{sec: parity synchronization between adjacent mesoscopic slices}.
Section~\ref{sec: columnar order} completes the proof of discotic columnar order, in particular showing that there is no other extremal and periodic Gibbs measure besides the twelve constructed in the previous sections.

\section{Preliminaries}
\label{sec:preliminaries}

\subsection{Basic definitions}
\label{sec:basic definitions}

\paragraph{Graphical constructions}

In this paper, we will consider three graphical constructions based on the (simple) cubic lattice $\Z^3$:
\begin{enumerate}
      \item The \textbf{plaquette graph} of $\Z^3$ has vertices corresponding to the plaquettes (faces) of $\Z^3$, and two vertices are connected by an edge if and only if their corresponding plaquettes intersect at a vertex or an edge.
      By an abuse of notation, we denote both the plaquette graph and its vertex set by $\F$, and identify each plaquette with its geometric center in $\R^3$, so that 
      \begin{equation}
            \label{eqn: identification of plaquettes with geometric centers}
            \F\equiv 
            \brackets[\big]{\parentheses[\big]{\tfrac12+\Z}\times\parentheses[\big]{\tfrac12+\Z}\times\Z}
            \cup \brackets[\big]{\Z\times\parentheses[\big]{\tfrac12+\Z}\times\parentheses[\big]{\tfrac12+\Z}}
            \cup \brackets[\big]{\parentheses[\big]{\tfrac12+\Z}\times\Z\times\parentheses[\big]{\tfrac12+\Z}}.
      \end{equation}
      \item The \textbf{attraction graph}, denoted by $\bA$, has vertex set $\F$, and two vertices are connected by an edge if and only if their corresponding plaquettes are coplanar and their geometric centers differ by $\pm2\vec{e}_i$ or $\pm2\vec{e}_j$, where $\vec{e}_i$ and $\vec{e}_j$ are the unit vectors of $\R^3$ parallel to the common plane of the plaquettes.
      \item An \emph{augmentation} of the plaquette graph of $\Z^3$, denoted by $\F^\ddag$, is obtained from $\F$ by adding an edge between each pair of plaquettes of $\Z^3$ whose corresponding vertices in the attraction graph $\bA$ are adjacent.
      We use the symbol $\ddag$ to denote the connectivity on $\F^\ddag$.
\end{enumerate}

Thus, each $t\in\F$ may be viewed both as a plaquette of $\Z^3$ and as a vertex of $\F$ and $\F^\ddag$.
Unless otherwise noted, we adopt the former viewpoint from here on, so, in particular, the terms ``vertex'' and ``edge'' refer to a vertex and an edge of $\Z^3$, respectively.
The main motivation for introducing the \emph{augmented plaquette graph} $\F^\ddag$ is that $\ddag$-connectivity makes the monomer-tetramer model studied in this paper into a Markov random field~\cite[(1.6)]{van1994disagreement} on $\F$.

We will also consider two connectivities on $\Z^d$, $d=2,3$.
The first, denoted by the symbol $\Box$, is the standard connectivity on $\Z^d$ induced by the $1$-norm on $\R^d$.
The second, denoted by the symbol $\boxtimes$, is induced by the sup-norm on $\R^d$ and is particularly useful for characterizing the connectivity of boundaries of $\Box$-connected subsets of $\Z^d$; see~\cite{timar2013boundary}.

\paragraph{Boxes}

In this paper, the term \textbf{box} refers to a closed, axis-aligned rectangular region in $\R^3$ with corners at the geometric centers of the cells of $\Z^3$: for $v_1,v_2,v_3\in 1/2+\Z$ and $L_1,L_2,L_3\in\Z_{>0}$, denote
\begin{equation}
    \rect{L_1\times L_2\times L_3}{(v_1,v_2,v_3)}\coloneqq[v_1,v_1+L_1]\times[v_2,v_2+L_2]\times[v_3,v_3+L_3]\subseteq\R^3.
\end{equation}
In particular, we write
\begin{equation}
      \mathrm{R}_{L_1\times L_2\times L_3}\coloneqq\rect{L_1\times L_2\times L_3}{(-1/2,-1/2,-1/2)}.
\end{equation}
A box is said to be \textbf{even} if its side lengths are all even.
If $R$ is a box, we write $\abs{R}$ for its volume, which coincides with the number of vertices of $\Z^3$ it contains.

\paragraph{Configurations}

A \textbf{tetramer} is a plaquette $t\in\F$ (so named since a plaquette has \textit{four} vertices), and its \textbf{orientation} is the unique label $o\in\set{x_1x_2,x_2x_3,x_3x_1}$ such that $t$ is parallel to the $o$-plane.
Two tetramers $t,t'\in\F$ are \textbf{incident} if they intersect at a vertex or an edge of $\Z^3$.
A \textbf{configuration} on $\F$ is a function $\sigma:\F\rightarrow\set{0,1}$ with the property that no two plaquettes in $\sigma^{-1}(\set{1})$ are incident.
We identify the function $\sigma$ with the set $\sigma^{-1}(\set{1})$.
Denote the set of all configurations on $\F$ by $\Omega$.
Equip $\Omega$ with the subspace topology inherited from the product topology on $\set{0,1}^\F$ and with the corresponding Borel measurable structure; denote the resulting $\sigma$-algebra by $\sF$.

Let $\Lambda\subseteq\R^3$ be a bounded region.
We introduce restricted sets of configurations given by various boundary conditions on $\Lambda$ (or, more precisely, on $\F\cap\Lambda$; see~\eqref{eqn: identification of plaquettes with geometric centers}):
\begin{enumerate}
    \item \emph{Prescribed boundary conditions}.
    Let $\rho\in\Omega$.
    Define the set of configurations in $\Lambda$ with $\rho$-boundary condition as
    \begin{equation}
        \label{eqn: prescribed boundary condition}
        \Omega^\rho_\Lambda\coloneqq\set{\sigma\in\Omega\mid\text{for all }t\in\F\setminus\Lambda,\,\sigma(t)=\rho(t)}.
    \end{equation}
    \item \emph{Periodic boundary conditions}.
    Suppose that $\Lambda$ is a box.
    Define the set of $\Lambda$-periodic configurations as
    \begin{equation}
        \Omega^\per_\Lambda\coloneqq\set[\big]{\sigma\in\Omega\mid\text{for all }t\in\F\text{ and }i\in\set{1,2,3},\,\sigma(t)=\sigma(t+\Side_i(\Lambda)\vec{e}_i)},
    \end{equation}
    where $\Side_i(\Lambda)$ is the $i$th side length of $\Lambda$.
\end{enumerate}

Finally, given a set $A\subseteq\R^3$ and a measurable function $f:\Omega\to\R$, we say that $f$ is \textbf{$A$-local} if the value of $f$ depends only on the restriction of a configuration to $A\cap\F$.

\subsection{The Hamiltonian and Gibbs measures}
\label{sec:the Hamiltonian}

Let $\sigma\in\Omega$, and recall that we have identified $\sigma\equiv\sigma^{-1}(\set{1})$.
We say that two tetramers $t,t'\in\sigma$ are \textbf{linked} if their corresponding vertices in the attraction graph $\bA$ are adjacent; in that case, we write $t\sim t'$.
Let $\beta>0$ be the inverse temperature, $\mu\in\R$ be the (tetramer) chemical potential, and $a>0$ be the strength of attraction between linked tetramers.
In the remainder of the paper, we will consider $\mu$ and $a$ to be fixed such that $\mu+2a>0$, and suppress them from the notation.
The formal Hamiltonian of the monomer-tetramer model, referred to as Model~III by Heilmann--Lieb~\cite{heilmann1979lattice}, is
\begin{equation}
    H(\sigma)\coloneqq-\mu\abs{\sigma}-a\sum_{\set{t,t'}\subseteq\sigma}\indicator{t\sim t'},
\end{equation}
where $\abs{\sigma}$ is the number of tetramers in the configuration $\sigma$.

In this paper, we study this model in the perturbative regime $\beta\gg1$.
To exploit the largeness of $\beta$, it is advantageous to shift the Hamiltonian by a constant multiple of the volume, as exemplified in the following formal rewriting, whose adaptation to finite volumes is straightforward:
\begin{equation}
    \label{eqn:shifted Hamiltonian monomer-tetramer}
    \frac{1}{4}(\mu+2a)\abs{\Z^3}+H(\sigma)
    =\frac{1}{4}(\mu+2a)(\abs{\Z^3}-4\abs{\sigma})+\frac{1}{2}a\sum_{t\in\sigma}\parentheses[\Bigg]{4-\sum_{t'\in\sigma}\indicator{t'\sim t}},
\end{equation}
where the terms on the right-hand side have the following intuitive interpretations:
\begin{enumerate}
    \item $\abs{\Z^3}-4\abs{\sigma}$ is the number of vacant vertices, or \textbf{vacancies} for short, which are vertices of $\Z^3$ not incident to any tetramer in $\sigma$.
    \item For each $t\in\sigma$, $4-\sum_{t'\in\sigma}\indicator{t'\sim t}$ is the number of sides of $t$ not linked to another tetramer in $\sigma$.
    Formally, we define a \textbf{broken link} as an edge in $\bA$ which is incident to exactly one tetramer in $\sigma$.
    Hence, $\sum_{t\in\sigma}\left(4-\sum_{t'\in\sigma}\indicator{t'\sim t}\right)$ is the total number of broken links of $\sigma$.
\end{enumerate}

For clarity, we recast~\eqref{eqn:shifted Hamiltonian monomer-tetramer} systematically using the standard notion of potentials (see~\cite[Section 6.3.2]{friedli2017statistical} or~\cite[(2.2)]{georgii2011gibbs}).
Let $B\subseteq\F$ be bounded (or $B\Subset\F$ for short).
\begin{enumerate}
    \item If $B=\set{p\mid p\text{ is incident to }v}$ for some vertex $v\in\Z^3$, define the $12$-body vacancy potential
    \begin{equation}
        \label{eqn:vacancy potential}
        \Phi_B(\sigma)\coloneqq\frac{1}{4}(\mu+2a)\indicator{v\text{ is a vacancy of }\sigma}.
    \end{equation}
    \item If $B$ has the form $\set{t_1,t_2}$ where $t_1$ and $t_2$ are adjacent in $\bA$, define the two-body broken-link potential
    \begin{equation}
        \label{eqn:broken-link potential}
        \Phi_B(\sigma)\coloneqq\frac{1}{2}a\indicator{\sigma(t_1)+\sigma(t_2)=1}.
    \end{equation}
    \item Otherwise, set $\Phi_B(\sigma)\coloneqq0$.
\end{enumerate}
Thus, the shifted formal Hamiltonian can be expressed as
\begin{equation}
    \sum_{B\Subset\F}\Phi_B(\sigma),
\end{equation}
whose adaptation to finite volumes is standard (see~\cite[(6.24)]{friedli2017statistical} or~\cite[(2.3)]{georgii2011gibbs}): if $\Lambda\subseteq\R^3$ is bounded and $\#$ denotes a boundary condition for $\Lambda$, then the finite-volume Hamiltonian for configurations in $\Lambda$ with $\#$-boundary condition is
\begin{equation}
    \label{eqn:finite-volume Hamiltonian}
    H^\#_{\Lambda}(\sigma)\coloneqq\sum_{\substack{B\Subset\F\\ B\cap\Lambda\ne\emptyset}}\Phi_B(\sigma).
\end{equation}
As an exception, we note that if $\#$ denotes the periodic boundary condition, then the sum over $B\Subset\F$ is instead restricted to a complete set of representatives of the equivalence classes of all $B\Subset\F$ under translations generated by $\set{\Side_i(\Lambda)\vec{e}_i\mid i\in\set{1,2,3}}$.

\paragraph{Gibbs measures}

Let $\Lambda\subseteq\R^3$ be bounded and $\#$ denote a boundary condition on $\Lambda$.
The weight of a configuration $\sigma\in\Omega^\#_\Lambda$ is defined as
\begin{equation}
    w^\#_{\Lambda;\beta}(\sigma)\coloneqq e^{-\beta H^\#_{\Lambda}(\sigma)}.
\end{equation}
Hence, each vacancy carries a penalty factor of $e^{-\frac{1}{4}\beta(\mu+2a)}$, whereas each broken link carries a penalty factor of $e^{-\frac{1}{2}\beta a}$.
The finite-volume Gibbs measure on $\Lambda$ with $\#$-boundary condition is given by
\begin{equation}
    \P^\#_{\Lambda;\beta}(\sigma)\coloneqq\frac{w^\#_{\Lambda;\beta}(\sigma)}{Z^\#_{\Lambda;\beta}}\quad\text{for }\sigma\in\Omega^\#_\Lambda,
\end{equation}
where $Z^\#_{\Lambda;\beta}\coloneqq\sum_{\sigma\in\Omega^\#_\Lambda}w^\#_{\Lambda;\beta}(\sigma)$ is the partition function.
Furthermore, given $E\subseteq\Omega$, we define the $E$-constrained partition function as
\begin{equation}
    \label{eqn: constrained partition function}
    Z^\#_{\Lambda;\beta}(E)\coloneqq\sum_{\sigma\in\Omega^\#_\Lambda\cap E}w^\#_{\Lambda;\beta}(\sigma).
\end{equation}
Lastly, we define infinite-volume Gibbs measures via the standard Dobrushin--Lanford--Ruelle formalism; see~\cite[Section 6.2.1]{friedli2017statistical} or~\cite[(2.9)]{georgii2011gibbs} for details.

\subsection{Reflection positivity and the chessboard estimate}
\label{sec:reflection positivity}

In this subsection, we formulate the reflection positivity of the monomer-tetramer model and use it to set up the chessboard estimate.
We will not give any proof below because the results presented here are direct adaptations of those in~\cite[Section~3]{hadas2026columnar} and~\cite[Section~2.3]{he2025extended} and the proofs transfer with little modification.

We begin with some notation.
Let $R=\rect{L_1\times L_2\times L_3}{z}$ be a box.
Define the \textbf{grid} of $R$ as $G^R\coloneqq(z_1+L_1\Z)\times(z_2+L_2\Z)\times(z_3+L_3\Z)$, and denote its origin-shifted version by $\cL^R\coloneqq L_1\Z\times L_2\Z\times L_3\Z$.
Let $T^R$ be the group generated by the reflections of $R$ across the axis-aligned planes that pass through points in $G^R$.
Thus, for each $v\in G^R$, there exists a unique isometry in $T^R$ that maps $R$ to $\rect{L_1\times L_2\times L_3}{v}$; we denote this isometry by $\tau_{R,v}$.
Given $f:\Omega\to\R$ and $\tau\in T^R$, define $(\tau f)(\sigma)\coloneqq f(\sigma\circ \tau)$.
Given a Gibbs measure $\P$ and an isometry $\tau:\R^3\to\R^3$, denote by $\tau\P$ the push-forward of $\P$ under $\tau$.
Given $z\in\Z^3$, let $\eta_{z}:\R^3\to\R^3$ denote the translation $v\mapsto v+z$.

It turns out that the monomer-tetramer model is \emph{reflection-positive} with respect to reflections across the edges of $\Z^3$~\cite[Section~3]{heilmann1979lattice}; for a proof of this fact in the setting of an analogously shifted Hamiltonian for a monomer-dimer model introduced in~\cite{heilmann1979lattice}, see~\cite[Proof of Lemma~2.2]{he2025extended}.

\begin{lemma}[Reflection positivity]
    Let $R=\rect{L_1\times L_2\times L_3}{z}$ be a box and $f:\Omega\to\R$ be $R$-local.
    If
    \begin{equation}
        (\Lambda,\tau)\in\set[\big]{
            (\mathrm{R}_{2L_1\times L_2\times L_3},\tau_{R,z+L_1\vec{e}_1}),
            (\mathrm{R}_{L_1\times 2L_2\times L_3},\tau_{R,z+L_2\vec{e}_2}),
            (\mathrm{R}_{L_1\times L_2\times 2L_3},\tau_{R,z+L_3\vec{e}_3})
        },
    \end{equation}
    then
    \begin{equation}
        \P_{\Lambda;\beta}^\per(f\cdot\tau f)\ge 0.
    \end{equation}
\end{lemma}

If $\Lambda$ is a box such that $2\Side_i(R)\mid\Side_i(\Lambda)$ for all $i\in\set{1,2,3}$, then we say that $R$ is a \textbf{block} of $\Lambda$.
In that case, set $T^R_\Lambda\coloneqq T^R/\cL^\Lambda$.
Given an $R$-local function $f:\Omega\to\R$, define its \textbf{$(R,\Lambda)$-chessboard seminorm} by 
\begin{equation}
    \label{eqn:finite-volume chessboard seminorm}
    \norm{f}_{R\mid\Lambda}\coloneqq\P_{\Lambda;\beta}^\per\parentheses[\Bigg]{\prod_{\tau\in T_\Lambda^R}\tau f}^{1/\abs{T_\Lambda^R}};
\end{equation}
the name ``seminorm'' is justified by the fact that $\norm{\cdot}_{R\mid\Lambda}$ indeed satisfies the standard properties of a seminorm~\cite[Proposition 3.3]{hadas2026columnar}.
A standard consequence of reflection positivity is the \emph{chessboard estimate}, which asserts that the expectation of a product of local functions of a particular form under $\P_{\Lambda;\beta}^\per$ factorizes as an upper bound into the product of their chessboard seminorms.

It is straightforward to extend the chessboard estimate to apply to finite products of \textit{periodic} Gibbs measures of the monomer-tetramer model by adapting the argument of Hadas--Peled~\cite[Proof of Proposition 3.8]{hadas2026columnar}, which has been done in~\cite{he2025extended} for a monomer-dimer model;
here, we say that a Gibbs measure is \textbf{periodic} if it is invariant under translations by a full-rank sublattice of $\Z^3$.
The results are as follows.

Let $k\ge 1$.
An element $(\sigma_i)_{i=1}^k\in\Omega^k$ may be thought of as a joint configuration that naturally specifies a map $\F\to\set{0,1}^k$ via $t\mapsto(\sigma_i(t))_{i=1}^k$.
Thus, given a box $R$, the notion of $R$-locality and transformation under each isometry $\tau\in T^R$ extend naturally to functions $\Omega^k\rightarrow\R$.
We equip $\Omega^k$ with the product $\sigma$-algebra $\sF^{\otimes k}$, and use the symbol $\otimes$ to denote the product of probability measures on $(\Omega,\sF)$.

Let $\Lambda$ be a box, $R$ be a block of $\Lambda$, and $f:\Omega^k\to\R$ be $R$-local.
Define
\begin{equation}
    \norm{f}_{R\mid\Lambda}\coloneqq(\otimes_{i=1}^k \P_{\Lambda;\beta}^\per)\parentheses[\Bigg]{\prod_{\tau\in T_\Lambda^R}\tau f}^{1/\abs{T_\Lambda^R}}.
\end{equation}
We note that the product measure $\otimes_{i=1}^k\P_{\Lambda;\beta}^\per$ remains reflection-positive with respect to reflections across the edges of $\Z^3$, so the quantity above satisfies the same seminorm properties as before.
Define the \textbf{$R$-chessboard seminorm} of $f$ by
\begin{equation}
    \label{eqn:infinite-volume chessboard seminorm}
    \norm{f}_R\coloneqq\limsup_{n\to\infty}\norm{f}_{R\mid\mathrm{R}_{n!\times n!\times n!}}.
\end{equation}

\begin{proposition}[Infinite-volume chessboard estimate]
    \label{prop: infinite-volume chessboard estimate}
    Let $R$ be a box. 
    Let $A\subseteq T^R$ be finite and $(f_\tau)_{\tau\in A}$ be $R$-local functions from $\Omega^k$ to $\R$.
    Then, given periodic Gibbs measures $(\P_i)_{i=1}^k$,
    \begin{equation}
        (\otimes_{i=1}^k \P_i)\parentheses[\Bigg]{\prod_{\tau\in A}\tau f_\tau}\le\prod_{\tau\in A}\norm{f_\tau}_R.
    \end{equation}
\end{proposition}

\begin{proposition}[Recursive chessboard estimate in infinite volume]
    \label{prop: infinite-volume recursive chessboard estimate}
    Let $R$ and $S$ be boxes, and suppose that the corners of $S$ are in $G^R$.
    Let $A\subseteq T^R$ be such that $\bigcup_{\tau\in A}\tau R\subseteq S$ and $(f_\tau)_{\tau\in A}$ be $R$-local functions.
    Then,
    \begin{equation}
        \norm{\prod_{\tau\in A}\tau f_\tau}_{S}\le\prod_{\tau\in A}\norm{f_\tau}_{R}.
    \end{equation}
\end{proposition}

\subsection{Strongly percolating sets}
\label{sec: strongly percolating sets}

A recurrent theme in our analysis is to take a local property of configurations and quantify its prevalence in typical configurations via percolation-type arguments.
In this subsection, we import the notion of strongly percolating sets from~\cite[Section 6]{hadas2026columnar}, which is well-suited for this purpose.
Some properties of these objects, also imported from previous works, may be found in Appendix~\ref{appx: properties of strongly percolating sets}.

We start with some notation.
Let $E\subseteq\Omega$ be an event, which should be thought of as a local property of configurations.
Define the random set $\es{E}:\Omega\to 2^{\Z^3}$ by 
\begin{equation}
    \es{E}(\sigma)\coloneqq\set{z\in\Z^3\mid\sigma\in\eta_{z}E},
\end{equation}
where we recall that $\eta_{z}$ denotes the translation by $z$.
Thus, $\es{E}$ is the random set of locations at which the local property $E$ holds.
Given $L_1,L_2,L_3\in\Z_{>0}$, define the random set
\begin{equation}
    \xre{L_1\times L_2\times L_3}{\es{E}}\coloneqq\set{z\in\Z^3\mid(L_1z_1,L_2z_2,L_3z_3)\in\es{E}},
\end{equation}
which tracks the manifestation of the local property $E$ on the scale of the rectangular grid $L_1\Z\times L_2\Z\times L_3\Z$.
The above definitions generalize naturally to \emph{joint} properties of multiple, independently sampled configurations, i.e., events $E\subseteq\Omega^k$, $k\ge 2$.

We now define strongly percolating sets.
Let $\eps_3$ be a small constant whose value we do not determine; see~\cite[Proof of Lemma~6.2]{hadas2026columnar} for its instance in two dimensions.
Thus, all the results in Appendix~\ref{appx: properties of strongly percolating sets} should be understood to hold provided that $\eps_3>0$ is sufficiently small.
Let $\eps\ge 0$.
A random set $B\subseteq\Z^3$ on a probability space $(\Omega,\sF,\P)$ is \textbf{$\eps$-rare} if, for all finite $A\subseteq\Z^3$, $\P(A\subseteq B)\le\eps^{\abs{A}}$.
A random set $\Psi\subseteq\Z^3$ on $(\Omega,\sF,\P)$ is \textbf{$\eps$-strongly percolating} if either $\eps\ge\eps_3$ or there exists an $\eps$-rare set $B$ such that $\Psi$ $\P$-almost surely contains an infinite $\Box$-component of $\Z^3\setminus B$.
Define
\begin{equation}
    \label{eqn: strong percolation parameter}
    \peierls{\P}(\Psi)\coloneqq\inf\set{\eps\ge 0\mid\Psi\text{ is }\eps\text{-strongly percolating}}.
\end{equation}
We mention here that the quantity $\peierls{\P}(\Psi)$ may be thought of as a measure of the size of $\Psi$, with smaller values indicating larger random sets; see~\cite[Section 6.1]{hadas2026columnar} for further discussion.

\subsection{Crossings in grid graphs}
\label{sec: crossings in grid graphs}

In this subsection, we formulate and prove three lemmas about crossings of grid graphs which will play a recurrent role in our analysis of the monomer-tetramer model.

Recall the construction of the \textbf{dual square lattice} $(\Z^2)^\ast$, which has vertices corresponding to the faces of $\Z^2$, and two vertices are connected by an edge if and only if their corresponding faces share an edge.
We note that $(\Z^2)^\ast$ coincides with the shifted square lattice $(1/2+\Z)^2$.
For $E\subseteq E(\Z^2)$, denote by $E^\ast$ the set of (dual) edges in $(\Z^2)^\ast$ bisecting the edges in $E$.

For $n\ge 1$, define the $n\times n$ \textbf{grid graph} $\sG_n$ as the induced subgraph of $\Z^2$ on the vertex set $\sV_n\coloneqq\set{0,\dots,n-1}^2$; denote its edge set by $\sE_n$.
Define the left and right \textbf{sides} of $\sG_n$ as the subsets of $\sV_n$ consisting of vertices with minimum and maximum $x_1$-coordinates, and define its bottom and top sides analogously.
A \textbf{horizontal crossing} of $\sG_n$ is a ($\Box$-)path in $\sG_n$ connecting the left to the right side, and a \textbf{vertical crossing} of $\sG_n$ is defined analogously.

\begin{lemma}
    \label{lem: vertex planarity}
    Let $n\ge 1$, $0\le r\le n-1$, and $V\subseteq \sV_n$.
    If no $\boxtimes$-component of $V$ contains more than $r$ vertices, then the following statements hold.
    \begin{enumerate}[label=(\roman*)]
        \item \label{itm: vertex planarity - crossings} There exists a (unique) $\Box$-component $C$ of $\sG_n\setminus V$ crossing $\sG_n$ in both directions.
        \item \label{itm: vertex planarity - cardinality} $\abs{C}\ge \abs{\sV_n}-r\abs{V}$.
    \end{enumerate} 
\end{lemma}

\begin{proof}
    We start with~\ref{itm: vertex planarity - crossings}.
    By~\cite[Lemma~5.9]{bollobas2006percolation}, if $\sG_n\setminus V$ does not contain a horizontal crossing of $\sG_n$, then $V$ contains a $\boxtimes$-path connecting the bottom and top sides of $\sG_n$, which is impossible since such a $\boxtimes$-path contains at least $n$ vertices.
    Hence, $\sG_n\setminus V$ contains a horizontal crossing.
    Similarly, it also contains a vertical crossing.
    Since a horizontal and a vertical crossing must intersect, these crossings belong to a common $\Box$-component $C$ of $\sG_n\setminus V$.
    The uniqueness of $C$ follows from the same intersection argument.

    We now prove~\ref{itm: vertex planarity - cardinality}.
    For each $\Box$-component $H$ of $\sV_n\setminus C$, let $\delta_C H\subseteq H$ consist of the vertices that are $\Box$-adjacent to $C$.
    Let $H$ be such a component.
    By the final assertion of~\cite[Lemma~2]{timar2013boundary}, $\delta_C H$ is $\boxtimes$-connected.
    Since $\delta_C H\subseteq V$, it follows that $\abs{\delta_C H}\le r$.
    Furthermore, it follows from the fact that $C$ contains crossings of $\sG_n$ in both directions that every row and column of $\sG_n$ that intersects $H$ also intersects $\delta_C H$, so the numbers of rows and columns intersecting $H$ are both at most $\abs{\delta_C H}$.
    Thus, $\abs{H}\le \abs{\delta_C H}^2\le r\abs{\delta_C H}$.
    Summing over all $\Box$-components $H$ of $\sV_n\setminus C$, we obtain
    \begin{equation}
        \abs{\sV_n\setminus C}\le r\,\sum_H \abs{\delta_C H}\le r\abs{V},
    \end{equation}
    as required.
\end{proof}

\begin{lemma}
    \label{lem: edge planarity}
    Let $n\ge 1$, $0\le r\le n-1$, and $E\subseteq \sE_n$.
    If every component $F^\ast$ of $E^\ast$ satisfies
    \begin{equation}
        \diam^\ast_\infty(F^\ast)
        \coloneqq\max_{u^\ast,v^\ast\in V(F^\ast)}\norm{u^\ast-v^\ast}_\infty\le r,
    \end{equation}
    then the following statements hold.
    \begin{enumerate}[label=(\roman*)]
        \item \label{itm: edge planarity - crossings} There exists a (unique) component $C$ of $\sG_n\setminus E$ crossing $\sG_n$ in both directions.
        \item \label{itm: edge planarity - cardinality} $\abs{C}\ge \abs{\sV_n}-\frac{r}{2}\abs{E}$.
    \end{enumerate} 
\end{lemma}

\begin{proof}
    We start with~\ref{itm: edge planarity - crossings}.
    By~\cite[Lemma~3]{bollobas2006short}, if $\sG_n\setminus E$ does not contain a horizontal crossing of $\sG_n$, then $E^\ast$ contains a component connecting the bottom and top sides of $\sG_n$, which is impossible since such a component has diameter $n$.
    Hence, $\sG_n\setminus E$ contains a horizontal crossing.
    Similarly, it also contains a vertical crossing.
    Since a horizontal and a vertical crossing must intersect, these crossings belong to a common $\Box$-component $C$ of $\sG_n\setminus E$.
    The uniqueness of $C$ follows from the same intersection argument.

    We now prove~\ref{itm: edge planarity - cardinality}.
    For each $\Box$-component $H$ of $\sG_n\setminus C$, let $\partial_C H\subseteq\sE_n$ consist of the edges with one endpoint in $H$ and the other in $C$.
    Let $H$ be such a component.
    It is straightforward that $\partial_C H$ is a minimal separating set of edges (in the sense of~\cite{timar2013boundary}) between $H$ and $C$.
    Observe that the cycle space of $\sG_n$ is generated by the unit squares of $\sG_n$.
    By~\cite[Lemma~1]{timar2013boundary}, for every partition of $\partial_C H$ into two nonempty sets, there exists a unit square that intersects both sets.
    Since the dual of every unit square is connected, it follows that $(\partial_C H)^\ast$ is connected.
    Since $\partial_C H\subseteq E$, we deduce that $\diam^\ast_\infty((\partial_C H)^\ast)\le r$.

    Let $w_H$ and $h_H$ be the numbers of columns and rows of $\sG_n$ intersecting $H$, respectively.
    It follows from the fact that $C$ contains crossings of $\sG_n$ in both directions that every row or column of $\sG_n$ that intersects $H$ also contains a horizontal or vertical edge of $\partial_C H$, so $\abs{\partial_C H}\ge w_H+h_H$ and $\max\set{w_H,h_H}\le r$.
    Thus,
    \begin{equation}
        \abs{H}\le w_H h_H\le \frac{1}{2}(w_H^2+h_H^2)\le \frac{r}{2}(w_H+h_H)\le \frac{r}{2}\abs{\partial_C H}.
    \end{equation}
    Summing over all $\Box$-components $H$ of $\sG_n\setminus C$, we obtain
    \begin{equation}
        \abs{\sV_n\setminus C}\le \frac{r}{2}\sum_H \abs{\partial_C H}\le \frac{r}{2}\abs{E},
    \end{equation}
    as required.
\end{proof}

\begin{lemma}
    \label{lem: vertex planarity component boundary}
    Let $n\ge 1$ and $V\subseteq \sV_n$.
    Suppose that $\sG_n\setminus V$ contains a $\Box$-component $C$ crossing $\sG_n$ in both directions.
    Let $H$ be a $\Box$-component of $V$, and let $\delta_C H\subseteq H$ consist of the vertices $\Box$-adjacent to the (unique) $\Box$-component of $\sG_n\setminus H$ containing $C$.
    Then, the following statements hold.
    \begin{enumerate}[label=(\roman*)]
        \item \label{itm: vertex planarity component boundary - connectivity} $\delta_C H$ is $\boxtimes$-connected.
        \item \label{itm: vertex planarity component boundary - cardinality} $\abs{\delta_C H}\ge 1+\diam_\infty(H)$, where $\diam_\infty(H)\coloneqq\max_{u,v\in H}\norm{u-v}_\infty$.
    \end{enumerate} 
\end{lemma}

\begin{proof}
    Statement~\ref{itm: vertex planarity component boundary - connectivity} follows immediately from~\cite[Lemma~2]{timar2013boundary} with $G=\sG_n$ and $G^+$ the graph on $\sV_n$ with $\boxtimes$-connectivity, applied to the $\Box$-component of $\sG_n\setminus H$ containing $C$ and any vertex of $H$.

    For~\ref{itm: vertex planarity component boundary - cardinality}, first observe that either the number of columns or the number of rows intersecting $H$ is equal to $1+\diam_\infty(H)$.
    Without loss of generality, suppose that it is the former.
    Since $C$ contains a horizontal crossing of $\sG_n$, every column of $\sG_n$ intersecting $H$ also intersects $C$, and so it contains a vertex of $\delta_C H$.
    Therefore, $\abs{\delta_C H}\ge 1+\diam_\infty(H)$.
\end{proof}

\section{Ground states and defects}
\label{sec: ground states and defects}

In this section, we classify the ground states of the monomer-tetramer model (in the sense of Footnote~\ref{fn: ground state}), as well as the minimum-energy excitations containing tetramers of a different orientation (in the sense made precise below).
These serve complementary roles: the ground states, of which there are uncountably many, constitute candidates for the low-temperature phases, while the excitations determine which ones are stabilized by thermal fluctuations.

\subsection{Classification of ground states}

First, we prove that every ground state of the monomer-tetramer model is layered, with each layer free to \textit{slide} within its plane independently of the others.
Thus, at zero temperature, energetic effects force a common orientation and, up to translation, fix the position of tetramers within each layer; however, they do not couple different layers.
This degeneracy is lifted only at positive temperature, through \textit{entropic} effects that align all adjacent layers (see Section~\ref{sec: parity synchronization between adjacent mesoscopic slices}).

\begin{proposition}
    \label{prop: every ground state is layered}
    For every ground state $\omega$, there exist $o=x_ix_{i+1}\in\cO$ and a function $\rho=(\rho_1,\rho_2):\Z\to\cP$ such that
    \begin{equation}
        \label{eqn: every ground state is layered}
        \omega=g_{o,\rho}\coloneqq\bigcup_{h\in\Z}\set[\big]{\brackets[\big]{2z_1+\rho_1(h)+\tfrac12}\vec{e}_i+\brackets[\big]{2z_2+\rho_2(h)+\tfrac12}\vec{e}_{i+1}+h\vec{e}_{i+2}\mid z_1,z_2\in\Z}.
    \end{equation}
    Conversely, every configuration of this form is a ground state.
\end{proposition}

\begin{proof}
    As the shifted Hamiltonian~\eqref{eqn:shifted Hamiltonian monomer-tetramer} makes clear, the ground states of the model are precisely the configurations without any vacancy or broken link.
    Suppose that $\omega$ is such a configuration.
    Then, $\omega$ contains a tetramer, which we may assume is $t=(1/2,1/2,0)$ by symmetry.
    Since $\omega$ has no broken links, $t$ must be linked to four tetramers in $\omega$, located at $\pm2\vec{e}_1$ and $\pm2\vec{e}_2$ from $t$, respectively.
    Repeating this argument shows that the plane $x_3=0$ is periodically packed with $x_1x_2$-oriented tetramers:
    \begin{equation}
        \label{eqn: ground states - layer 0}
        \set[\big]{\parentheses[\big]{2z_1+\tfrac12,2z_2+\tfrac12,0}\mid z_1,z_2\in\Z}\subseteq\omega.
    \end{equation}
    Now, the vertices $(0,0,\pm 1)$ are each incident to a tetramer in $\omega$, which is necessarily $x_1x_2$-oriented as otherwise~\eqref{eqn: ground states - layer 0} and the hard-core constraint would force $\omega$ to have a broken link.
    Repeating the same intraplane argument as before shows that the planes $x_3=\pm 1$ are each periodically packed with $x_1x_2$-oriented tetramers.
    Finally, repeating this argument for all $x_3\in\Z$ shows that $\omega$ is of the form~\eqref{eqn: every ground state is layered}.
    The converse is clear, as no configuration of the form~\eqref{eqn: every ground state is layered} has any vacancy or broken link.
\end{proof}

We state two simple consequences of this classification.

\begin{corollary}
    \label{cor: partition function periodic boundary}
    If $\Lambda$ is a box with all even side lengths, then $Z^\per_{\Lambda;\beta}\ge 1$.
\end{corollary}

\begin{proof}
    Since $\Lambda$ has all even side lengths, by Proposition~\ref{prop: every ground state is layered}, there exists at least one $\Lambda$-periodic ground state, which contributes a weight of $1$ to the partition function $Z^\per_{\Lambda;\beta}$.
\end{proof}

\begin{corollary}
    \label{cor: elementary defect chessboard seminorms}
    Let $v$ be a vertex of $\Z^3$ and $e$ be an edge of $\bA$.
    Let $R_v$ be the $1\times1\times1$ box centered at $v$, and $R_e$ be the $1\times2\times2$, $2\times1\times2$, or $2\times2\times1$ box centered at the midpoint of $e$ such that the short side of $R_e$ is perpendicular to the plaquettes at the endpoints of $e$.
    Then, 
    \begin{enumerate}[label=(\alph*)]
        \item \label{itm: vacancy chessboard seminorm} $\norm{v\text{ is a vacancy}}_{R_v}\le e^{-\frac{1}{4}\beta(\mu+2a)}$, and 
        \item \label{itm: broken link chessboard seminorm} there exists a universal constant $C_{\ref{cor: elementary defect chessboard seminorms}}>0$ such that $\norm{e\text{ is a broken link}}_{R_e}\le C_{\ref{cor: elementary defect chessboard seminorms}}e^{-\frac{1}{2}\beta a}$.
    \end{enumerate}
\end{corollary}

\begin{proof}
    Item~\ref{itm: vacancy chessboard seminorm} follows immediately from the definition of the chessboard seminorm and Corollary~\ref{cor: partition function periodic boundary}.
    For Item~\ref{itm: broken link chessboard seminorm}, note that there is a uniformly bounded number of configurations on $R_e$ that make $e$ into a broken link, each having $R_e$-chessboard seminorm no greater than $e^{-\frac{1}{2}\beta a}$.
    Item~\ref{itm: broken link chessboard seminorm} then follows by the subadditivity of the chessboard seminorm.
\end{proof}

\subsection{Defect components}
\label{sec: defect components}

Next, we characterize excitations above a fixed ground state $g$.
Specifically, we will group together local discrepancies between a configuration and $g$ into \textit{defect components}, and associate an energy to each defect component in such a way that the energy of a configuration is the sum of the energies of its defect components (see Proposition~\ref{prop: energy of defect components}).
Defect components containing only tetramers parallel to the layers of $g$ are confined to a single layer, while those containing \textit{transverse} tetramers straddle adjacent layers.
The latter so-called \textit{transverse defect components} are thus the only ones sensitive to the relative parities across different layers, and lay the cornerstone for the order-by-disorder mechanism treated in Section~\ref{sec: parity synchronization between adjacent mesoscopic slices}.

Formally, with respect to a reference ground state $g$, a \textbf{defect component} of a configuration $\sigma$ is a $\ddag$-component of the symmetric difference $g\triangle\sigma$.
Let $\Gamma^g(\sigma)$ be the set of defect components of $\sigma$ with respect to $g$.
For each $\gamma\in\Gamma^g(\sigma)$, distinguish the set $R_\gamma\coloneqq(g\setminus\sigma)\cap \gamma$ of \textbf{removed} tetramers from the set $I_\gamma\coloneqq(\sigma\setminus g)\cap \gamma$ of \textbf{inserted} tetramers; we interpret the latter as the set of tetramers \textit{contained} in $\gamma$.
If $\gamma$ is finite, define its \textbf{energy} as
\begin{equation}
    \label{eqn: energy of defect component}
    E(\gamma)\coloneqq(\mu+2a)(\abs{R_\gamma}-\abs{I_\gamma})+\frac{1}{2}a(\abs{\partial R_\gamma}+\abs{\partial I_\gamma}),
\end{equation}
where, for each $S\subseteq\F$, $\partial S$ denotes the edge boundary~\cite[Appendix~B.14]{friedli2017statistical} of $S$ in $\bA$.

\begin{lemma}
    \label{lem: partition of broken links}
    Let $g$ be a ground state, $\sigma\in\Omega$, and $\cA$ be the set of $\bA$-components of $R_\gamma$ and $I_\gamma$, where $\gamma$ ranges over $\Gamma^g(\sigma)$.
    Then, the collection $\set{\partial A\mid A\in\cA,\,\partial A\ne\emptyset}$ is a partition of the set of broken links of $\sigma$.
\end{lemma}

\begin{proof}
    It suffices to show that:
    \begin{enumerate}[label=(\roman*)]
        \item \label{itm: partition of broken links - forward inclusion} for all $A\in\cA$, every edge in $\partial A$ is a broken link of $\sigma$,
        \item \label{itm: partition of broken links - backward inclusion} every broken link of $\sigma$ is contained in $\partial A$ for some $A\in\cA$, and
        \item \label{itm: partition of broken links - disjointness} $\partial A\cap\partial A'=\emptyset$ for all distinct $A,A'\in\cA$.
    \end{enumerate}

    For~\ref{itm: partition of broken links - forward inclusion}, let $A\in\cA$ and $e=\set{t,t'}\in\partial A$, where $t\in A$ and $t'\not\in A$.
    There are two cases.
    First, if $t\in R_\gamma$ for some defect component $\gamma$, then $t'\in g$ by the definition of $\bA$-connectivity.
    In fact, $t'\in\sigma$, as otherwise $t'\in g\setminus\sigma$, contradicting the assumption that $t'\not\in A$.
    Second, if $t\in I_\gamma$ for some $\gamma$, then $t'\not\in\sigma$, as otherwise $t'\in A$.
    In both cases, one endpoint of $e$ is in $\sigma$ while the other one is not, so $e$ is a broken link of $\sigma$.
    
    For~\ref{itm: partition of broken links - backward inclusion}, suppose that $e=\set{t,t'}$ is a broken link of $\sigma$, where $t\in\sigma$ and $t'\not\in\sigma$.
    There are two cases.
    First, if $t\in g$, then $t'\in g\setminus\sigma$, so $t'\in R_\gamma$ for some defect component $\gamma$, and $e\in\partial R_\gamma$.
    Second, if $t\not\in g$, then $t\in I_\gamma$ for some $\gamma$, so $e\in\partial I_\gamma$.
    
    For~\ref{itm: partition of broken links - disjointness}, let $A,A'\in\cA$ be distinct.
    There are three cases.
    First, if $A$ and $A'$ are $\bA$-components of the same set $R_\gamma$ or $I_\gamma$ for some defect component $\gamma$, then the claim is clear.
    Second, if $A$ and $A'$ are, respectively, $\bA$-components of $R_\gamma$ and $R_{\gamma'}$, or of $I_\gamma$ and $I_{\gamma'}$, for distinct $\gamma$ and $\gamma'$, then $\partial A\cap\partial A'=\emptyset$, as otherwise $\gamma$ and $\gamma'$ would be $\bA$-adjacent, hence $\ddag$-adjacent.
    Third, $A$ and $A'$ are $\bA$-components of $R_\gamma$ and $I_{\gamma'}$ for not necessarily distinct $\gamma$ and $\gamma'$.
    Suppose that $\set{t,t'}\in\partial A\cap\partial A'$, where $t\in R_\gamma$ and $t'\in I_{\gamma'}$.
    Since $t\in g$, the definition of $\bA$-connectivity forces $t'\in g$, a contradiction.

    The proof is now complete.
\end{proof}

\begin{proposition}
    \label{prop: energy of defect components}
    Let $g=g_{o,\rho}$ be a ground state, $\Lambda\subseteq\R^3$ be bounded, and $\sigma\in\Omega_\Lambda^g$.
    Then, $\Gamma^g(\sigma)$ is finite, each $\gamma\in\Gamma^g(\sigma)$ is finite with $\abs{R_\gamma}\ge\abs{I_\gamma}$, and 
    \begin{equation}
        \label{eqn: energy of defect components}
        H_\Lambda^g(\sigma)=\sum_{\gamma\in\Gamma^g(\sigma)}E(\gamma).
    \end{equation}
\end{proposition}

\begin{proof}
    In this setting, the definition of the $g$-boundary condition~\eqref{eqn: prescribed boundary condition} implies that $g\triangle\sigma$ is finite, so $\Gamma^g(\sigma)$ is finite and each $\gamma\in\Gamma^g(\sigma)$ is finite.

    Next, fix $\gamma\in\Gamma^g(\sigma)$.
    By Proposition~\ref{prop: every ground state is layered}, every vertex incident to a tetramer $t\in I_\gamma$ is incident to a tetramer $t'\in g$.
    Then, $t$ and $t'$ are incident, so $t'\in R_\gamma$.
    Hence, the number of vertices incident to $I_\gamma$ is at most that of vertices incident to $R_\gamma$, and the difference is exactly $4(\abs{R_\gamma}-\abs{I_\gamma})$, which implies that $\abs{R_\gamma}\ge\abs{I_\gamma}$.
    Moreover, the number of vacancies in $\sigma$ is exactly $4\sum_{\gamma\in\Gamma^g(\sigma)}(\abs{R_\gamma}-\abs{I_\gamma})$.

    On the other hand, by Lemma~\ref{lem: partition of broken links}, the number of broken links of $\sigma$ is exactly $\sum_{\gamma\in\Gamma^g(\sigma)}(\abs{\partial R_\gamma}+\abs{\partial I_\gamma})$.
    We conclude that
    \begin{equation}
        H_\Lambda^g(\sigma)
        =\frac{1}{4}(\mu+2a)\sum_{\gamma\in\Gamma^g(\sigma)}4(\abs{R_\gamma}-\abs{I_\gamma})+\frac{1}{2}a\sum_{\gamma\in\Gamma^g(\sigma)}(\abs{\partial R_\gamma}+\abs{\partial I_\gamma})
        =\sum_{\gamma\in\Gamma^g(\sigma)}E(\gamma),
    \end{equation}
    as desired.
\end{proof}

With respect to $g_{o,\rho}$, we say that a tetramer $t\in\F$ is \textbf{transverse} if it is not $o$-oriented; by extension, a defect component $\gamma$ is \textbf{transverse} if it contains a transverse tetramer.
We have the following characterizations of transverse defect components.

\begin{figure}[tbp]
    \centering

    \begin{subfigure}[t]{0.32\textwidth}
        \centering
        \includegraphics[width=\linewidth]{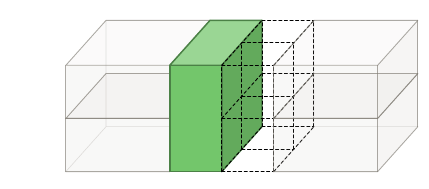}
        \caption{}
        \label{fig:defect-aligned-side}
    \end{subfigure}~
    \begin{subfigure}[t]{0.32\textwidth}
        \centering
        \includegraphics[width=\linewidth]{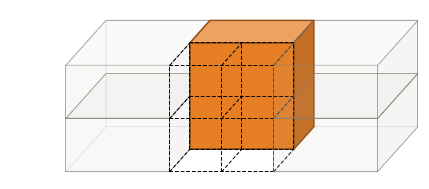}
        \caption{}
        \label{fig:defect-aligned-front}
    \end{subfigure}~
    \begin{subfigure}[t]{0.32\textwidth}
        \centering
        \includegraphics[width=\linewidth]{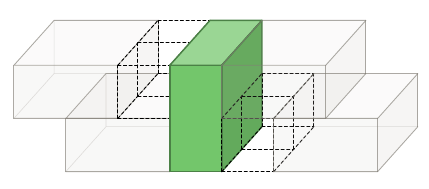}
        \caption{}
        \label{fig:defect-shifted}
    \end{subfigure}

    \par\medskip

    \begin{subfigure}[t]{0.32\textwidth}
        \centering
        \includegraphics[width=\linewidth]{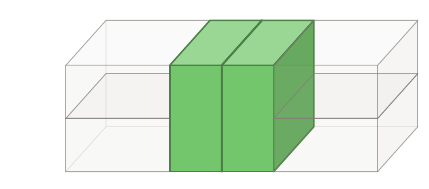}
        \caption{}
        \label{fig:defect-pair-side}
    \end{subfigure}~
    \begin{subfigure}[t]{0.32\textwidth}
        \centering
        \includegraphics[width=\linewidth]{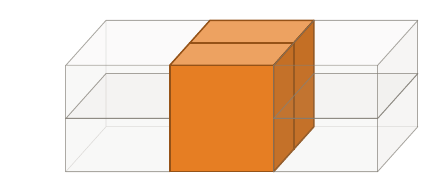}
        \caption{}
        \label{fig:defect-pair-front}
    \end{subfigure}

    \caption{Defect components of the monomer-tetramer model in which two reference tetramers in adjacent layers are replaced by \subref{fig:defect-aligned-side}--\subref{fig:defect-shifted}~one or \subref{fig:defect-pair-side},~\subref{fig:defect-pair-front}~two transverse tetramers.
    Among all defect components containing at least one transverse tetramer, these are, up to symmetry, the only ones of minimum energy: those with one transverse tetramer when $\mu\le 0$, and those with two when $\mu\ge 0$ (see Proposition~\ref{prop: minimum-energy defect components}).
    Transverse tetramers are shown in green and orange, vacancies as hollow cubes with dashed edges, and tetramers from the reference ground state in gray.}
    \label{fig: minimum energy defects}
\end{figure}

\begin{proposition}
    \label{prop: minimum-energy defect components}
    Let $g=g_{o,\rho}$ be a ground state, $\sigma\in\Omega$, and $\gamma\in\Gamma^g(\sigma)$ be bounded.
    Suppose that $\gamma$ is a transverse defect component.
    Then, the following statements hold.
    \begin{enumerate}[label=(\alph*)]
        \item \label{itm: minimum defect component - single} $\abs{\partial R_\gamma}+\abs{\partial I_\gamma}\ge 12$, with equality if and only if $\abs{R_\gamma}=2$ and $\abs{I_\gamma}=1$, in which case $E(\gamma)=E_1\coloneqq\mu+8a$; see Figures~\ref{fig:defect-aligned-side}--\ref{fig:defect-shifted}.
        \item \label{itm: minimum defect component - double} If $\abs{R_\gamma}=\abs{I_\gamma}$, then $\abs{\partial R_\gamma}+\abs{\partial I_\gamma}\ge 16$, with equality if and only if $R_\gamma$ and $I_\gamma$ are distinct pairs of opposite faces of the same cell of $\Z^3$, in which case $E(\gamma)=E_2\coloneqq8a$; see Figures~\ref{fig:defect-pair-side}--\ref{fig:defect-pair-front}.
        \item \label{itm: minimum defect component - gap} Write $E_0\coloneqq\min\set{E_1,E_2}$ and let
        \begin{equation}
            \kappa\coloneqq\begin{cases}
                \min\set{\mu,a} & \text{if }\mu>0 \\
                a & \text{if }\mu=0 \\
                \min\set{\mu+3a,-\mu} & \text{if }-2a<\mu<0
            \end{cases}
            >0.
        \end{equation}
        Every finite transverse defect component $\gamma$ satisfies $E(\gamma)\ge E_0+\kappa$, unless $\gamma$ realizes one of the equality cases in~\ref{itm: minimum defect component - single} or~\ref{itm: minimum defect component - double} whose energy equals $E_0$.
    \end{enumerate}
\end{proposition}

\begin{proof}
    Let $\cA_\gamma$ be the set of $\bA$-components of $R_\gamma$ and $I_\gamma$.
    Note that $\partial A\cap\partial A'=\emptyset$ for distinct $A,A'\in\cA_\gamma$ by Lemma~\ref{lem: partition of broken links}.
    Furthermore, since $\gamma$ is finite, each $A\in\cA_\gamma$ satisfies $\partial A\ne\emptyset$.
    In fact, it is easy to see that $\abs{\partial A}=4$ if and only if $A$ consists of a single tetramer, $\abs{\partial A}=6$ if and only if $A$ consists of two $\bA$-adjacent tetramers, and $\abs{\partial A}\ge 8$ otherwise.
    
    We start with~\ref{itm: minimum defect component - single}.
    Since $\gamma$ contains a transverse tetramer, $I_\gamma\ne\emptyset$, so $\abs{\partial I_\gamma}\ge 4$, with equality if and only if $\abs{I_\gamma}=1$.
    Moreover, this transverse tetramer intersects two layers of $g$ (the term \textit{layer} is understood in the context of Proposition~\ref{prop: every ground state is layered}), so $R_\gamma$ contains at least two tetramers in distinct layers of $g$.
    Hence, $\abs{\partial R_\gamma}\ge 8$, with equality only if $\abs{R_\gamma}=2$.
    Conversely, if $\abs{R_\gamma}=2$ and $\abs{I_\gamma}=1$, then, up to symmetry, $\gamma$ is necessarily of one of the types depicted in Figures~\ref{fig:defect-aligned-side}--\ref{fig:defect-shifted}, and in each case $\abs{\partial R_\gamma}+\abs{\partial I_\gamma}=12$.

    We now prove~\ref{itm: minimum defect component - double}.
    Suppose that $\abs{R_\gamma}=\abs{I_\gamma}$.
    By the previous paragraph, $\abs{R_\gamma}\ge 2$, so $\abs{I_\gamma}\ge 2$ as well.
    In the case that $\abs{R_\gamma}=\abs{I_\gamma}=2$, it is easy to check that the resulting defect component $\gamma$ is exactly as described in~\ref{itm: minimum defect component - double} and depicted in Figures~\ref{fig:defect-pair-side}--\ref{fig:defect-pair-front} and satisfies $\abs{\partial R_\gamma}+\abs{\partial I_\gamma}=16$ (which proves the converse).
    Otherwise, $\abs{R_\gamma}=\abs{I_\gamma}\ge 3$.
    Since $R_\gamma$ intersects at least two layers of $g$, it follows from the classification at the beginning of the proof that $\abs{\partial R_\gamma}\ge 4+6=10$.
    Similarly, $\abs{\partial I_\gamma}\ge 8$ when $\abs{I_\gamma}\ge 3$.
    Thus, $\abs{\partial R_\gamma}+\abs{\partial I_\gamma}\ge 18$ in this case, which completes the proof of the forward direction.

    Finally, we prove~\ref{itm: minimum defect component - gap}.
    Suppose that $\gamma$ is a finite transverse defect component.
    If $\gamma$ realizes one of the equality cases in~\ref{itm: minimum defect component - single} or~\ref{itm: minimum defect component - double} whose energy exceeds $E_0$, then $E(\gamma)-E_0=\abs{\mu}\ge\kappa$.
    Otherwise, suppose that $\gamma$ does not realize any of the equality cases.
    We split into three cases.
    \begin{enumerate}
        \item If $\abs{R_\gamma}=\abs{I_\gamma}$, then $\abs{\partial R_\gamma}+\abs{\partial I_\gamma}\ge 18$ by the proof of~\ref{itm: minimum defect component - double}.
        \item If $\abs{R_\gamma}=\abs{I_\gamma}+1$, then we must have that $\abs{I_\gamma}\ge 2$.
        Indeed, if $\abs{I_\gamma}=1$, then $\abs{R_\gamma}=2$, and $\gamma$ realizes the equality case in~\ref{itm: minimum defect component - single}, as argued in the proof of~\ref{itm: minimum defect component - single}.
        Since $\abs{I_\gamma}\ge 2$, $\abs{\partial I_\gamma}\ge 6$, and $\abs{R_\gamma}\ge 3$ so that $\abs{\partial R_\gamma}\ge 10$ as argued in the proof of~\ref{itm: minimum defect component - double}.
        Hence, $\abs{\partial R_\gamma}+\abs{\partial I_\gamma}\ge 16$.
        \item If $\abs{R_\gamma}\ge\abs{I_\gamma}+2$, then $\abs{R_\gamma}\ge 3$ and $\abs{\partial R_\gamma}\ge 10$ by the proof of~\ref{itm: minimum defect component - double}, and $\abs{\partial I_\gamma}\ge 4$ by the proof of~\ref{itm: minimum defect component - single}, so that $\abs{\partial R_\gamma}+\abs{\partial I_\gamma}\ge 14$.
    \end{enumerate}
    By applying the definition~\eqref{eqn: energy of defect component} of $E(\gamma)$, the proof is complete.
\end{proof}

\section{Orientational symmetry breaking}
\label{sec: orientational symmetry breaking}

\subsection{The Heilmann--Lieb approach}
\label{sec: Heilmann-Lieb approach}

First, we recast the Heilmann--Lieb characterization of orientational order in the monomer-tetramer model~\cite[Section~5]{heilmann1979lattice} under the strong-percolation framework of Section~\ref{sec: strongly percolating sets}.
In this model, the mechanism for orientational order is energetic: a $2\times2\times2$ cube intersecting tetramers of different orientations must contain a vacancy or a broken link local to it, which the chessboard estimate shows is exponentially unlikely at low temperature.
While the conclusion of this subsection is essentially the same as that of Heilmann--Lieb, its formulation in terms of the strong-percolation framework assigns an orientation label to each periodic-ergodic Gibbs measure, in a form suited for the arguments that follow.

Note that, by compactness, the monomer-tetramer model admits at least one $\Z^3$-invariant Gibbs measure $\P$, obtained as the weak limit of a sequence of finite-volume Gibbs measures with periodic boundary conditions, so that none of the statements below is vacuous.

Consider the $2\times2\times2$ cube $R=\mathrm{R}_{2\times 2\times 2}$. 
We say that $R$ is \textbf{bad} with respect to a configuration $\sigma$ if either (a) it contains a vacancy of $\sigma$, or (b) it does not contain any vacancy but intersects tetramers in $\sigma$ of different orientations, and denote by $Q$ the event that $R$ is bad.
It is easy to show by a case analysis that, in the latter case, at least one broken link of $\sigma$ is local to $R$, namely, there exists an edge $e\in\bA$ such that the event that $e$ is a broken link is $R$-local.
Applying a union bound over the finitely many restrictions to $R$ of configurations in which $R$ is bad, we find that there exists a universal constant $C_{\eqref{eqn: bad square chessboard seminorm}}\ge 1$ such that
\begin{equation}
    \label{eqn: bad square chessboard seminorm}
    \norm{Q}_R < C_{\eqref{eqn: bad square chessboard seminorm}} e^{-\beta\min\set{\mu+2a,2a}/4}.
\end{equation}

If $R$ is not bad, then every vertex of $\Z^3$ in $R$ is incident to a tetramer and these tetramers all have the same orientation $o\in\cO$.
Denote this $R$-local event by $G_o$, and introduce the random sets
\begin{equation}
    \label{eqn: Heilmann-Lieb good events}
    \Psi_{o}\coloneqq\es{G_{o}}=\xre{1\times1\times 1}{\es{G_{o}}}\quad\text{for }o\in\cO,\quad\text{and }
    \Psi\coloneqq\bigcup_{o\in\cO}\Psi_{o}.
\end{equation}

\begin{proposition}[Three orientations]
    \label{prop: three orientations}
    For all $\eps\in(0,\eps_3)$, there exists $\beta_0=\beta_0(\eps)>0$ such that, for all $\beta\ge\beta_0$ and periodic-ergodic Gibbs measures $\P$, the following holds.
    There exists a unique orientation label $\Ori{\P}\in\cO$ such that, out of the random sets $\set{\Psi_o\mid o\in\cO}$, $\Psi_{\Ori{\P}}$ is $\eps$-strongly percolating, while the others have only finite $\boxtimes$-components $\P$-a.s.
\end{proposition}

\begin{proof}
    By Lemma~\ref{lem:strongly percolating grid events}\ref{itm:strongly percolating grid events - chessboard bound}, for all periodic Gibbs measures $\P$,
    \begin{equation}
        \label{eqn: HL strong percolation parameter}
        \peierls{\P}(\Psi)\le \sqrt[8]{\norm{Q}_R}.
    \end{equation}
    On the other hand, notice that if $v\in\Psi_{o}$ and $w\in\Psi_{o'}$ for $o\ne o'$, then $v$ and $w$ are not $\square$-adjacent.
    Thus, by Lemma~\ref{lem:splitting strongly percolating sets}, for all periodic-ergodic Gibbs measures $\P$,
    \begin{equation}
        \min_{o\in\cO}\peierls{\P}(\Psi_{o})\le\peierls{\P}(\Psi).
    \end{equation}
    Let $\beta_0>0$ be such that $\sqrt[8]{C_{\eqref{eqn: bad square chessboard seminorm}}e^{-\beta_0\min\set{\mu+2a,2a}/4}}=\eps$.
    Combining~\eqref{eqn: bad square chessboard seminorm} and the above displays, we get that, for all $\beta\ge\beta_0$,
    \begin{equation}
        \label{eqn: at least one good event strongly percolates}
        \min_{o\in\cO}\peierls{\P}(\Psi_{o}) <\eps<\eps_3.
    \end{equation}
    Assume without loss of generality that $\peierls{\P}(\Psi_{x_1x_2})<\eps$, as the other cases are analogous.
    We need to show that $\Psi_{x_2x_3}$ and $\Psi_{x_3x_1}$ have finite $\boxtimes$-components $\P$-a.s.
    Fix $\eps'\in(\eps,\eps_3)$.
    By~\eqref{eqn: strong percolation parameter} and Lemma~\ref{lem:Peierls argument}, there exists an $\eps'$-rare set $B$ such that $\Psi_{x_1x_2}$ contains the unique infinite $\square$-component $I$ of $\Z^3\setminus B$, and each $\boxtimes$-component of $\Z^3\setminus I$ is finite, $\P$-a.s.
    Noting that $\Psi_{x_2x_3}\cup\Psi_{x_3x_1}\subseteq\Z^3\setminus\Psi_{x_1x_2}\subseteq\Z^3\setminus I$ completes the proof.
\end{proof}

For each periodic-ergodic Gibbs measure $\P$, denote by $\Ori{\P}\in\cO$ the orientation label on $\P$ supplied by Proposition~\ref{prop: three orientations}.
Note that the coordinate permutation $\varphi:(x_1,x_2,x_3)\mapsto(x_3,x_1,x_2)$ naturally induces an automorphism of $\cO$ given by $\varphi\cdot x_ix_{i+1}=x_{i+1}x_{i+2}$.
It is clear that if $\Ori{\P}=o$, then $\Ori{\varphi\P}=\varphi\cdot o$ and $\Ori{\varphi^2\P}=\varphi^2\cdot o$.

\subsection{Mesoscopic slabs have a dominant orientation}
\label{sec: mesoscopic slabs have a dominant orientation}

Next, we upgrade the Heilmann--Lieb characterization of periodic-ergodic Gibbs measures to a mesoscopic one.
This step is necessary because the order-by-disorder mechanism that we will establish in Section~\ref{sec: parity synchronization between adjacent mesoscopic slices} is weak, and only becomes effective on length scales $\ell$ with $\ell^2e^{-\beta E_0}\gg1$.
Specifically, inspired by an event considered by Barlow~\cite[(2.1)]{barlow2004random} in the context of high-density Bernoulli site percolation and itself attributed to Deuschel--Pisztora~\cite[Theorem~1.1]{deuschel1996surface}, we will prove the prevalence of mesoscopic slabs containing a high-density crossing component of good cubes.
While we specialize to the case of $x_1x_2$-oriented phases here, the results transfer immediately to the other orientations by coordinate permutation.
Throughout, let $\ell\in\Z_{>0}$ be a (large) mesoscopic length scale to be chosen later (see~\eqref{eqn: mesoscopic length scale}).

For each $\sigma\in\Omega$, define
\begin{equation}
    \cB_\sigma\coloneqq\set[\big]{(z_1,z_2)\in \Z^2\mid \eta_{{(z_1,z_2,0)}}\mathrm{R}_{2\times2\times2}\text{ is bad with respect to }\sigma}.
\end{equation}
Thus, each $\Box$-connected subset of $\Z^2\setminus\cB_\sigma$ inherits a unique orientation label from its associated \textit{good} cubes.
Consider the mesoscopic \textbf{slab} $S=\mathrm{R}_{2\ell\times2\ell\times 2}$.
Recall the grid graphs defined in Section~\ref{sec: crossings in grid graphs}.
We say that $S$ is \textbf{orientation-regular} with respect to $\sigma\in\Omega$ if there exists a $\Box$-component $C_\sigma$ of $\sG_{2\ell-1}\setminus\cB_\sigma$ with the following properties:
\begin{enumerate}
    \item \label{itm: orientational regularity - density} $\abs{C_\sigma}\ge \frac{63}{64}\abs{\sV_{2\ell-1}}$.
    \item \label{itm: orientational regularity - crossings} $C_\sigma$ contains both horizontal and vertical crossings of $\sG_{2\ell-1}$.
\end{enumerate}
By Condition~\ref{itm: orientational regularity - density}, $C_\sigma$ is unique if it exists.
In that case, define the \textbf{dominant orientation} in $S$ to be the orientation of $C_\sigma$.
For each $o\in\cO$, define the event
\begin{equation}
    G^\reg_o\coloneqq\set{\sigma\in\Omega\mid S\text{ is orientation-regular with dominant orientation }o},
\end{equation}
and write $G^\reg\coloneqq\bigcup_{o\in\cO}G^\reg_o$.
The main result of this subsection is as follows.

\begin{proposition}
    \label{prop: mesoscopic slabs have a dominant orientation}
    For all integers $1\le r\le\ell/2$, there exists $\beta_0=\beta_0(r)>0$ such that, for all $\beta\ge\beta_0$ and periodic Gibbs measures $\P$,
    \begin{equation}
        \begin{multlined}
            \peierls{\P}\parentheses{\xre{\ell\times \ell\times 1}{\es{G^\reg}}}
            \le \eps_\reg(\beta,r,\ell)
            \coloneqq(2\ell-1)^2 64^r 
            \parentheses[\big]{C_{\eqref{eqn: bad square chessboard seminorm}}e^{-\beta\min\set{\mu+2a,2a}/4}}^{r/32}
            \\
            +(64er)^{\ceil{(2\ell-1)^2/(64r)}}
            \parentheses[\big]{C_{\eqref{eqn: bad square chessboard seminorm}}e^{-\beta\min\set{\mu+2a,2a}/4}}^{(2\ell-1)^2/(2048r)}.
        \end{multlined}
    \end{equation}
\end{proposition}

Before delving into its proof, we note the following consequence of the proposition, which clarifies that orientation-regular slabs in an $o$-oriented phase typically have dominant orientation $o$, as is expected.

\begin{corollary}
    \label{cor: mesoscopic slabs have a dominant orientation}
    In the setting of Proposition~\ref{prop: mesoscopic slabs have a dominant orientation}, suppose that $\eps_\reg(\beta,r,\ell)<\eps_3$.
    Then, for all $\ell\Z^3$-ergodic Gibbs measures $\P$,
    \begin{equation}
        \peierls{\P}\parentheses{\xre{\ell\times \ell\times 1}{\es{G^\reg_{\Ori{\P}}}}}
        \le \eps_\reg(\beta,r,\ell).
    \end{equation}
    Moreover, for each $o\ne\Ori{\P}$, the random set $\xre{\ell\times \ell\times 1}{\es{G^\reg_{o}}}$ has finite $\boxtimes$-components $\P$-a.s.
\end{corollary}

\begin{proof}
    Let $o,o'\in\cO$ be distinct.
    It follows immediately from Condition~\ref{itm: orientational regularity - density} that $G^\reg_{o}\cap G^\reg_{o'}=\emptyset$ and $G^\reg_{o}\cap\eta_{\ell\vec{e}_i}G^\reg_{o'}=\eta_{\ell\vec{e}_i}G^\reg_{o}\cap G^\reg_{o'}=\emptyset$ for $i=1,2$, and from Condition~\ref{itm: orientational regularity - crossings} that $G^\reg_{o}\cap\eta_{\vec{e}_3}G^\reg_{o'}=\eta_{\vec{e}_3}G^\reg_{o}\cap G^\reg_{o'}=\emptyset$ as well.
    Hence, by Lemma~\ref{lem:splitting strongly percolating sets},
    \begin{equation}
        \label{eqn: splitting orientation regular events}
        \min_{o\in\cO}\peierls{\P}\parentheses{\xre{\ell\times \ell\times 1}{\es{G^\reg_o}}} \le \peierls{\P}\parentheses{\xre{\ell\times \ell\times 1}{\es{G^\reg}}}.
    \end{equation}
    Moreover, by the assumption $\eps_\reg(\beta,r,\ell)<\eps_3$, the same argument as in the proof of Proposition~\ref{prop: three orientations} shows that the minimum in~\eqref{eqn: splitting orientation regular events} is attained by a unique label $o\in\cO$.

    It remains to identify $o$ as $\Ori{\P}$.
    We temporarily introduce the shorthand $\Pi\coloneqq\xre{\ell\times \ell\times 1}{\es{G^\reg_o}}$.
    For each $\sigma\in\Omega$ and $u\in\Pi(\sigma)$, the slab $S$ is, by definition, orientation-regular with respect to the translated configuration $\eta_{(-\ell u_1,-\ell u_2,-u_3)}\sigma$; denote by $C_{\sigma,u}$ the corresponding $\Box$-component supplied by the definition of orientation regularity.
    Write $D_{\sigma,u}\coloneqq\eta_{(\ell u_1,\ell u_2,u_3)}(C_{\sigma,u}\times\set{0})$, and consider the random set
    \begin{equation}
        D_\sigma\coloneqq\bigcup_{u\in\Pi(\sigma)} D_{\sigma,u}.
    \end{equation}
    Note that $D_\sigma\subseteq\Psi_o(\sigma)$ (cf.~\eqref{eqn: Heilmann-Lieb good events}).
    Using Condition~\ref{itm: orientational regularity - density} as before, we find that $D_{\sigma,u}\cap D_{\sigma,v}\ne\emptyset$ for all $\set{u,v}\subseteq\Pi(\sigma)$ such that $u-v\in\set{\vec{e}_1,\vec{e}_2}$.
    On the other hand, if $u-v=\vec{e}_3$, we have that $C_{\sigma,u}\cap C_{\sigma,v}\ne\emptyset$ by Condition~\ref{itm: orientational regularity - crossings}, so $D_{\sigma,u}$ and $D_{\sigma,v}$ contain $\Box$-adjacent vertices.
    Hence, on the probability-one event that $\Pi(\sigma)$ contains an infinite $\Box$-component, so too does $D_\sigma$, but this forces $o=\Ori{\P}$ as a consequence of Proposition~\ref{prop: three orientations}.
\end{proof}

The rest of the subsection is devoted to the proof of Proposition~\ref{prop: mesoscopic slabs have a dominant orientation}.

\begin{lemma}
    \label{lem: defects associated with orientational irregularity}
    Let $1\le r\le\ell$ be an integer.
    If $S$ is not orientation-regular with respect to $\sigma\in\Omega$, then at least one of the following holds:
    \begin{enumerate}[label=(\roman*)]
        \item \label{itm: orientational irregularity - connected} $\cB_\sigma\cap \sV_{2\ell-1}$ contains a $\boxtimes$-connected subset of cardinality $r$.
        \item \label{itm: orientational irregularity - density} $\abs{\cB_\sigma\cap \sV_{2\ell-1}}\ge \abs{\sV_{2\ell-1}}/(64r)$.
    \end{enumerate}
\end{lemma}

\begin{proof}
    By contraposition, suppose that neither~\ref{itm: orientational irregularity - connected} nor~\ref{itm: orientational irregularity - density} holds.
    In the setting of Lemma~\ref{lem: vertex planarity}, let $n=2\ell-1$ and $V=\cB_\sigma\cap\sV_n$.
    The negation of~\ref{itm: orientational irregularity - connected} asserts that every $\boxtimes$-component $\gamma$ of $V$ satisfies $\abs{\gamma}\le r-1$.
    Hence, by Lemma~\ref{lem: vertex planarity}\ref{itm: vertex planarity - crossings}, $\sG_{n}\setminus V$ has a unique $\Box$-component $C_\sigma$ crossing $\sG_n$ in both directions.
    In particular, $C_\sigma\subseteq\sG_{n}\setminus\cB_\sigma$, as otherwise the horizontal crossing alone would contradict the negation of~\ref{itm: orientational irregularity - connected}; thus, Condition~\ref{itm: orientational regularity - crossings} is verified.
    On the other hand, by Lemma~\ref{lem: vertex planarity}\ref{itm: vertex planarity - cardinality} and the negation of~\ref{itm: orientational irregularity - density},
    \begin{equation}
        \abs{C_\sigma}
        \ge \abs{\sV_n}-r\abs{V}
        \ge \abs{\sV_n}-r\,\frac{\abs{\sV_n}}{64r}=\frac{63}{64}\abs{\sV_n},
    \end{equation}
    which verifies Condition~\ref{itm: orientational regularity - density}.
    We conclude that $S$ is orientation-regular, as required.
\end{proof}

We now deduce Proposition~\ref{prop: mesoscopic slabs have a dominant orientation}.

\begin{proof}[Proof of Proposition~\ref{prop: mesoscopic slabs have a dominant orientation}]
    By Lemma~\ref{lem:strong percolation of complement of rare set}, it suffices to prove that $\xre{\ell\times\ell\times 1}{\es{(G^\reg)^c}}$ is $\eps_\reg$-rare.
    Let $\sigma\in\Omega$ and $A\subseteq\Z^3$ be a finite set.
    Suppose that
    \begin{equation}
        \label{eqn: mesoscopic slabs have a dominant orientation - rare set containment}
        A\subseteq\xre{\ell\times\ell\times 1}{\es{(G^\reg)^c}}(\sigma).
    \end{equation}
    Let $0\le s\le\abs{A}$ be the number of points $u\in A$ such that Case~\ref{itm: orientational irregularity - connected} of Lemma~\ref{lem: defects associated with orientational irregularity} holds with respect to the translated configuration $\eta_{(-\ell u_1,-\ell u_2,-u_3)}\sigma$.
    In view of Lemma~\ref{lem: defects associated with orientational irregularity}, the inclusion~\eqref{eqn: mesoscopic slabs have a dominant orientation - rare set containment} dictates at least $sr+(\abs{A}-s)\abs{\sV_{2\ell-1}}/(64r)$ cubes to be bad, counting multiplicity, along with at most
    \begin{equation}
        \brackets[\Big]{\abs{\sV_{2\ell-1}} 8^{2r}}^s
        \binom{\abs{\sV_{2\ell-1}}}{\ceil{\abs{\sV_{2\ell-1}}/(64r)}}^{\abs{A}-s}
    \end{equation}
    choices for these cubes.
    Fix one such choice of cubes and remove all the duplicates.
    Since each $2\times2\times2$ cube is contained in $\eta_{(\ell v_1,\ell v_2,v_3)}S$ for at most four distinct points $v\in\Z^3$, at least $1/4$ of the original number of cubes remain.
    By the pigeonhole principle, there exists $z\in\set{0,1}^3$ such that at least $1/8$ of the remaining cubes have corners in $\eta_{z}(1/2+2\Z)^3$.
    Choose such a $z$ deterministically, and consider the corresponding subfamily of cubes.
    The event that every originally chosen cube is bad implies that every cube in the subfamily is bad.
    Hence, applying Proposition~\ref{prop: infinite-volume chessboard estimate} to the latter event, using~\eqref{eqn: bad square chessboard seminorm}, and applying a union bound over all original choices of cubes, we get that
    \begin{equation}
        \begin{multlined}
            \P\parentheses[\big]{A\subseteq\xre{\ell\times\ell\times 1}{\es{(G^\reg)^c}}}
            \le \sum_{s=0}^{\abs{A}}
            \binom{\abs{A}}{s}
            \brackets[\Big]{\abs{\sV_{2\ell-1}} 8^{2r}}^s
            \binom{\abs{\sV_{2\ell-1}}}{\ceil{\abs{\sV_{2\ell-1}}/(64r)}}^{\abs{A}-s}
            \\
            \times\parentheses[\big]{C_{\eqref{eqn: bad square chessboard seminorm}} e^{-\beta\min\set{\mu+2a,2a}/4}}^{\frac{1}{32}\brackets[\big]{sr+(\abs{A}-s)\frac{1}{64r}\abs{\sV_{2\ell-1}}}}.
        \end{multlined}
    \end{equation}
    Using the elementary inequality $\binom{n}{k}\le\parentheses{en/k}^k$ to bound the second bracket and applying the binomial theorem, we conclude that $\xre{\ell\times\ell\times 1}{\es{(G^\reg)^c}}$ is $\eps_\reg$-rare, as required.
\end{proof}

\section{Partial translational symmetry breaking}
\label{sec: partial translational symmetry breaking}

\subsection{Mesoscopic slices have a dominant parity}
\label{sec: mesoscopic slices have a dominant parity}

As a first step towards establishing partial translational symmetry breaking, we prove here that a slice in an orientation-regular slab with dominant orientation $x_1x_2$ is typically dominated by $x_1x_2$-oriented tetramers of a single parity.
As with orientational order, the mechanism is energetic: within the same layer, tetramers of different parities must be separated by broken links.
As before, the results transfer to the other orientations by coordinate permutation.

Consider the mesoscopic \textbf{slice} $L\coloneqq\mathrm{R}_{2\ell\times2\ell\times 1}$.
For $p\in\cP$, define $G^\maj_p$ as the event that at least $7/8$ of the $x_1x_2$-oriented tetramers of parity $p$ that intersect $L$ are present.
Define $G^\dom_p\subseteq G^\maj_p$ as the event that, in addition, the set of vertices in $L$ covered by such tetramers has a $\Box$-component that crosses $\sG_{2\ell}$ in both directions (in the sense of Section~\ref{sec: crossings in grid graphs}).
Such a component is unique if it exists, and, in that case, define the \textbf{dominant parity} in $L$ to be $p$.
Write $G^\dom\coloneqq\bigcup_{p\in\cP}G^\dom_p$ and $B_0\coloneqq G^\reg_{x_1x_2}\setminus G^\dom$.
The main result of this subsection is as follows.

\begin{proposition}
    \label{prop: mesoscopic slices have a dominant parity}
    There exists a constant $\beta_0>0$ such that, for all $\beta\ge\beta_0$, integers $32\le \ell\le e^{\beta(E_0+\kappa/4)}$ and $1\le r\le\ell$, and periodic Gibbs measures $\P$,
    \begin{equation}
        \begin{multlined}
            \peierls{\P}\parentheses{\xre{\ell\times \ell\times 1}{\es{B_0^c}}}
            \le\eps_\dom(\beta,r,\ell)
            \coloneqq4\ell(2\ell-1) 36^r 
            \parentheses[\big]{2^{32}C_{\ref{cor: elementary defect chessboard seminorms}}e^{-\beta a/2}}^{r/32}
            \\
            +(16er)^{\ceil{\ell(2\ell-1)/(4r)}}
            \parentheses[\big]{2^{32}C_{\ref{cor: elementary defect chessboard seminorms}}e^{-\beta a/2}}^{\ell(2\ell-1)/(128r)}.
        \end{multlined}
    \end{equation}
\end{proposition}

Before turning to its proof, we briefly comment on the role of Proposition~\ref{prop: mesoscopic slices have a dominant parity} in the proof of partial translational symmetry breaking.
The density condition defining $G^\maj_p$ implies that if $u,v\in\xre{\ell\times\ell\times 1}{\es{B_0^c}}$ differ by one of $\pm\vec{e}_1,\pm\vec{e}_2$, then the dominant parities in their respective slices, whenever both exist, must agree.
In view of Lemma~\ref{lem:splitting strongly percolating sets}, the existence of four phases within the $x_1x_2$-oriented phase, distinguished by their dominant parities, would therefore follow if this agreement also held for points that differ by $\pm\vec{e}_3$.
By the end of this subsection, the agreement property, or \textit{synchronization}, of dominant parities between vertically adjacent slices will be the only missing ingredient, which will then be addressed in Section~\ref{sec: parity synchronization between adjacent mesoscopic slices}.

To prove Proposition~\ref{prop: mesoscopic slices have a dominant parity}, we will first use Lemma~\ref{lem: edge planarity} to translate the failure of parity dominance into two geometric alternatives.
To this end, given $\sigma\in\Omega$, label each vertex of $\Z^2\times\set{0}$ by the parity of the $x_1x_2$-oriented tetramer in $\sigma$ that covers it, if one exists, or by the fixed label ``else'' otherwise.
Let $\cE_\sigma$ be the set of edges in $\Z^2\times\set{0}$ whose endpoints have different labels.
By construction, every edge in $\cE_\sigma$ connects a vertex covered by an $x_1x_2$-oriented tetramer to one that is not covered by an $x_1x_2$-oriented tetramer of the same parity (and possibly not by an $x_1x_2$-oriented tetramer at all), and thus corresponds to a broken link of $\sigma$.
More precisely, for each $e\in\cE_\sigma$, let $u_1$ be its lexicographically first endpoint that is covered by an $x_1x_2$-oriented tetramer $t\in\sigma$, and let $u_2$ be its other endpoint; we associate with $e$ the broken link $\set{t,t+2(u_2-u_1)}\in\bA$.
Note that at most two edges in $\cE_\sigma$ are associated with the same broken link.

We have the following analogue of Lemma~\ref{lem: defects associated with orientational irregularity}.

\begin{lemma}
    \label{lem: defects associated with parity non-dominance}
    Let $\ell\ge 32$ and $1\le r\le\ell$ be integers. 
    If $\sigma\in G^\reg_{x_1x_2}\setminus G^\dom$, then at least one of the following holds:
    \begin{enumerate}[label=(\roman*)]
        \item \label{itm: parity non-dominance - connected} $\cE_\sigma\cap\sE_{2\ell}$ contains $r$ edges whose dual edges form a connected graph.
        \item \label{itm: parity non-dominance - density} $\abs{\cE_\sigma\cap\sE_{2\ell}}\ge\abs{\sE_{2\ell}}/(16r)$.
    \end{enumerate}
\end{lemma}

\begin{proof}
    By contraposition, suppose that neither~\ref{itm: parity non-dominance - connected} nor~\ref{itm: parity non-dominance - density} holds.
    Our goal is to show that $\sigma\in (G^\reg_{x_1x_2})^c\cup G^\dom$.
    Suppose that $\sigma\in G^\reg_{x_1x_2}$.
    We need to show that $\sigma\in G^\dom$.

    In the setting of Lemma~\ref{lem: edge planarity}, let $n=2\ell$ and $E=\cE_\sigma\cap\sE_n$.
    The negation of~\ref{itm: parity non-dominance - connected} asserts that every subset $F\subseteq E$ with connected dual satisfies $\abs{F}\le r-1$, which implies that $\diam^\ast_\infty(F^\ast)\le r$.
    Hence, by Lemma~\ref{lem: edge planarity}, $\sG_{n}\setminus E$ has a unique $\Box$-component $C_\sigma$ crossing $\sG_n$ in both directions, and 
    \begin{align}
        \abs{C_\sigma}
        \ge{}&\abs{\sV_n}-\frac{r}{2}\abs{E}
        {}&&
        \nonumber
        \\
        \ge{}& \abs{\sV_n}-\frac{r}{2}\cdot\frac{\abs{\sE_n}}{16r}
        {}&& \text{by the negation of~\ref{itm: parity non-dominance - density}}
        \nonumber
        \\
        \label{eqn: parity non-dominance - density lower bound}
        \ge{}& \frac{15}{16}\abs{\sV_n}
        {}&& \text{since }\abs{\sE_n}=2n(n-1)\le 2\abs{\sV_n}.
    \end{align}
    By construction, all the vertices in $C_\sigma$ have the same label.
    Since $\sigma\in G^\reg_{x_1x_2}$, no more than $1/64$ of the vertices in $\sV_n$ have the label ``else'', so~\eqref{eqn: parity non-dominance - density lower bound} forces this common label to be some $p\in\cP$.
    Now, the number of $x_1x_2$-oriented tetramers of parity $p$ that intersect $L$ is at most $(\ell+1)^2$, while~\eqref{eqn: parity non-dominance - density lower bound} implies that at least $\abs{C_\sigma}/4\ge (15/64)\abs{\sV_n}$ of them are present in $\sigma$.
    Since
    \begin{equation}
        \frac{15\abs{\sV_n}}{64(\ell+1)^2}
        =\frac{15(2\ell)^2}{64(\ell+1)^2}
        \ge\frac{7}{8}
        \quad
        \text{for all }\ell\ge 32,
    \end{equation}
    we conclude that $\sigma\in G^\dom_p\subseteq G^\dom$, as desired.
\end{proof}

We now deduce Proposition~\ref{prop: mesoscopic slices have a dominant parity}.

\begin{proof}[Proof of Proposition~\ref{prop: mesoscopic slices have a dominant parity}]
    By Lemma~\ref{lem:strong percolation of complement of rare set}, it suffices to prove that $\xre{\ell\times\ell\times 1}{\es{B_0}}$ is $\eps_\dom$-rare.
    Let $\sigma\in\Omega$ and $A\subseteq\Z^3$ be a finite set.
    Suppose that
    \begin{equation}
        \label{eqn: mesoscopic slices have a dominant parity - rare set containment}
        A\subseteq\xre{\ell\times\ell\times 1}{\es{B_0}}(\sigma).
    \end{equation}
    Let $0\le s\le\abs{A}$ be the number of points $u\in A$ such that Case~\ref{itm: parity non-dominance - connected} of Lemma~\ref{lem: defects associated with parity non-dominance} holds with respect to the translated configuration $\eta_{(-\ell u_1,-\ell u_2,-u_3)}\sigma$.
    In view of Lemma~\ref{lem: defects associated with parity non-dominance}, the inclusion~\eqref{eqn: mesoscopic slices have a dominant parity - rare set containment} dictates at least $sr+(\abs{A}-s)\abs{\sE_{2\ell}}/(16r)$ edges whose endpoints have different labels, counting multiplicity, along with at most
    \begin{equation}
        \brackets[\Big]{\abs{\sE_{2\ell}} 6^{2r}}^s
        \binom{\abs{\sE_{2\ell}}}{\ceil{\abs{\sE_{2\ell}}/(16r)}}^{\abs{A}-s}
    \end{equation}
    choices for these edges.
    Fix one such choice of edges and remove all the duplicates.
    Since each of these edges is contained in $\eta_{(\ell v_1,\ell v_2,v_3)}L$ for at most four distinct points $v\in\Z^3$, at least $1/4$ of the original number of edges remain.
    Moreover, since all the broken links in question are between $x_1x_2$-oriented plaquettes, at each remaining edge, there are two choices for the side of the edge on which the broken link lies.
    Fix such a choice for each edge.
    Since at most two edges in $\cE_\sigma$ are associated with the same broken link, the number of distinct associated broken links is at least $1/8$ of the original number of edges.
    
    Recall the notation $R_e$ from the statement of Corollary~\ref{cor: elementary defect chessboard seminorms}.
    By the pigeonhole principle, there exists $z\in\set{0,1}^2\times\set{0}$ such that at least $1/4$ of the forced broken links $e$ are such that $R_e$ has corners in $\eta_{z}[(1/2+2\Z)^2\times(1/2+\Z)]$.
    Choose such a $z$ deterministically, and consider the resulting subfamily of broken links.
    Applying Proposition~\ref{prop: infinite-volume chessboard estimate} to the event that every $\bA$-edge in this subfamily is a broken link, using Corollary~\ref{cor: elementary defect chessboard seminorms}\ref{itm: broken link chessboard seminorm}, and applying a union bound over all original choices of edges as well as the sides of each edge on which the broken links lie, we get that
    \begin{equation}
        \begin{multlined}
            \P\parentheses[\big]{A\subseteq\xre{\ell\times\ell\times 1}{\es{B_0}}}
            \le \sum_{s=0}^{\abs{A}}
            \binom{\abs{A}}{s}
            \brackets[\Big]{\abs{\sE_{2\ell}} 6^{2r}}^s
            \brackets[\bigg]{\binom{\abs{\sE_{2\ell}}}{\ceil{\abs{\sE_{2\ell}}/(16r)}}}^{\abs{A}-s}
            \\
            \times 2^{sr+(\abs{A}-s)\abs{\sE_{2\ell}}/(16r)}\parentheses[\big]{C_{\ref{cor: elementary defect chessboard seminorms}}e^{-\beta a/2}}^{\frac{1}{32}\brackets[\big]{sr+(\abs{A}-s)\frac{1}{16r}\abs{\sE_{2\ell}}}}.
        \end{multlined}
    \end{equation}
    Using the elementary inequality $\binom{n}{k}\le\parentheses{en/k}^k$ to bound the second bracket and applying the binomial theorem, we conclude that, provided that $\beta$ is large enough so that $2^{32}C_{\ref{cor: elementary defect chessboard seminorms}}e^{-\beta a/2}\le 1$, the random set $\xre{\ell\times\ell\times 1}{\es{B_0}}$ is $\eps_\dom$-rare, as required.
\end{proof}

\subsection{Parity synchronization between adjacent mesoscopic slices}
\label{sec: parity synchronization between adjacent mesoscopic slices}

As anticipated in Section~\ref{sec: mesoscopic slices have a dominant parity}, we now carry out the most delicate step of the whole argument: proving that dominant parities in vertically adjacent slices, whenever both exist, typically agree.
Unlike orientational order and parity dominance, this synchronization is not driven by the energy, since parallel tetramers in different layers do not interact.
Instead, it is accomplished through an \textbf{order-by-disorder} effect: having matching dominant parities permits \textit{more} of the minimum-energy transverse defect components identified in Proposition~\ref{prop: minimum-energy defect components}.
Although each such component carries a penalty factor of $e^{-\beta E_0}$, the \textit{choice} of including or excluding it in adjacent slices with matching parities contributes a factor of $1+e^{-\beta E_0}$.
Heuristically, repeating these choices independently across the two slices yields a gain factor of order $\exp{C\ell^2 e^{-\beta E_0}}$, which diverges as $\beta\to\infty$ provided that $\ell^2 e^{-\beta E_0}\to\infty$.

Informed by this heuristic, we choose the \textit{mesoscopic} length scale
\begin{equation}
    \label{eqn: mesoscopic length scale}
    \ell\coloneqq24\ceil[\bigg]{\frac{e^{\beta(E_0+\kappa/8)}}{24}},
\end{equation}
on which the minimum-energy defect components contribute significantly, while the energy gap $\kappa$ ensures that higher-energy components contribute an exponentially smaller correction.
Moreover, we set, in the contexts of Corollary~\ref{cor: mesoscopic slabs have a dominant orientation} and Proposition~\ref{prop: mesoscopic slices have a dominant parity}, the \textit{microscopic} length scale
\begin{equation}
    \label{eqn: microscopic length scale}
    r\coloneqq\floor[\bigg]{\frac{1024(E_0+\kappa/8)}{\min\set{\mu+2a,2a}}}+1,
\end{equation}
so that $\lim_{\beta\to\infty}\eps_\reg(\beta,r,\ell)=\lim_{\beta\to\infty}\eps_\dom(\beta,r,\ell)=0$, which will be important in Section~\ref{sec: four parities}.

Recall the slab $S=\mathrm{R}_{2\ell\times 2\ell\times 2}$ introduced in Section~\ref{sec: mesoscopic slabs have a dominant orientation}.
For distinct $p,p'\in\cP$, define the event $M_{p,p'}\coloneqq G_p^\maj\cap\eta_{\vec{e}_3}G_{p'}^\maj$.
The main result of this subsection is as follows.

\begin{theorem}[Synchronization estimate]
    \label{thm: synchronization estimate}
    There exists a constant $\beta_0>0$ such that, for all $\beta\ge\beta_0$,
    \begin{equation}
        \norm{M_{p,p'}}_S
        \le \exp\set[\Big]{-e^{-\beta E_0}\ell^2/3}
        +8\ell(2\ell-1) 6^{2r} \parentheses[\big]{C_{\ref{cor: elementary defect chessboard seminorms}}e^{-\beta a/2}}^{r/8}.
    \end{equation}
\end{theorem}

We record the following consequence of the theorem for later use.
Define $B_3\coloneqq\bigcup_{p\ne p'\in\cP}M_{p,p'}$.

\begin{corollary}
    \label{cor: parity synchronization}
    There exists a constant $\beta_0>0$ such that, for all $\beta\ge\beta_0$ and periodic Gibbs measures $\P$,
    \begin{equation}
        \peierls{\P}\parentheses{\xre{\ell\times \ell\times 1}{\es{B_3^c}}} 
        \le \eps_{\syn}(\beta,r,\ell)
        \coloneqq\parentheses[\Big]{12\exp\set[\Big]{-e^{-\beta E_0}\ell^2/3}
        +96\ell(2\ell-1) 6^{2r} \parentheses[\big]{C_{\ref{cor: elementary defect chessboard seminorms}}e^{-\beta a/2}}^{r/8}}^{1/8}.
    \end{equation}
\end{corollary}

\begin{proof}
    Note that the event $B_3$ is invariant under reflections across all three axis-aligned planes passing through the center of $\mathrm{R}_{2\ell\times 2\ell\times 2}$.
    Thus,
    \begin{align}
        {}& \peierls{\P}\parentheses{\xre{\ell\times \ell\times 1}{\es{B_3^c}}} 
        {}&&
        \nonumber
        \\
        \le {}& \norm{B_3}_S^{1/8}
        {}&& \text{by Lemma~\ref{lem:strongly percolating grid events}}
        \nonumber
        \\
        \le {}& \parentheses[\bigg]{\sum_{p\ne p'\in\cP}\norm{M_{p,p'}}_S}^{1/8}
        {}&& \text{by subadditivity}
        \nonumber
        \\
        \le {}& \parentheses[\Big]{12\exp\set[\Big]{-e^{-\beta E_0}\ell^2/3}
        +96\ell(2\ell-1) 6^{2r} \parentheses[\big]{C_{\ref{cor: elementary defect chessboard seminorms}}e^{-\beta a/2}}^{r/8}}^{1/8}
        {}&& \text{by Theorem~\ref{thm: synchronization estimate}}
        ,
    \end{align}
    as required.
\end{proof}

The rest of this subsection is dedicated to the proof of Theorem~\ref{thm: synchronization estimate}.

\subsubsection{Excluding long chains of broken links (Step 0)}
\label{sec: long chains of broken links}

Fix $p\in\cP$.
Given $\sigma\in\Omega$, label each vertex of $\Z^2\times\set{0}$ by $p$ if it is covered by an $x_1x_2$-oriented tetramer of parity $p$, or by the fixed label ``else'' otherwise.
Let $\cE_{\sigma,p}$ be the set of edges in $\Z^2\times\set{0}$ whose endpoints have different labels.
For each $e\in\cE_{\sigma,p}$, let $u_1$ be the endpoint of $e$ labeled $p$, $u_2$ the other endpoint, and $t\in\sigma$ the tetramer covering $u_1$; we associate with $e$ the broken link $\set{t,t+2(u_2-u_1)}$, noting that this association (which is different from the one in Section~\ref{sec: mesoscopic slices have a dominant parity}) is also at most two-to-one.
Recall the mesoscopic slice $L$ introduced in Section~\ref{sec: mesoscopic slices have a dominant parity}, and define the $L$-local event
\begin{equation}
    \label{eqn: exclusion of long chains of broken links}
    B^{\mathrm{long}}_{p,r}\coloneqq\set{\sigma\in\Omega\mid\text{there exists }F\subseteq\cE_{\sigma,p}\cap\sE_{2\ell}\text{ with }\abs{F}=r\text{ and }F^\ast\text{ connected}}.
\end{equation}

\begin{proposition}
    \label{prop: long chains of broken links are rare}
    For all $\beta>0$, integers $1\le r\le\ell$, and $p\in\cP$,
    \begin{equation}
        \dbars[\big]{B^{\mathrm{long}}_{p,r}}_L
        \le 4\ell(2\ell-1) 6^{2r} \parentheses[\big]{C_{\ref{cor: elementary defect chessboard seminorms}}e^{-\beta a/2}}^{r/8}.
    \end{equation}
\end{proposition}

\begin{proof}
    The proof is a slight variation of that of Proposition~\ref{prop: mesoscopic slices have a dominant parity}.
    Let $n$ be sufficiently large so that $L$ is a block of $\Lambda_n\coloneqq\mathrm{R}_{n!\times n!\times n!}$ (see Section~\ref{sec:reflection positivity}).
    By definition, 
    \begin{equation}
        \dbars[\big]{B^{\mathrm{long}}_{p,r}}_{L\mid\Lambda_n}
        =\P^\per_{\Lambda_n;\beta}\parentheses[\Bigg]{\bigcap_{\tau\in T_{\Lambda_n}^L}\tau B^{\mathrm{long}}_{p,r}}^{1/\bars{T_{\Lambda_n}^L}}.
    \end{equation}
    The disseminated event $\bigcap_{\tau\in T_{\Lambda_n}^L}\tau B^{\mathrm{long}}_{p,r}$ dictates at least $r\bars{T_{\Lambda_n}^L}$ \textit{distinct} edges whose endpoints have different labels, along with at most $\parentheses{\abs{\sE_{2\ell}}6^{2r}}^{\bars{T_{\Lambda_n}^L}}$ choices for these edges.
    Fix one such choice of edges.
    Since at most two edges in $\cE_{\sigma,p}$ are associated with the same broken link, the number of distinct associated broken links is at least $1/2$ of the original number of edges.
    Moreover, since all the broken links are between $x_1x_2$-oriented plaquettes, by the pigeonhole principle, there exists $z\in\set{0,1}^2\times\set{0}$ such that at least $1/4$ of the forced broken links $e$ are such that $R_e$ has corners in $\eta_{z}[(1/2+2\Z)^2\times(1/2+\Z)]$.
    Choose such a $z$ deterministically, and consider the corresponding subfamily of broken links.
    Applying the finite-volume version of Proposition~\ref{prop: infinite-volume chessboard estimate} to the event that every $\bA$-edge in this subfamily is a broken link, using the finite-volume version of Corollary~\ref{cor: elementary defect chessboard seminorms}\ref{itm: broken link chessboard seminorm}, and applying a union bound over all original choices of edges, we get that
    \begin{equation}
        \P^\per_{\Lambda_n;\beta}\parentheses[\Bigg]{\bigcap_{\tau\in T_{\Lambda_n}^L}\tau B^{\mathrm{long}}_{p,r}}
        \le \parentheses[\big]{\abs{\sE_{2\ell}}6^{2r}}^{\bars{T_{\Lambda_n}^L}}
        \parentheses[\big]{C_{\ref{cor: elementary defect chessboard seminorms}}e^{-\beta a/2}}^{\frac{1}{8}r\bars{T_{\Lambda_n}^L}}.
    \end{equation}
    Taking the limit superior of the $\bars{T_{\Lambda_n}^L}$-th root of both sides as $n\to\infty$ completes the proof.
\end{proof}

For distinct $p,p'\in\cP$, let
\begin{equation}
    M^{\mathrm{short}}_{p,p'}\coloneqq M_{p,p'}\setminus\parentheses[\big]{B^{\mathrm{long}}_{p,r}\cup\eta_{\vec{e}_3}B^{\mathrm{long}}_{p',r}}.
\end{equation}
Since
\begin{equation}
    M_{p,p'}\subseteq M^{\mathrm{short}}_{p,p'}
    \cup B^{\mathrm{long}}_{p,r}
    \cup\eta_{\vec{e}_3}B^{\mathrm{long}}_{p',r},
\end{equation}
we have that
\begin{equation}
    \dbars[\big]{M_{p,p'}}_S
    \le \dbars[\big]{M^{\mathrm{short}}_{p,p'}}_S+8\ell(2\ell-1) 6^{2r} \parentheses[\big]{C_{\ref{cor: elementary defect chessboard seminorms}}e^{-\beta a/2}}^{r/8}
\end{equation}
by Propositions~\ref{prop: infinite-volume recursive chessboard estimate} and~\ref{prop: long chains of broken links are rare}.
Therefore, to prove Theorem~\ref{thm: synchronization estimate}, it suffices to show that
\begin{equation}
    \dbars[\big]{M^{\mathrm{short}}_{p,p'}}_S\le \exp\set[\Big]{-e^{-\beta E_0}\ell^2/3}.
\end{equation}

\subsubsection{The setup}
\label{sec: synchronization - setup}

For the rest of the subsection, fix distinct $p,p'\in\cP$, write $\Lambda_n\coloneqq\mathrm{R}_{n!\times n!\times n!}$, and assume that $n$ is sufficiently large so that $S$ is a block of $\Lambda_n$.
Let 
\begin{equation}
    \widetilde{T}^S_{\Lambda_n}\coloneqq\set[\big]{\tau_{S,v},\,\tau_{S,v}\eta_{\vec{e}_3}\mid v\in G_S,\,\tau_{S,v}S\subseteq\Lambda_n}
\end{equation}
be the set of transformations that map $L$ onto the slices of the reflected copies of $S$ in $\Lambda_n$.
By the definition~\eqref{eqn:infinite-volume chessboard seminorm} of the chessboard seminorm, the synchronization estimate boils down to bounding a ratio of constrained (recall~\eqref{eqn: constrained partition function}) and unconstrained partition functions on $\Lambda_n$ as $n\to\infty$:
\begin{equation}
    \label{eqn: synchronization - setup - chessboard seminorm}
    \dbars[\big]{M^{\mathrm{short}}_{p,p'}}_S
    =\limsup_{n\to\infty} \brackets[\Bigg]{\frac{Z^\per_{\Lambda_n}\parentheses[\big]{D_n\parentheses[\big]{M^{\mathrm{short}}_{p,p'}}}}{Z^\per_{\Lambda_n}}}^{\abs{S}/\abs{\Lambda_n}},
\end{equation}
where we have suppressed $\beta$ from the notation for the partition function and introduced the shorthand 
\begin{equation}
    \label{eqn: synchronization - setup - disseminated event}
    D_n\parentheses[\big]{M^{\mathrm{short}}_{p,p'}}\coloneqq\bigcap_{\substack{v\in G_S\\\tau_{S,v}S\subseteq\Lambda_n}}\tau_{S,v} M^{\mathrm{short}}_{p,p'}
\end{equation}
for the disseminated version\footnote{
    In view of the definition~\eqref{eqn:finite-volume chessboard seminorm} of the (finite-volume) chessboard seminorm, it may seem more natural to define the disseminated event as $\bigcap_{\tau\in T^S_{\Lambda_n}}\tau M^{\mathrm{short}}_{p,p'}$.
    However, since $T^S_{\Lambda_n}=T^S/\cL^{\Lambda_n}$ consists of \textit{cosets} of transformations, the intersection over $\tau\in T^S_{\Lambda_n}$ is only meaningful for $\Lambda_n$-periodic configurations, where the choice of coset representatives is irrelevant.
    Later, we will need to apply the disseminated event to configurations that are not $\Lambda_n$-periodic, which is why we use the definition~\eqref{eqn: synchronization - setup - disseminated event} instead.
    Under the periodic boundary condition on $\Lambda_n$, the two definitions agree, so the identity~\eqref{eqn: synchronization - setup - chessboard seminorm} still holds.
} of $M^{\mathrm{short}}_{p,p'}$; for later use, we also define $D_n(M_{p,p'})$ analogously.
The dissemination assigns a unique \textbf{reference parity} to each $x_1x_2$-oriented layer of $\Z^3$ via the periodic boundary condition on $\Lambda_n$, and these parities uniquely determine a (layered) ground state $g$.
The repair map we will use is most naturally defined for configurations with the $g$-boundary condition rather than the periodic one.
To switch to the former, we use the following analogue of the map in~\cite[Proposition~2.1]{hadas2026columnar}.
Given a box $R$ and $\rho\in\Omega$, define $m^\rho_R:\Omega\to\Omega^\rho_R$ as follows: let $m^\rho_R(\sigma)$ be obtained from $\sigma$ by first setting $\sigma\equiv\rho$ on $\F\setminus R$, and then removing all tetramers in $R$ that are incident to another tetramer.

\begin{lemma}
    \label{lem: boundary modification preserves bulk constraint}
    For all $n$ such that $S$ is a block of $\Lambda_n$, the map $m^g_{\Lambda_n}$ satisfies
    \begin{equation}
        m^g_{\Lambda_n}\parentheses[\big]{D_n\parentheses[\big]{M^{\mathrm{short}}_{p,p'}}}\subseteq D_n\parentheses[\big]{M^{\mathrm{short}}_{p,p'}}.
    \end{equation}
\end{lemma}

\begin{proof}
    It is clear that the map $m^g_{\Lambda_n}$ preserves the majority parity in each disseminated slice $\tau L$, $\tau\in\widetilde{T}_{\Lambda_n}^S$.
    Thus,
    \begin{equation}
        m^g_{\Lambda_n}\parentheses[\big]{D_n\parentheses[\big]{M_{p,p'}}}\subseteq D_n\parentheses[\big]{M_{p,p'}}.
    \end{equation}
    Hence, it suffices to show that, for all $\tau\in\widetilde{T}_{\Lambda_n}^S$ and $\sigma\in D_n(M_{p,p'})$, the set of edges $e\in\sE_{2\ell}$ whose endpoints have different labels with respect to $\tau^{-1}m^g_{\Lambda_n}(\sigma)$ is contained in the corresponding set for $\tau^{-1}\sigma$.
    Let $e$ be in the former set.
    Since $m^g_{\Lambda_n}$ preserves every tetramer in $\sigma\cap g$ by construction, the map preserves every label equal to the reference parity in the layer intersecting $\tau L$, so the endpoints of $e$ cannot both carry this label with respect to $\tau^{-1}\sigma$.
    On the other hand, since every tetramer in $m^g_{\Lambda_n}(\sigma)\setminus\sigma$ belongs to $g\cap(\F\setminus\Lambda_n)$ and covers no vertex in $\Lambda_n$, the map also preserves the label ``else'', so the endpoints of $e$ cannot both be labeled ``else'' with respect to $\tau^{-1}\sigma$ either.
    Therefore, the endpoints of $e$ have different labels with respect to $\tau^{-1}\sigma$, as required.
\end{proof}

By Lemma~\ref{lem: boundary modification preserves bulk constraint} and a standard comparison of boundary conditions (for instance, see~\cite[Proposition~2.1]{hadas2026columnar}, whose proof applies here with minimal changes), we may replace the periodic boundary condition in the constrained partition function with the $g$-boundary condition, up to a boundary error that vanishes in the limit $n\to\infty$.
We now seek an upper bound on the ratio
\begin{equation}
    \label{eqn: synchronization estimate - goal}
    \frac{Z^g_{\Lambda_n}\parentheses[\big]{D_n\parentheses[\big]{M^{\mathrm{short}}_{p,p'}}}}{Z^\per_{\Lambda_n}}.
\end{equation}

\begin{figure}[tbp]
    \centering

    \begin{subfigure}[t]{0.45\textwidth}
        \centering
        \includegraphics[width=\linewidth]{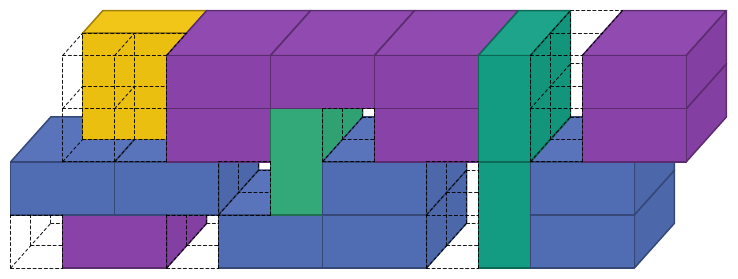}
        \caption{}
        \label{fig: repair map - original}
    \end{subfigure}
    \hspace{0.02\textwidth}
    \begin{subfigure}[t]{0.45\textwidth}
        \centering
        \includegraphics[width=\linewidth]{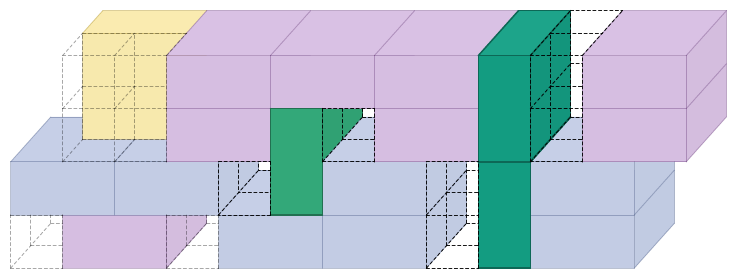}
        \caption{}
        \label{fig: repair map - obstructions}
    \end{subfigure}

    \par\medskip

    \begin{subfigure}[t]{0.45\textwidth}
        \centering
        \includegraphics[width=\linewidth]{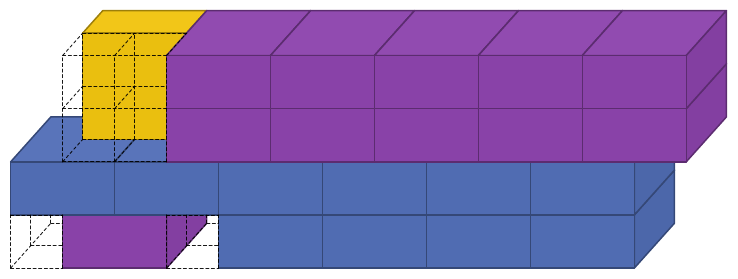}
        \caption{}
        \label{fig: repair map - erased}
    \end{subfigure}
    \hspace{0.02\textwidth}
    \begin{subfigure}[t]{0.45\textwidth}
        \centering
        \includegraphics[width=\linewidth]{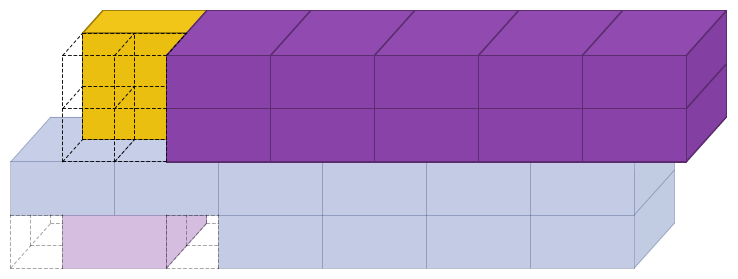}
        \caption{}
        \label{fig: repair map - moving layers}
    \end{subfigure}

    \par\medskip

    \begin{subfigure}[t]{0.45\textwidth}
        \centering
        \includegraphics[width=\linewidth]{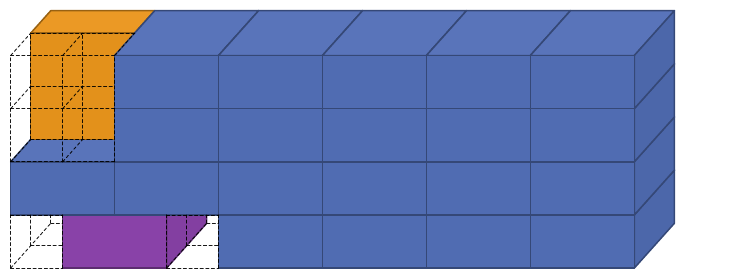}
        \caption{}
        \label{fig: repair map - translated}
    \end{subfigure}
    \hspace{0.02\textwidth}
    \begin{subfigure}[t]{0.45\textwidth}
        \centering
        \includegraphics[width=\linewidth]{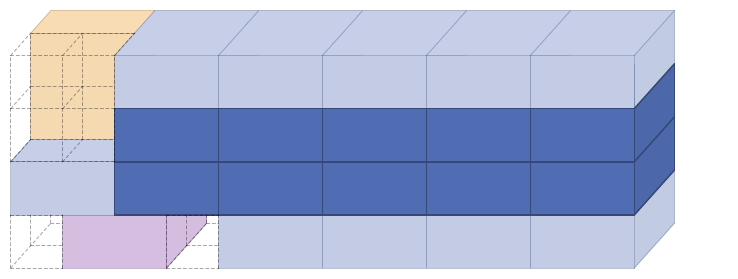}
        \caption{}
        \label{fig: repair map - available}
    \end{subfigure}

    \par\medskip

    \begin{subfigure}[t]{0.45\textwidth}
        \centering
        \includegraphics[width=\linewidth]{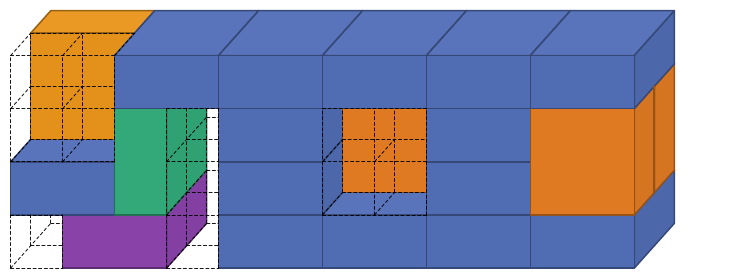}
        \caption{}
        \label{fig: repair map - replacement 1}
    \end{subfigure}
    \hspace{0.02\textwidth}
    \begin{subfigure}[t]{0.45\textwidth}
        \centering
        \includegraphics[width=\linewidth]{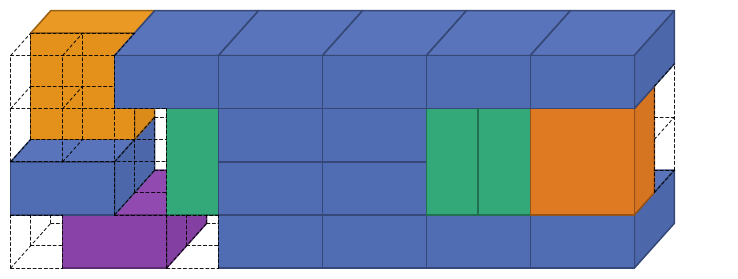}
        \caption{}
        \label{fig: repair map - replacement 2}
    \end{subfigure}

    \caption{The three steps of the repair map $\fR=\fI\circ\fT\circ\fE$, illustrated on \subref{fig: repair map - original}~a portion of four adjacent layers of a configuration $\sigma$, in which the bottom two layers have reference parity $p=(0,1)$ and the top two $p'=(1,1)$.
    In Step~\ref{itm: step - remove transverse components}, the obstructive defect components, one of Type~\ref{itm: obstruction - incompatible translations} and one of Type~\ref{itm: obstruction - attractive link}, are \subref{fig: repair map - obstructions}~identified and \subref{fig: repair map - erased}~erased, yielding $\fE(\sigma)$.
    In Step~\ref{itm: step - synchronize parities}, the tetramers of $\fE(\sigma)$ intersecting layers of reference parity $p'$ are \subref{fig: repair map - moving layers}~identified and \subref{fig: repair map - translated}~translated by $(p-p',0)$, yielding $(\fT\circ\fE)(\sigma)$.
    In Step~\ref{itm: step - insert minimum-energy defects}, the available cells, which straddle the interface between formerly mismatched layers, are \subref{fig: repair map - available}~identified, and \subref{fig: repair map - replacement 1}, \subref{fig: repair map - replacement 2}~minimum-energy transverse defect components are optionally and independently inserted into each one, yielding the set of configurations $\fR(\sigma)$.
    Here, we take $\mu=0$, so that each cell may receive one or two transverse tetramers.}
    \label{fig: repair map}
\end{figure}

Our strategy for estimating~\eqref{eqn: synchronization estimate - goal} combines ideas from the works of Duminil-Copin--Peled--Samotij--Spinka~\cite{duminil2017exponential} on the loop $O(n)$ model (see also the lecture notes by Peled--Spinka~\cite{peled2019lectures}) and Galvin--Kahn~\cite{galvin2004phase} and Peled--Samotij~\cite{peled2014odd} on the hard-core model on $\Z^d$.
At the heart of the argument is a \textit{repair map} that acts on a configuration $\sigma\in \Omega^g_{\Lambda_n}\cap D_n\parentheses[\big]{M^{\mathrm{short}}_{p,p'}}$ in three steps:
\begin{enumerate}
    \item \label{itm: step - remove transverse components} for a subset of transverse defect components $\gamma$, it removes $I_\gamma$ from $\sigma$ and reinstates $R_\gamma$;
    \item \label{itm: step - synchronize parities} it translates all layers of reference parity $p'$ by $(p-p',0)$; and
    \item \label{itm: step - insert minimum-energy defects} it optionally inserts the minimum-energy defect components identified in Proposition~\ref{prop: minimum-energy defect components}.
\end{enumerate}
See Figure~\ref{fig: repair map} for a step-by-step illustration.
We note that Steps~\ref{itm: step - remove transverse components} and~\ref{itm: step - synchronize parities} have counterparts in the repair map constructed in~\cite[Section~2.3]{duminil2017exponential}, while the optional insertions in Step~\ref{itm: step - insert minimum-energy defects} parallel the operations defined in~\cite[Section~2.5]{galvin2004phase} and~\cite[Sections~4.4.1--4.4.2]{peled2014odd}.
Historically, the idea of using a repair map to facilitate the comparison of partition functions has a precedent in the work of Dobrushin on the hard-core model~\cite[Sections~5--6]{dobrushin1968problem}.

\subsubsection{Repairing obstructive defects (Step~\ref{itm: step - remove transverse components})}

Let $\sigma\in\Omega^g_{\Lambda_n}$.
The $g$-boundary condition ensures that $\Gamma^g(\sigma)$ is finite and every $\gamma\in\Gamma^g(\sigma)$ is finite.
We call a transverse defect component $\gamma$ (with respect to $g$; see Section~\ref{sec: defect components}) \textbf{obstructive} if at least one of the following conditions holds:
\begin{enumerate}
    \item \label{itm: obstruction - incompatible translations} there exists a transverse tetramer $t\in I_\gamma$ intersecting layers of both reference parities;
    \item \label{itm: obstruction - attractive link} there exists a transverse tetramer $t\in I_\gamma$ such that $t'=t+2\vec{e}_3\in I_\gamma$.
\end{enumerate}
Let $\Gamma^g_\obs(\sigma)\subseteq\Gamma^g(\sigma)$ consist of the obstructive defect components.
Each $\gamma\in\Gamma^g_\obs(\sigma)$ \textit{obstructs} the layer-wise translations in Step~\ref{itm: step - synchronize parities} as follows.
If $\gamma$ is of Type~\ref{itm: obstruction - incompatible translations}, the tetramer $t$ intersects two layers, only one of which receives a nontrivial translation, leaving no consistent way to translate $t$ itself.
If $\gamma$ is of Type~\ref{itm: obstruction - attractive link} but not Type~\ref{itm: obstruction - incompatible translations}, the tetramers $t$ and $t'$ each receive a well-defined translation, but the translations sever the attractive link between them and might create new broken links.

To remove these obstructions, we erase each $\gamma\in\Gamma^g_\obs(\sigma)$ by removing $I_\gamma$ from $\sigma$ and reinstating $R_\gamma$.
Formally, we define the \textbf{erasure map} 
\begin{equation}
    \fE:\Omega_{\Lambda_n}^g\to\Omega_{\Lambda_n}^g,\quad
    \fE(\sigma)\coloneqq\brackets[\bigg]{\sigma\setminus\bigcup_{\gamma\in\Gamma^g_\obs(\sigma)}I_\gamma}\cup\bigcup_{\gamma\in\Gamma^g_\obs(\sigma)}R_\gamma.
\end{equation}
The layer-wise translations in Step~\ref{itm: step - synchronize parities} can then be applied to $\fE(\sigma)$ unambiguously, without introducing any additional broken links.
See Figures~\ref{fig: repair map - original}--\ref{fig: repair map - erased}.

Write
\begin{equation}
    \sR^g_n\coloneqq\set[\big]{\Gamma^g_\obs(\sigma)\mid\sigma\in\Omega^g_{\Lambda_n}\cap D_n\parentheses[\big]{M^{\mathrm{short}}_{p,p'}}}.
\end{equation}
Let $N^{\min}_n$ be the number of distinct obstructive defect components $\gamma$ with $E(\gamma)=E_0$ occurring among all $\Gamma\in\sR^g_n$.
The main result of this sub-subsection is as follows.

\begin{proposition}
    \label{prop: information erased by repair}
    There exist constants $\beta_0,C_{\ref{prop: information erased by repair}}>0$ such that, for all $\beta\ge\beta_0$, $\ell\ge r$, and sufficiently large $n$,
    \begin{equation}
        \sum_{\Gamma\in\sR^g_n}\prod_{\gamma\in\Gamma}e^{-\beta E(\gamma)}
        \le \parentheses[\big]{1+e^{-\beta E_0}}^{N_n^{\min}}\exp{C_{\ref{prop: information erased by repair}} e^{-\beta(E_0+\kappa/2)}\abs{\Lambda_n}}.
    \end{equation}
\end{proposition}

The rest of this sub-subsection is dedicated to the proof of Proposition~\ref{prop: information erased by repair}.
The following lemma uses the absence of long chains of broken links, as excluded in Section~\ref{sec: long chains of broken links}, to bound the size of each transverse defect component in terms of the size of its $\bA$-boundary.
Since every boundary $\bA$-edge is a broken link, this leads to a \textit{Peierls condition} for all transverse defect components.

\begin{lemma}
    \label{lem: obstructive defect components}
    If $\sigma\in\Omega^g_{\Lambda_n}\cap D_n\parentheses[\big]{M^{\mathrm{short}}_{p,p'}}$ and $\gamma\in\Gamma^g(\sigma)$, then the following hold:
    \begin{enumerate}[label=(\roman*)]
        \item \label{itm: obstructive defect components - geometry} $\abs{R_\gamma}+\abs{I_\gamma}\le 2r\abs{\partial R_\gamma}$.
        \item \label{itm: obstructive defect components - Peierls} \emph{Peierls condition:} $E(\gamma)\ge\frac{a}{4r}(\abs{R_\gamma}+\abs{I_\gamma})$.
    \end{enumerate}
\end{lemma}

\begin{proof}[Proof of~\ref{itm: obstructive defect components - geometry}]
    By replacing $\sigma$ with the partially repaired configuration
    \begin{equation}
        \sigma'\coloneqq\brackets[\bigg]{\sigma\setminus\bigcup_{\substack{\gamma'\in\Gamma^g(\sigma)\\\gamma'\ne \gamma}}I_{\gamma'}}\cup\bigcup_{\substack{\gamma'\in\Gamma^g(\sigma)\\\gamma'\ne \gamma}}R_{\gamma'}\in\Omega^g_{\Lambda_n}\cap D_n\parentheses[\big]{M^{\mathrm{short}}_{p,p'}}
    \end{equation}
    if necessary, we may assume that $\Gamma^g(\sigma)=\set{\gamma}$.
    Let $V(R_\gamma)$ be the set of vertices of $\Z^3$ covered by $R_\gamma$.
    Fix $\tau\in\widetilde{T}^S_{\Lambda_n}$ (see Section~\ref{sec: synchronization - setup}).
    In the setting of Lemma~\ref{lem: edge planarity}, let $n=2\ell$ and $E_\tau\subseteq\sE_n$ consist of those edges whose endpoints have different labels with respect to $\tau^{-1}\sigma$.
    By the definition of $M^{\mathrm{short}}_{p,p'}$, every subset $F\subseteq E_\tau$ with connected dual satisfies $\abs{F}\le r-1$, which implies that $\diam^\ast_\infty(F^\ast)\le r$.
    Hence, by Lemma~\ref{lem: edge planarity}, $\sG_{n}\setminus E_\tau$ has a unique $\Box$-component $C_{\tau}$ crossing $\sG_n$ in both directions, and $\abs{\sV_n\setminus C_{\tau}}\le\frac{r}{2}\abs{E_\tau}$.

    By construction, every vertex of $C_\tau$ has the same label with respect to $\tau^{-1}\sigma$, which is either the majority parity in $L$ or the label ``else''.
    In the former case, since every vertex of $\tau^{-1}V(R_\gamma)\cap L$ is labeled ``else'', we have that
    \begin{equation}
        \abs{\tau^{-1}V(R_\gamma)\cap L} 
        \le \abs{\sV_n\setminus C_\tau}.
    \end{equation}
    Suppose now that we are in the latter case.
    Since at most $(\ell+1)^2$ tetramers in $g$ intersect $\tau L$, at most $1/8$ of which are missing from $\sigma$ by the majority constraint, we have that
    \begin{equation}
        \label{eqn: obstructive defect components - majority constraint implication}
        \abs{\tau^{-1}V(R_\gamma)\cap L}\le 4\cdot\frac{1}{8}(\ell+1)^2 \le \frac{1}{2}\abs{\sV_n}.
    \end{equation}
    Thus,
    \begin{align}
        \abs{\tau^{-1}V(R_\gamma)\cap L} 
        {}&\le \abs{\sV_n}-\abs{\tau^{-1}V(R_\gamma)\cap L} 
        {}&&\text{by~\eqref{eqn: obstructive defect components - majority constraint implication}}
        \nonumber
        \\
        {}&\le \abs{\sV_n}-\abs{C_\tau} 
        {}&&\begin{subarray}{t}
            \text{since every vertex of $\tau^{-1}V(R_\gamma)\cap L$}
            \\
            \text{is labeled ``else''}
        \end{subarray}
        \nonumber
        \\
        {}&= \abs{\sV_n\setminus C_\tau}.
    \end{align}
    In both cases, we now have that
    \begin{equation}
        \label{eqn: obstructive defect components - case analysis conclusion}
        \abs{V(R_\gamma)\cap \tau L}
        =\abs{\tau^{-1}V(R_\gamma)\cap L} 
        \le \abs{\sV_n\setminus C_\tau}
        \le \frac{r}{2}\abs{E_\tau}.
    \end{equation}
    Summing over $\tau\in\widetilde{T}^S_{\Lambda_n}$, we deduce that
    \begin{align}
        \abs{R_\gamma}
        {}&\le \sum_{\tau\in\widetilde{T}^S_{\Lambda_n}} \abs{V(R_\gamma)\cap \tau L}
        {}&& \begin{subarray}{t}
            \text{since every $t\in R_\gamma$ intersects $\Lambda_n$ and hence}
            \\
            \text{some $\tau L$, by the $g$-boundary condition}
        \end{subarray}
        \nonumber 
        \\
        {}&\le r\cdot\frac{1}{2}\sum_{\tau\in\widetilde{T}^S_{\Lambda_n}} \abs{E_\tau}
        {}&&\text{by~\eqref{eqn: obstructive defect components - case analysis conclusion}}
        \nonumber 
        \\
        {}&\le r\abs{\partial R_\gamma}
        {}&&\begin{subarray}{t}
            \text{by the at most two-to-one association between edges}
            \\
            \text{and broken links described in Section~\ref{sec: long chains of broken links}}
        \end{subarray}
        .
    \end{align}
    After noting that $\abs{I_\gamma}\le\abs{R_\gamma}$ by Proposition~\ref{prop: energy of defect components}, the proof is complete.
\end{proof}

\begin{proof}[Proof of~\ref{itm: obstructive defect components - Peierls}]
    Recall the definition of $E(\gamma)$ from~\eqref{eqn: energy of defect component}.
    Bounding $\abs{R_\gamma}\ge\abs{I_\gamma}$ by Proposition~\ref{prop: energy of defect components} and $\abs{\partial I_\gamma}\ge 0$ and applying~\ref{itm: obstructive defect components - geometry}, we obtain the desired bound.
\end{proof}

We now deduce Proposition~\ref{prop: information erased by repair}.

\begin{proof}[Proof of Proposition~\ref{prop: information erased by repair}]
    By Proposition~\ref{prop: minimum-energy defect components}, the energy of a finite transverse defect component is either equal to $E_0$ or at least $E_0+\kappa$.
    Let $(\sR_n^g)^\ast\subseteq\sR_n^g$ consist of those $\Gamma\in\sR_n^g$ such that $E(\gamma)\ge E_0+\kappa$ for all $\gamma\in\Gamma$, and write $(\sC_n^g)^\ast\coloneqq\bigcup_{\Gamma\in (\sR_n^g)^\ast} \Gamma$.
    Then, we may bound
    \begin{equation}
        \sum_{\Gamma\in\sR^g_n}\prod_{\gamma\in\Gamma}e^{-\beta E(\gamma)}
        \le \parentheses[\big]{1+e^{-\beta E_0}}^{N_n^{\min}}\sum_{\Gamma\in(\sR^g_n)^\ast}\prod_{\gamma\in\Gamma}e^{-\beta E(\gamma)}.
    \end{equation}

    Next, observe that, for each $\gamma\in(\sC^g_n)^\ast$,
    \begin{align}
        E(\gamma)
        {}&\ge \frac{E_0+\frac{\kappa}{2}}{E_0+\kappa}\,(E_0+\kappa)+\parentheses[\bigg]{1-\frac{E_0+\frac{\kappa}{2}}{E_0+\kappa}}E(\gamma)
        {}&& \begin{subarray}{t}
            \text{since $E(\gamma)\ge E_0+\kappa$}
            \\
            \text{and $E_0,\kappa>0$}
        \end{subarray}
        \nonumber
        \\
        \label{eqn: information erased by repair - energy bound}
        {}&\ge E_0+\frac{\kappa}{2}+\frac{a\kappa}{8r(E_0+\kappa)}(\abs{R_\gamma}+\abs{I_\gamma})
        {}&&\text{by Lemma~\ref{lem: obstructive defect components}\ref{itm: obstructive defect components - Peierls}}
        .
    \end{align}
    Furthermore, note that, given a reference ground state $g$, a defect component $\gamma$ is completely determined by the $\ddag$-connected set $R_\gamma\cup I_\gamma$.
    Since $\F^\ddag$ is a regular graph, there exists a universal constant $C_1>0$ such that, uniformly in the transverse tetramer $t\in\F$,
    \begin{align}
        \sum_{\substack{\gamma\in(\sC^g_n)^\ast\\R_\gamma\cup I_\gamma\ni t}} e^{-\beta E(\gamma)}
        ={}&\sum_{s=3}^\infty\sum_{\substack{\gamma\in(\sC^g_n)^\ast\\R_\gamma\cup I_\gamma\ni t\\\abs{R_\gamma\cup I_\gamma}=s}} e^{-\beta E(\gamma)}
        {}&&\begin{subarray}{t}
            \text{since $\abs{R_\gamma\cup I_\gamma}\ge 3$; see}
            \\
            \text{the proof of Proposition~\ref{prop: minimum-energy defect components}}
        \end{subarray}
        \nonumber
        \\
        \le{}& e^{-\beta(E_0+\kappa/2)}\sum_{s=3}^\infty C_1^s \exp{-\frac{\beta a\kappa s}{8r(E_0+\kappa)}}
        {}&&\text{by~\eqref{eqn: information erased by repair - energy bound}}
        \nonumber
        \\
        \le{}& e^{-\beta(E_0+\kappa/2)}
        {}&&\begin{subarray}{t}
            \text{provided that}
            \\
            C_1 \exp{-\frac{\beta a\kappa}{8r(E_0+\kappa)}}\le\frac{1}{2}
        \end{subarray}
        .
    \end{align}
    Now, there exists a universal constant $C_2>0$ such that no more than $C_2\abs{\Lambda_n}$ transverse tetramers $t\in\F$ intersect $\Lambda_n$.
    Hence,
    \begin{equation}
        \sum_{\Gamma\in(\sR^g_n)^\ast}\prod_{\gamma\in\Gamma}e^{-\beta E(\gamma)}
        \le\prod_{\substack{t\in\F:\,t\cap\Lambda_n\ne\emptyset\\\text{transverse}}}
        \brackets[\bigg]{1+\sum_{\substack{\gamma\in(\sC^g_n)^\ast\\R_\gamma\cup I_\gamma\ni t}} e^{-\beta E(\gamma)}}
        \le \brackets[\Big]{1+e^{-\beta(E_0+\kappa/2)}}^{C_2\abs{\Lambda_n}}.
    \end{equation}
    After using the elementary inequality $1+x\le e^x$ for $x>0$, the proof is complete.
\end{proof}

Before proceeding, we record the following estimate on $N_n^{\min}$ for later use.

\begin{lemma}
    \label{lem: bound on number of erased minimum-energy defects}
    For all $n$ such that $S$ is a block of $\Lambda_n$,
    \begin{equation}
        N_n^{\min}\begin{cases}
            =0 & \text{if }\mu>0\text{ or $p,p'$ differ in both coordinates}\\
            \le \frac{(n!)^3}{4}+\frac{(n!)^2}{2} & \text{otherwise}
        \end{cases}
        .
    \end{equation}
\end{lemma}

\begin{proof}
    Suppose that $\gamma$ is an obstructive transverse defect component with energy $E_0$.
    By Proposition~\ref{prop: minimum-energy defect components}, $\gamma$ cannot satisfy Condition~\ref{itm: obstruction - attractive link}.
    Moreover, Condition~\ref{itm: obstruction - incompatible translations} fails for $\gamma$ if either (a) $\mu>0$, which requires $p=p'$ (see Figures~\ref{fig:defect-pair-side}--\ref{fig:defect-pair-front}), or (b) $p$ and $p'$ differ in both coordinates, in which case no such $\gamma$ exists.
    In both cases, $N_n^{\min}=0$.

    Suppose now that $\mu\le0$ and that $p$ and $p'$ agree in at least one coordinate.
    As established above, every obstructive defect component $\gamma$ with energy $E_0$ is of the type shown in Figure~\ref{fig:defect-shifted}; in particular, it intersects two (adjacent) layers of different reference parities.
    Within $\Lambda_n$, there are $n!/2$ choices for these two layers.
    Fix such a choice.
    Within these two layers, there are at most $n!/2+1$ choices for the position of $\gamma$ in the direction corresponding to the coordinate in which $p$ and $p'$ agree, and at most $n!$ choices for its position in the remaining direction.
    We conclude that, in this case,
    \begin{equation}
        N_n^{\min}\le \frac{n!}{2}\parentheses[\bigg]{\frac{n!}{2}+1}n!,
    \end{equation}
    as desired.
\end{proof}

\subsubsection{Synchronizing parities and inserting minimum-energy defects (Steps~\ref{itm: step - synchronize parities}--\ref{itm: step - insert minimum-energy defects})}

By the definition of $g$ (see Section~\ref{sec: synchronization - setup}), a layer $\set{x_3=j}$ has reference parity $p$ precisely when $j\equiv 0,3\pmod{4}$.
Let $\sigma\in\Omega^g_{\Lambda_n}$.
Since $\fE$ erases every defect component in $\Gamma^g(\sigma)$ satisfying Condition~\ref{itm: obstruction - incompatible translations}, each tetramer in $\fE(\sigma)$ intersects layers of the same reference parity (see Figure~\ref{fig: repair map - erased}).
Let $(\fT\circ\fE)(\sigma)$ be the configuration obtained from $\fE(\sigma)$ by translating every tetramer in $\fE(\sigma)$ that intersects a layer of reference parity $p'$ by $(p-p',0)$ and leaving the others in place.
In particular, $\fT(g)$ is the $x_1x_2$-oriented ground state with parity $p$ in every layer.
Since $\fE(\sigma)$ coincides with $g$ outside of $\Lambda_n$ and the translations move the tetramers in $\fE(\sigma)$ by at most one unit in each coordinate direction, we have thus obtained a well-defined \textbf{translation map}
\begin{equation}
    \fT:\fE(\Omega^g_{\Lambda_n})\to\Omega^{\fT(g)}_{\widetilde{\Lambda}_n},\quad\text{where }\widetilde{\Lambda}_n\coloneqq\eta_{(-1,-1,-1)}\mathrm{R}_{(n!+2)\times(n!+2)\times(n!+2)}.
\end{equation}
See Figures~\ref{fig: repair map - erased}--\ref{fig: repair map - translated}.

Next, consider the cells of $\Z^3$ whose two $x_1x_2$-oriented faces belong to $\fT(g)\cap\Lambda_n$ and lie in layers of different reference parities with respect to $g$.
By construction, these cells are disjoint.
We call such a cell \textbf{available} with respect to $\sigma\in\Omega_{\Lambda_n}^g$ if both $x_1x_2$-oriented faces of the cell are occupied in $(\fT\circ\fE)(\sigma)$ (see Figure~\ref{fig: repair map - available}); denote the set of available cells by $\cQ_n(\sigma)$.
Define the \textbf{insertion map}
\begin{equation}
    \fI:(\fT\circ\fE)(\Omega_{\Lambda_n}^g)\to\sP\parentheses[\Big]{\Omega_{\widetilde{\Lambda}_n}^{\fT(g)}}
\end{equation}
by letting $(\fI\circ\fT\circ\fE)(\sigma)$ consist of all configurations obtained from $(\fT\circ\fE)(\sigma)$ by choosing any $Q\subseteq\cQ_n(\sigma)$ and, independently in each cell of $Q$, replacing the pair of $x_1x_2$-oriented tetramers with one or two transverse tetramers occupying
\begin{enumerate}
    \item if $\mu>0$, either pair of opposite vertical faces of the cell;
    \item if $\mu=0$, any one of its four vertical faces or either pair of opposite vertical faces;
    \item if $-2a<\mu<0$, any one of its four vertical faces.
\end{enumerate}
See Figures~\ref{fig: repair map - translated}--\ref{fig: repair map - replacement 2}.
Denote the number of allowable replacements in each cell by
\begin{equation}
    \label{eqn: number of allowable replacements}
    \delta(\mu)\coloneqq\begin{cases}
        2 & \text{if }\mu>0 \\
        6 & \text{if }\mu=0 \\
        4 & \text{if }-2a<\mu<0
    \end{cases}
    .
\end{equation}

Define the \textbf{repair map}
\begin{equation}
    \fR:\Omega_{\Lambda_n}^g\to\sP\parentheses[\Big]{\Omega_{\widetilde{\Lambda}_n}^{\fT(g)}},\quad
    \fR\coloneqq\fI\circ\fT\circ\fE.
\end{equation}
The main result of this sub-subsection is as follows.

\begin{proposition}
    \label{prop: synchronization estimate - partition function comparison}
    For all $n$ such that $S$ is a block of $\Lambda_n$,
    \begin{equation}
        \label{eqn: synchronization estimate - partition function comparison}
        Z^g_{\Lambda_n}\parentheses[\big]{D_n\parentheses[\big]{M^{\mathrm{short}}_{p,p'}}}
        \le \frac{\sum_{\Gamma\in\sR^g_n}\prod_{\gamma\in\Gamma}e^{-\beta E(\gamma)}}{\parentheses[\big]{1+\delta(\mu)e^{-\beta E_0}}^{\frac{1}{12}(n!)^3}}\,
        Z^{\fT(g)}_{\widetilde{\Lambda}_n}
        .
    \end{equation}
\end{proposition}

The rest of this sub-subsection is dedicated to the proof of Proposition~\ref{prop: synchronization estimate - partition function comparison}.

\begin{lemma}
    \label{lem: repair map weight comparison}
    For all $n$ such that $S$ is a block of $\Lambda_n$ and all $\sigma\in\Omega_{\Lambda_n}^g$,
    \begin{equation}
        w^g_{\Lambda_n}(\sigma)\le \frac{\exp\set[\big]{-\beta\sum_{\gamma\in\Gamma^g_\obs(\sigma)}E(\gamma)}}{\parentheses[\big]{1+\delta(\mu)e^{-\beta E_0}}^{\abs{\cQ_n(\sigma)}}}\sum_{\sigma'\in\fR(\sigma)} w^{\fT(g)}_{\widetilde{\Lambda}_n}(\sigma').
    \end{equation}
\end{lemma}

\begin{proof}
    By Proposition~\ref{prop: energy of defect components} (specifically~\eqref{eqn: energy of defect components}),
    \begin{equation}
        w_{\Lambda_n}^g(\sigma)
        =\exp\set[\bigg]{-\beta\sum_{\gamma\in\Gamma^g_\obs(\sigma)}E(\gamma)}\,w_{\Lambda_n}^g(\fE(\sigma)).
    \end{equation}
    Since $\fT$ preserves the total number of vacancies and does not increase the number of broken links,
    \begin{equation}
        w_{\Lambda_n}^g(\fE(\sigma))
        \le w_{\widetilde{\Lambda}_n}^{\fT(g)}((\fT\circ\fE)(\sigma)).
    \end{equation}
    Starting from $(\fT\circ\fE)(\sigma)$, we may carry out the replacements that produce any given configuration in $\fR(\sigma)$ in any order, since the available cells are disjoint.
    Fix such an order.
    In each allowable replacement, removing the two $x_1x_2$-oriented tetramers creates at most eight broken links and exactly eight vacancies, while each transverse tetramer inserted creates at most four more broken links (though it may also decrease the number of broken links) and decreases the number of vacancies by exactly four.
    Hence, replacing the pair of $x_1x_2$-oriented tetramers by one transverse tetramer increases the energy by at most $12\cdot\frac{a}{2}+4\cdot\frac{\mu+2a}{4}=8a+\mu$, while replacing them by two transverse tetramers increases the energy by at most $16\cdot\frac{a}{2}=8a$.
    Thus, each allowable replacement increases the energy by at most $E_0$, so
    \begin{equation}
        \parentheses[\big]{1+\delta(\mu)e^{-\beta E_0}}^{\abs{\cQ_n(\sigma)}}
        w_{\widetilde{\Lambda}_n}^{\fT(g)}((\fT\circ\fE)(\sigma))
        \le \sum_{\sigma'\in\fR(\sigma)} w_{\widetilde{\Lambda}_n}^{\fT(g)}(\sigma').
    \end{equation}
    Combining the above displayed equations completes the proof.
\end{proof}

\begin{lemma}
    \label{lem: available cells lower bound}
    For all $\sigma\in\Omega_{\Lambda_n}^g\cap D_n\parentheses[\big]{M^{\mathrm{short}}_{p,p'}}$, $\abs{\cQ_n(\sigma)}\ge\frac{1}{12}(n!)^3$.
\end{lemma}

\begin{proof}
    There are at least $(n!/2)^3$ cells of $\Z^3$ whose two $x_1x_2$-oriented faces belong to $\fT(g)\cap\Lambda_n$ and lie in layers of different reference parities with respect to $g$.
    By the $g$-boundary condition, the disseminated majority constraint $D_n(M_{p,p'})$, and overcounting the missing reference tetramers (with respect to $g$) at the intersections of images of slices under $\widetilde{T}^S_{\Lambda_n}$, we have that 
    \begin{equation}
        \abs{g\setminus\sigma}
        \le \frac{(n!)^3}{4\ell^2}\cdot \frac{1}{8}(\ell+1)^2.
    \end{equation}
    We conclude that
    \begin{equation}
        \abs{\cQ_n(\sigma)}
        \ge\parentheses[\bigg]{\frac{n!}{2}}^3-\frac{(n!)^3}{4\ell^2}\cdot \frac{1}{8}(\ell+1)^2
        \ge \frac{(n!)^3}{12},
    \end{equation}
    having used that $\ell\ge 24$ (see~\eqref{eqn: mesoscopic length scale}) in the last inequality.
\end{proof}

\begin{lemma}
    \label{lem: repair map disjointness}
    If $\sigma,\sigma'\in\Omega_{\Lambda_n}^g$ are distinct and $\Gamma^g_\obs(\sigma)=\Gamma^g_\obs(\sigma')$, then $\fR(\sigma)\cap\fR(\sigma')=\emptyset$.
\end{lemma}

\begin{proof}
    By contraposition, suppose that $\sigma,\sigma'\in\Omega_{\Lambda_n}^g$ are such that $\Gamma^g_\obs(\sigma)=\Gamma^g_\obs(\sigma')$ and $\fR(\sigma)\cap\fR(\sigma')\ne\emptyset$.
    We need to show that $\sigma=\sigma'$.
    Let $\omega\in\fR(\sigma)\cap\fR(\sigma')$.
    By construction, any transverse tetramer in $\omega$ that intersects layers of both reference parities with respect to $g$ must have been inserted in Step~\ref{itm: step - insert minimum-energy defects}.
    Hence, removing all such tetramers from $\omega$ and restoring the pairs of $x_1x_2$-oriented tetramers in the corresponding available cells recovers $(\fT\circ\fE)(\sigma)=(\fT\circ\fE)(\sigma')$.
    Since $\fT$ is invertible, we have that $\fE(\sigma)=\fE(\sigma')$.
    Finally, since $\Gamma^g_\obs(\sigma)=\Gamma^g_\obs(\sigma')$, we have that 
    \begin{equation}
        \sigma
        =\brackets[\bigg]{\fE(\sigma)\setminus\bigcup_{\gamma\in\Gamma^g_\obs(\sigma)}R_\gamma}\cup\bigcup_{\gamma\in\Gamma^g_\obs(\sigma)}I_\gamma
        =\brackets[\bigg]{\fE(\sigma')\setminus\bigcup_{\gamma\in\Gamma^g_\obs(\sigma')}R_\gamma}\cup\bigcup_{\gamma\in\Gamma^g_\obs(\sigma')}I_\gamma
        =\sigma',
    \end{equation}
    as desired.
\end{proof}

We now deduce Proposition~\ref{prop: synchronization estimate - partition function comparison}.

\begin{proof}[Proof of Proposition~\ref{prop: synchronization estimate - partition function comparison}]
    By summing over $\sigma\in\Omega^g_{\Lambda_n}\cap D_n\parentheses[\big]{M^{\mathrm{short}}_{p,p'}}$ in Lemma~\ref{lem: repair map weight comparison}, we get that
    \begin{equation}
        Z^g_{\Lambda_n}\parentheses[\big]{D_n\parentheses[\big]{M^{\mathrm{short}}_{p,p'}}}
        \le \sum_{\sigma\in\Omega^g_{\Lambda_n}\cap D_n(M_{p,p'}^{\mathrm{short}})}
        \frac{\exp\set[\big]{-\beta\sum_{\gamma\in\Gamma^g_\obs(\sigma)}E(\gamma)}}{\parentheses[\big]{1+\delta(\mu)e^{-\beta E_0}}^{\abs{\cQ_n(\sigma)}}}\sum_{\sigma'\in\fR(\sigma)} w^{\fT(g)}_{\widetilde{\Lambda}_n}(\sigma').
    \end{equation}
    After applying Lemma~\ref{lem: available cells lower bound} to the denominator, we bound
    \begin{align}
        {}& \sum_{\sigma\in\Omega^g_{\Lambda_n}\cap D_n(M_{p,p'}^{\mathrm{short}})}
        \exp\set[\bigg]{-\beta\sum_{\gamma\in\Gamma^g_\obs(\sigma)}E(\gamma)}
        \sum_{\sigma'\in\fR(\sigma)} w^{\fT(g)}_{\widetilde{\Lambda}_n}(\sigma')
        \nonumber
        \\
        ={}& \sum_{\Gamma\in\sR^g_n}
        \exp\set[\bigg]{-\beta\sum_{\gamma\in\Gamma}E(\gamma)}
        \sum_{\substack{\sigma\in\Omega^g_{\Lambda_n}\cap D_n(M_{p,p'}^{\mathrm{short}})\\\Gamma^g_\obs(\sigma)=\Gamma}}
        \sum_{\sigma'\in\fR(\sigma)} w^{\fT(g)}_{\widetilde{\Lambda}_n}(\sigma')
        \nonumber
        \\
        \le{}& \sum_{\Gamma\in\sR^g_n}
        \exp\set[\bigg]{-\beta\sum_{\gamma\in\Gamma}E(\gamma)}\,
        Z^{\fT(g)}_{\widetilde{\Lambda}_n},
    \end{align}
    having used Lemma~\ref{lem: repair map disjointness} in the last step.
    The proof is complete.
\end{proof}

\subsubsection{Proof of Theorem~\ref{thm: synchronization estimate}}
\label{sec: proof of synchronization estimate}

As noted in Sections~\ref{sec: long chains of broken links} and~\ref{sec: synchronization - setup}, it suffices to estimate the ratio~\eqref{eqn: synchronization estimate - goal}. 
Using Propositions~\ref{prop: synchronization estimate - partition function comparison} and~\ref{prop: information erased by repair}, we bound
\begin{equation}
    Z_{\Lambda_n}^g\parentheses[\big]{D_n\parentheses[\big]{M^{\mathrm{short}}_{p,p'}}}
    \le \frac{\parentheses[\big]{1+e^{-\beta E_0}}^{N_n^{\min}}\exp\set[\big]{C_{\ref{prop: information erased by repair}} e^{-\beta(E_0+\kappa/2)}\abs{\Lambda_n}}}{\parentheses[\big]{1+\delta(\mu)e^{-\beta E_0}}^{\frac{1}{12}(n!)^3}}\,
    Z^{\fT(g)}_{\widetilde{\Lambda}_n}.
\end{equation}
After using Lemma~\ref{lem: bound on number of erased minimum-energy defects} to bound $N_n^{\min}$ and recalling~\eqref{eqn: number of allowable replacements}, we bound
\begin{equation}
    \label{eqn: motivating problem manifestation}
    \frac{1}{12}\log\parentheses[\big]{1+\delta(\mu)e^{-\beta E_0}}
    -\frac{\indicator{\mu\le 0}}{4}\log\parentheses[\big]{1+e^{-\beta E_0}}
    -C_{\ref{prop: information erased by repair}} e^{-\beta(E_0+\kappa/2)}
    \ge \frac{1}{24}e^{-\beta E_0}
\end{equation}
for sufficiently large $\beta$.
Furthermore, by another comparison of boundary conditions, there exists $C_{\eqref{eqn: synchronization estimate - boundary condition comparison}}>0$ such that, for all $n$ such that $S$ is a block of $\Lambda_n$,
\begin{equation}
    \label{eqn: synchronization estimate - boundary condition comparison}
    Z_{\widetilde{\Lambda}_n}^{\fT(g)} \le e^{C_{\eqref{eqn: synchronization estimate - boundary condition comparison}}(n!)^2}Z_{\Lambda_n}^{\mathrm{per}}.
\end{equation}
Combining the above estimates yields
\begin{equation}
    \frac{Z_{\Lambda_n}^g\parentheses[\big]{D_n\parentheses[\big]{M^{\mathrm{short}}_{p,p'}}}}{Z_{\Lambda_n}^\per}
    \le \exp\set[\bigg]{-\frac{1}{24}e^{-\beta E_0}(n!)^3+C_{\eqref{eqn: synchronization estimate - boundary condition comparison}}(n!)^2},
\end{equation}
as required.

\subsection{Four parities}
\label{sec: four parities}

We conclude this section by establishing partial translational symmetry breaking within the three orientational phases of Section~\ref{sec: orientational symmetry breaking}, splitting each into four.
Recall from Proposition~\ref{prop: every ground state is layered} that the ground states of a given orientation form an uncountable family: while the tetramers are arranged periodically within each layer, different layers are free to slide relative to one another.
Thermal fluctuations lift this degeneracy through the order-by-disorder mechanism of Section~\ref{sec: parity synchronization between adjacent mesoscopic slices}.
Namely, since adjacent layers of matching parity admit more of the minimum-energy transverse defect components identified in Proposition~\ref{prop: minimum-energy defect components}, the system favors, for each orientation, the four ground states in which all layers have the same parity (whereby the tetramers \textit{stack} into columns).
Thus, these twelve ground states give rise to twelve phases of the model, labeled by $\cO\times\cP$.

As before, we specialize to the $x_1x_2$-oriented phases.
For $p\in\cP$, define the random set
\begin{equation}
    \label{eqn: parity random set}
    \Theta_p\coloneqq\xre{\ell\times \ell\times 1}{\es{G^\reg_{x_1x_2}\cap G^\dom_p}}\cap\bigcap_{k=0,3}\xre{\ell\times \ell\times 1}{\es{B_k^c}},
\end{equation}
and write $\Theta\coloneqq\bigcup_{p\in\cP}\Theta_p$.
The main result of this subsection is as follows.

\begin{proposition}[Four parities]
    \label{prop: four parities}
    For all $\eps\in(0,\eps_3)$, there exists $\beta_0>0$ such that the following holds for all $\beta\ge\beta_0$.
    For all $\ell\Z^3$-ergodic Gibbs measures $\P$ with $\Ori{\P}=x_1x_2$, there exists a unique parity label $\Par{\P}\in\cP$ such that, out of the random sets $\set{\Theta_p\mid p\in\cP}$, $\Theta_{\Par{\P}}$ is $\eps$-strongly percolating, while the others have only finite $\boxtimes$-components $\P$-a.s.
\end{proposition}

\begin{proof}
    Observe that
    \begin{equation}
        \Theta=\xre{\ell\times \ell\times 1}{\es{G^\reg_{x_1x_2}}}\cap\bigcap_{k=0,3}\xre{\ell\times \ell\times 1}{\es{B_k^c}}.
    \end{equation}
    By Lemma~\ref{lem:intersection of strongly percolating sets}, Corollary~\ref{cor: mesoscopic slabs have a dominant orientation}, Proposition~\ref{prop: mesoscopic slices have a dominant parity}, and Corollary~\ref{cor: parity synchronization}, we have that
    \begin{equation}
        \label{eqn: four parities - cubic bound}
        \peierls{\P}(\Theta)\le 3\sqrt[3]{\max\set*{\eps_\reg(\beta,r,\ell),\eps_\dom(\beta,r,\ell),\eps_\syn(\beta,r,\ell)}}.
    \end{equation}
    On the other hand, observe that $\rmd_\Box(\Theta_p,\Theta_{p'})\ge 2$ for $p\ne p'$, so Lemma~\ref{lem:splitting strongly percolating sets} yields
    \begin{equation}
        \min_{p\in\cP}\peierls{\P}(\Theta_p)\le\peierls{\P}(\Theta).
    \end{equation}
    Arguing as in the proof of Proposition~\ref{prop: three orientations} completes the proof.
\end{proof}

We extend the parity label to the other two orientations by coordinate permutation.
Specifically, given $i\in\Z/3\Z$ and an $\ell\Z^3$-ergodic Gibbs measure $\P$ with $\Ori{\P}=x_ix_{i+1}$, we set 
\begin{equation}
    \label{eqn: definition of parity label for other orientations}
    \Par{\P}\coloneqq\Par{\varphi^{1-i}\P}.
\end{equation}
Note that this is a tautology when $i=1$ and otherwise well-defined by Proposition~\ref{prop: four parities}, since $\varphi^{1-i}(\ell\Z^3)=\ell\Z^3$ and $\varphi^{1-i}\cdot x_ix_{i+1}=x_1x_2$.

\section{Discotic columnar order}
\label{sec: columnar order}

\subsection{Twelve phases}
\label{sec: twelve phases}

We now prove that the orientation and parity labels of Sections~\ref{sec: orientational symmetry breaking} and~\ref{sec: partial translational symmetry breaking}, together, classify the periodic Gibbs measures of the model at low temperature.
The bulk of the work lies in showing that each phase label in $\cO\times\cP$ corresponds to a \textit{unique} Gibbs measure.
We accomplish this using the disagreement percolation method of van den Berg~\cite{van1993uniqueness}, which consists in comparing independent samples from two Gibbs measures with the same phase label and showing that, almost surely, there is no infinite path of plaquettes on which the two samples disagree.
The mechanism at work is again energetic: since a disagreement component never intersects a vertex covered by a reference tetramer (that is, a tetramer in the crystalline ground state corresponding to the phase label) present in both samples, within any layer that the component intersects, its cross-sections are fully surrounded by broken links.
The crossings in the definition of parity dominance enter essentially here by allowing us to pay for each cross-section with the broken links enclosing it.
As a result, the probability that two distant plaquettes are connected by a disagreement path decays exponentially in their distance, which verifies the assumption of disagreement percolation and yields uniqueness.

For each Gibbs measure $\P$ with both an orientation and a parity label, define its phase label by
\begin{equation}
    \Phase{\P}\coloneqq(\Ori{\P},\Par{\P}).
\end{equation}
The main result of this subsection is as follows.

\begin{proposition}[Twelve phases]
    \label{prop:twelve phases}
    There exists a constant $\beta_0>0$ such that the following holds for all $\beta\ge\beta_0$.
    For each $\pi=(x_ix_{i+1},p)\in\cO\times\cP$, there exists a unique $\ell\Z^3$-ergodic Gibbs measure $\P_\pi$ such that $\Phase{\P_\pi}=\pi$. 
    Moreover:
    \begin{enumerate}[label=(\alph*)]
        \item \label{itm:twelve phases - extremal} For each $\pi\in\cO\times\cP$, $\P_\pi$ is extremal.
        \item \label{itm:twelve phases - invariance} $\P_\pi$ is $2\Z\vec{e}_i\oplus 2\Z\vec{e}_{i+1}\oplus\Z\vec{e}_{i+2}$-invariant when $\pi\in\set{x_ix_{i+1}}\times\cP$.
        \item \label{itm:twelve phases - transformation} For all $o=x_ix_{i+1}\in\cO$, $p\in\cP$, $z\in\Z^3$, and $k\in\Z$,
        \begin{equation}
            \varphi^k\eta_z\P_{o,p}=\P_{\varphi^k\cdot o,\,p+(z_i,z_{i+1})}.
        \end{equation}
        \item \label{itm:twelve phases - decomposition} Every periodic Gibbs measure is a convex combination of $(\P_\pi)_{\pi\in\cO\times\cP}$.
    \end{enumerate}
\end{proposition}

Thus, we establish Theorem~\ref{thm:main}\ref{itm:main - invariance and extremality}, \ref{itm:main - transformation}, and~\ref{itm:main - classification of periodic Gibbs measures}.
The proof of Proposition~\ref{prop:twelve phases} is by implementing the disagreement percolation method introduced by van den Berg~\cite{van1993uniqueness}.
We quote the following formulation of the method from~\cite{hadas2026columnar}.

\begin{theorem}[{Disagreement percolation,~\cite[Theorem 8.2]{hadas2026columnar}}]
    \label{thm:disagreement percolation}
    Let $\P,\P'$ be Gibbs measures of a Markov random field. 
    Let $\sigma,\sigma'$ be independent samples from $\P$ and $\P'$, respectively, and define their \textbf{disagreement set} by $\Delta_{\sigma,\sigma'}\coloneqq\set{t\in\F\mid\sigma(t)\ne\sigma'(t)}$.
    If 
    \begin{equation}
        \label{eqn:disagreement percolation - no infinite component}
        (\P\otimes\P')(\Delta_{\sigma,\sigma'}\text{ has an infinite connected component})=0,
    \end{equation}
    then the following hold:
    \begin{enumerate}[label=(\alph*)]
        \item $\P=\P'$ and $\P$ is extremal;
        \item Let $A,B\subseteq\F$ and $f,g:\Omega\to[-1,1]$ be measurable, $A$- and $B$-local functions, respectively.
        Then,
        \begin{equation}
            \abs{\Cov_\P(f,g)}\le 2(\P\otimes\P')(A\text{ and }B\text{ are connected by a path in }\Delta_{\sigma,\sigma'}).
        \end{equation}
    \end{enumerate}
\end{theorem}

Recall from Section~\ref{sec:basic definitions} that the monomer-tetramer model is a Markov random field under the $\ddag$-connectivity on $\F$.
To verify the condition~\eqref{eqn:disagreement percolation - no infinite component}, we will prove the following proposition.

\begin{proposition}
    \label{prop:disagreement percolation decay}
    There exist constants $\beta_0,C,c>0$ such that, for all $\beta\ge\beta_0$, if $\P$ and $\P'$ are $\ell\Z^3$-ergodic Gibbs measures with $\Phase{\P}=\Phase{\P'}=(x_1x_2,(0,0))$ and $\sigma,\sigma'$ are independent samples from $\P$ and $\P'$, respectively, then, for all $A,B\subseteq\F$,
    \begin{equation}
        (\P\otimes\P')(A\text{ and }B\text{ are connected by a $\ddag$-path in }\Delta_{\sigma,\sigma'})\le \sum_{s\in A}\sup_{t\in B}\alpha_{x_1x_2}(s,t),
    \end{equation}
    where 
    \begin{equation}
        \label{eqn: Heilmann-Lieb decay}
        \alpha_{x_1x_2}(s,t)\coloneqq C\exp\set[\bigg]{-c\brackets[\bigg]{\frac{\abs{s_1-t_1}+\abs{s_2-t_2}}{\ell}+\abs{s_3-t_3}}}.
    \end{equation}
\end{proposition}

Note that this establishes Theorem~\ref{thm:main}\ref{itm:main - decay of correlations}.
Once Proposition~\ref{prop:disagreement percolation decay} is established, Proposition~\ref{prop:twelve phases} readily follows by using the following auxiliary lemma.

\begin{lemma}
    \label{lem: phase label transformation}
    Let $\eps\in(0,\eps_3/2)$ and $\beta\ge\beta_0$, where $\beta_0=\beta_0(\eps)>0$ is as in Proposition~\ref{prop: four parities}.
    Let $i\in\set{1,2,3}$ and $\P$ be an $\ell\Z^3$-ergodic Gibbs measure with $\Ori{\P}=x_ix_{i+1}$.
    Then, the following statements hold.
    \begin{enumerate}[label=(\roman*)]
        \item \label{itm: phase label transformation - permutation} $\Phase{\varphi\P}=(\varphi\cdot\Ori{\P},\Par{\P})$, where $\varphi$ is the coordinate permutation from Section~\ref{sec: Heilmann-Lieb approach}.
        \item \label{itm: phase label transformation - translation} For all $z\in\Z^3$, $\Phase{\eta_z\P}=(\Ori{\P},\Par{\P}+(z_i,z_{i+1}))$.
    \end{enumerate}
\end{lemma}

\begin{proof}[Proof of~\ref{itm: phase label transformation - permutation}]
    To see that $\Ori{\varphi\P}=x_{i+1}x_{i+2}$, we compute
    \begin{align}
        {}& (\varphi\P)(\Psi_{x_{i+1}x_{i+2}}(\sigma)\text{ has an infinite $\Box$-component})
        {}&&
        \nonumber
        \\
        ={}&\P(\Psi_{x_{i+1}x_{i+2}}(\sigma\circ\varphi^{-1})\text{ has an infinite $\Box$-component})
        {}&& \begin{subarray}{t}
            \text{by the definition of}
            \\
            \text{the push-forward $\varphi\P$}
        \end{subarray}
        \nonumber
        \\
        ={}&\P(\Psi_{x_ix_{i+1}}(\sigma)\text{ has an infinite $\Box$-component})
        {}&& \begin{subarray}{t}
            \text{by the rotation covariance}
            \\
            \text{of the percolation event}
        \end{subarray}
        \nonumber
        \\
        ={}&1
        {}&&\text{by Proposition~\ref{prop: three orientations}}
        .
    \end{align}
    By Proposition~\ref{prop: three orientations}, this forces $\Ori{\varphi\P}=x_{i+1}x_{i+2}=\varphi\cdot\Ori{\P}$.
    Furthermore, we have that
    \begin{align}
        \Par{\varphi\P}
        {}&=\Par{\varphi^{1-(i+1)}(\varphi\P)}
        {}&&\text{by~\eqref{eqn: definition of parity label for other orientations}, since }\Ori{\varphi\P}=x_{i+1}x_{i+2}
        \nonumber
        \\
        {}&=\Par{\varphi^{1-i}\P}
        {}&&
        \nonumber
        \\
        {}&=\Par{\P}
        {}&&\text{by~\eqref{eqn: definition of parity label for other orientations}, since }\Ori{\P}=x_{i}x_{i+1},
    \end{align}
    as desired.
\end{proof}

\begin{proof}[Proof of~\ref{itm: phase label transformation - translation}]
    We start by checking that $\Ori{\eta_z\P}=\Ori{\P}$.
    Recalling the random sets $\Psi_o$, $o\in\cO$, from Section~\ref{sec: Heilmann-Lieb approach}, we compute
    \begin{align}
        {}& (\eta_z\P)(\Psi_{x_ix_{i+1}}(\sigma)\text{ has an infinite $\Box$-component})
        {}&&
        \nonumber
        \\
        ={}&\P(\Psi_{x_ix_{i+1}}(\sigma\circ\eta_z^{-1})\text{ has an infinite $\Box$-component})
        {}&& \begin{subarray}{t}
            \text{by the definition of}
            \\
            \text{the push-forward $\eta_z\P$}
        \end{subarray}
        \nonumber
        \\
        ={}&\P(\Psi_{x_ix_{i+1}}(\sigma)\text{ has an infinite $\Box$-component})
        {}&& \begin{subarray}{t}
            \text{by the translation invariance}
            \\
            \text{of the percolation event}
        \end{subarray}
        \nonumber
        \\
        ={}&1
        {}&&\text{by Proposition~\ref{prop: three orientations}}
        .
    \end{align}
    By Proposition~\ref{prop: three orientations}, this forces $\Ori{\eta_z\P}=x_ix_{i+1}=\Ori{\P}$.

    It remains to show that $\Par{\eta_z\P}=\Par{\P}+(z_i,z_{i+1})$, which constitutes the bulk of the proof.
    We start by making two reductions.
    First, since
    \begin{align}
        \Par{\eta_z\P}
        {}&=\Par{\varphi^{1-i}(\eta_z\P)}
        {}&&\text{by~\eqref{eqn: definition of parity label for other orientations}, since }\Ori{\eta_z\P}=x_{i}x_{i+1}
        \nonumber
        \\
        {}&=\Par{\eta_{\varphi^{1-i}z}(\varphi^{1-i}\P)}
        {}&&\text{using that $\varphi^j\eta_z=\eta_{\varphi^j z}\varphi^j$}
    \end{align}
    and $\Par{\P}=\Par{\varphi^{1-i}\P}$, where $\Ori{\varphi^{1-i}\P}=x_1x_2$ by~\ref{itm: phase label transformation - permutation}, we may assume without loss of generality that $i=1$.
    Second, by composing translations, we may further assume that $z\in\set{\pm\vec{e}_k\mid k\in\set{1,2,3}}$.

    Recall the random sets $\Theta_p$, $p\in\cP$, from Section~\ref{sec: four parities}; in particular, recall that each $\Theta_p$ is defined with respect to the grid $\ell\Z\times\ell\Z\times\Z$ (see~\eqref{eqn: parity random set}).
    We now split into two cases: $z\in\set{\pm\vec{e}_3}$ and $z\in\set{\pm\vec{e}_1,\pm\vec{e}_2}$.
    If $z\in\set{\pm\vec{e}_3}$, we compute
    \begin{align}
        {}& (\eta_z\P)(\Theta_{\Par{\P}}(\sigma)\text{ has an infinite $\Box$-component})
        {}&&
        \nonumber
        \\
        ={}&\P(\Theta_{\Par{\P}}(\sigma\circ\eta_z^{-1})\text{ has an infinite $\Box$-component})
        {}&& \begin{subarray}{t}
            \text{by the definition of}
            \\
            \text{the push-forward $\eta_z\P$}
        \end{subarray}
        \nonumber
        \\
        ={}&\P(\Theta_{\Par{\P}}(\sigma)\text{ has an infinite $\Box$-component})
        {}&& \begin{subarray}{t}
            \text{by the invariance of the percolation event}
            \\
            \text{under translations in the $\vec{e}_3$-direction}
        \end{subarray}
        \nonumber
        \\
        ={}&1
        {}&&\text{by Proposition~\ref{prop: four parities}}
        .
        \label{eqn: phase label transformation - translation - parity percolation}
    \end{align}
    By Proposition~\ref{prop: four parities}, this forces $\Par{\eta_z\P}=\Par{\P}$.
    If $z\in\set{\pm\vec{e}_1,\pm\vec{e}_2}$, we need a different argument since the percolation event considered in~\eqref{eqn: phase label transformation - translation - parity percolation} is, a priori, only invariant under translations by $\ell\Z\times\ell\Z\times\Z$.
    Following similar arguments in~\cite[Section~7.2]{hadas2026columnar}, let us suppose by contradiction that $\Par{\eta_z\P}\ne\Par{\P}+(z_1,z_2)$.
    By Proposition~\ref{prop: four parities}, we have that
    \begin{equation}
        \max\set*{\peierls{\P}(\Theta_{\Par{\P}}),\peierls{\eta_z\P}(\Theta_{\Par{\eta_z\P}})} \le \eps.
    \end{equation}
    Using Lemma~\ref{lem:quantitative Peierls argument} with $d=0$ and recalling~\eqref{eqn: parity random set}, we get that
    \begin{equation}
        \max\set*{\P(\Omega\setminus E_{\Par{P}}),(\eta_z\P)(\Omega\setminus E_{\Par{\eta_z\P}})} 
        \le \eps/\eps_3
        <1/2,
    \end{equation}
    where we introduced the shorthand $E_p\coloneqq G^\reg_{x_1x_2}\cap G^\dom_{p}\cap B_0^c\cap B_3^c$ for $p\in\cP$.
    Notice that
    \begin{equation}
        (\eta_z\P)(\Omega\setminus E_{\Par{\eta_z\P}})
        =\P(\Omega\setminus \eta_{-z}E_{\Par{\eta_z\P}}).
    \end{equation}
    Thus, 
    \begin{equation}
        \P(E_{\Par{\P}}\cap\eta_{-z}E_{\Par{\eta_z\P}})
        \ge 1-\P(\Omega\setminus E_{\Par{\P}})-\P(\Omega\setminus \eta_{-z}E_{\Par{\eta_z\P}})
        >0,
    \end{equation}
    which implies that, with nonzero probability, the slice $L$ has majority parity $\Par{\P}$ while the slice $\eta_{-z}L$ has majority parity $\Par{\eta_z\P}-(z_1,z_2)\ne \Par{\P}$, but this is impossible given the overlap between $L$ and $\eta_{-z}L$.
\end{proof}

We now deduce Proposition~\ref{prop:twelve phases} from Proposition~\ref{prop:disagreement percolation decay}.

\begin{proof}[Proof of~\ref{itm:twelve phases - extremal}]
    We first prove the existence of $\P_\pi$.
    Let $\P$ be a $\Z^3$-invariant Gibbs measure supplied by the discussion in Section~\ref{sec: Heilmann-Lieb approach}.
    By the ergodic decomposition theorem~\cite[Theorem (14.17)]{georgii2011gibbs}, $\P$ decomposes into a convex combination of $\ell\Z^3$-ergodic Gibbs measures.
    Let $\nu$ be any $\ell\Z^3$-ergodic Gibbs measure in this decomposition.
    By Proposition~\ref{prop: three orientations}, $\Ori{\nu}$ is well-defined.
    Write $\Ori{\nu}=x_kx_{k+1}$.
    By Proposition~\ref{prop: four parities} and~\eqref{eqn: definition of parity label for other orientations}, $\Par{\nu}$ is also well-defined.
    By Lemma~\ref{lem: phase label transformation}, we may take 
    \begin{equation}
        \P_\pi=\varphi^{i-k}\eta_{(p_1-\Par{\nu}_1)\vec{e}_k+(p_2-\Par{\nu}_2)\vec{e}_{k+1}}\nu,
    \end{equation}
    which is $\ell\Z^3$-ergodic with phase label $\pi$, proving existence.

    For uniqueness, suppose that $\P_\pi$ and $\P'_\pi$ are two such measures.
    Define
    \begin{equation}
        \label{eqn: twelve phases - extremal - reference measure}
        \P_0\coloneqq \varphi^{1-i}\eta_{-p_1\vec{e}_i-p_2\vec{e}_{i+1}}\P_\pi
    \end{equation}
    and $\P'_0$ with $\P_\pi'$ in place of $\P_\pi$.
    By Lemma~\ref{lem: phase label transformation}, $\Phase{\P_0}=\Phase{\P'_0}=(x_1x_2,(0,0))$.
    Let $\sigma,\sigma'$ be independent samples from $\P_0$ and $\P'_0$, respectively.
    By Proposition~\ref{prop:disagreement percolation decay}, the condition~\eqref{eqn:disagreement percolation - no infinite component} is satisfied, so $\P_0=\P'_0$ and $\P_0$ is extremal.
    Upon undoing the transformations in~\eqref{eqn: twelve phases - extremal - reference measure}, we deduce that $\P_\pi=\P_\pi'$ and $\P_\pi$ is extremal, as required.
\end{proof}

\begin{proof}[Proof of~\ref{itm:twelve phases - invariance}]
    Let $\P_\pi$ be as in the proof of~\ref{itm:twelve phases - extremal}.
    Let $z\in2\Z\vec{e}_i\oplus 2\Z\vec{e}_{i+1}\oplus\Z\vec{e}_{i+2}$.
    By Lemma~\ref{lem: phase label transformation}\ref{itm: phase label transformation - translation}, 
    \begin{equation}
        \Phase{\eta_z\P_\pi}
        =(\Ori{\P_\pi},\Par{\P_\pi}+(z_i,z_{i+1}))
        =(\Ori{\P_\pi},\Par{\P_\pi})
        =\Phase{\P_\pi}.
    \end{equation}
    By the uniqueness proven in~\ref{itm:twelve phases - extremal}, we conclude that $\eta_z\P_\pi=\P_\pi$.
\end{proof}

\begin{proof}[Proof of~\ref{itm:twelve phases - transformation}]
    We have that
    \begin{align}
        \Phase{\varphi^k\eta_z\P_{o,p}} ={}& (\varphi^k\cdot\Ori{\eta_z\P_{o,p}},\Par{\eta_z\P_{o,p}})
        {}&& \text{by iterating Lemma~\ref{lem: phase label transformation}\ref{itm: phase label transformation - permutation}}
        \nonumber
        \\
        ={}& (\varphi^k\cdot o,p+(z_i,z_{i+1}))
        {}&& \text{by Lemma~\ref{lem: phase label transformation}\ref{itm: phase label transformation - translation}}.
    \end{align}
    By the uniqueness proven in~\ref{itm:twelve phases - extremal}, we conclude that $\varphi^k\eta_z\P_{o,p}=\P_{\varphi^k\cdot o,\,p+(z_i,z_{i+1})}$.
\end{proof}

\begin{proof}[Proof of~\ref{itm:twelve phases - decomposition}]
    Let $\P$ be a periodic Gibbs measure and $\cL\subseteq\Z^3$ be a full-rank sublattice such that $\P$ is $\cL$-invariant.
    By taking its intersection with $\ell\Z^3$ if necessary, we may assume that $\cL\subseteq\ell\Z^3$.
    By the ergodic decomposition theorem~\cite[Theorem (14.17)]{georgii2011gibbs}, $\P$ is a convex combination of $\cL$-ergodic Gibbs measures.
    Let $\nu$ be any $\cL$-ergodic Gibbs measure, and consider the $\ell\Z^3$-invariant measure
    \begin{equation}
        \label{eqn: twelve phases - decomposition - auxiliary measure}
        \tilde{\nu}\coloneqq\frac{1}{\abs{\ell\Z^3/\cL}}\sum_{\eta\in\ell\Z^3/\cL}\eta\nu.
    \end{equation}
    We claim that $\tilde{\nu}$ is $\ell\Z^3$-ergodic.
    Indeed, if $A$ is an $\ell\Z^3$-invariant event, then
    \begin{align}
        \tilde{\nu}(A)
        {}&=\frac{1}{\abs{\ell\Z^3/\cL}}\sum_{\eta\in\ell\Z^3/\cL}\nu(\eta^{-1}A)
        {}&&\begin{subarray}{t}
            \text{by the definition of}
            \\
            \text{the push-forward $\eta\nu$}
        \end{subarray}
        \nonumber
        \\
        {}&=\frac{1}{\abs{\ell\Z^3/\cL}}\sum_{\eta\in\ell\Z^3/\cL}\nu(A)
        {}&&\text{since $A$ is $\ell\Z^3$-invariant}
        \nonumber
        \\
        {}&=\nu(A)\in\set{0,1}
        {}&&\begin{subarray}{t}
            \text{since $\nu$ is $\cL$-ergodic and $A$ is}
            \\
            \text{$\cL$-invariant, using that $\cL\subseteq\ell\Z^3$.}
        \end{subarray}
    \end{align}
    By the uniqueness proven in~\ref{itm:twelve phases - extremal}, $\tilde{\nu}=\P_\pi$ for some $\pi\in\cO\times\cP$.
    Since $\P_\pi$ is extremal, it follows from the decomposition~\eqref{eqn: twelve phases - decomposition - auxiliary measure} that $\nu=\P_\pi$.
    The proof is complete.
\end{proof}

The rest of the subsection is dedicated to the proof of Proposition~\ref{prop:disagreement percolation decay}.

\subsubsection{Sealed boxes}

Here, we distinguish \textit{sealed} boxes in which a joint configuration $(\sigma,\sigma')\in\Omega^2$ exhibits desirable properties that rule out the existence of large, sprawling disagreement components nearby.

Let $\fm\ge 2$ be an integer.
Write $R=\mathrm{R}_{\ell\times\ell\times\fm}$, $S=\eta_{(-\ell,-\ell,-\fm)}\mathrm{R}_{3\ell\times3\ell\times3\fm}$, and, for $v\in\Z^3$, $t_v$ for the unique $x_1x_2$-oriented plaquette of parity $(0,0)$ covering $v$.
Define the events
\begin{equation}
    \Sigma_0\coloneqq\bigcap_{i,j=-1}^0\bigcap_{k=-\fm}^{2\fm-1}\eta_{(\ell i,\ell j,k)}G^\dom_{(0,0)}\subseteq\Omega
\end{equation}
and
\begin{equation}
    \begin{multlined}
        \Sigma_1\coloneqq\set[\big]{(\sigma,\sigma')\in\Omega^2\mid\text{every $\Box$-path connecting $R$ to $\R^3\setminus S$}
        \\
        \text{contains a vertex $v\in S$ such that $t_v\in\sigma\cap\sigma'$}
        }.
    \end{multlined}
\end{equation}
We say that the box $\mathrm{R}_{\ell\times\ell\times\fm}$ is \textbf{sealed} in $(\sigma,\sigma')$ if the event $\Sigma_1$ holds.
Write $\Pi:=\xre{\ell\times\ell\times\fm}{\es{\Sigma_1}}$.

\begin{proposition}
    \label{prop: sealed boxes - favorable environment}
    For every $\eps>0$ and integer $\fm\ge 2$, there exists $\beta_0>0$ such that, for all $\beta\ge\beta_0$ and $\ell\Z^3$-ergodic Gibbs measures $\P$ with $\Phase{\P}=(x_1x_2,0,0)$,
    \begin{equation}
        \peierls{\P}(\xre{\ell\times\ell\times\fm}{\es{\Sigma_0}})
        \le \eps.
    \end{equation}
\end{proposition}

\begin{proof}
    Write
    \begin{equation}
        \label{eqn: rewriting of Sigma_0}
        \Sigma_0=\bigcap_{k_1,k_2=-1}^0 \bigcap_{p_3=-1}^1 \eta_{(\ell k_1,\ell k_2,\fm p_3)} \bigcap_{r_3=0}^{\fm-1} \eta_{r_3\vec{e}_3} G^\dom_{(0,0)}
        .
    \end{equation}
    By Lemma~\ref{lem:strongly percolating grid events}\ref{itm:strongly percolating grid events - subgrid over grid}, we have that 
    \begin{equation}
        \peierls{\P}\parentheses[\Bigg]{\xre{\ell\times\ell\times\fm}{\es{\bigcap\nolimits_{r_3=0}^{\fm-1} \eta_{r_3\vec{e}_3} G^\dom_{(0,0)}}}}
        \le \fm\,\peierls{\P}(\xre{\ell\times\ell\times 1}\es{G^\dom_{(0,0)}}).
    \end{equation}
    Thus, by Lemma~\ref{lem:intersection of strongly percolating sets}, 
    \begin{equation}
        \peierls{\P}\parentheses{\xre{\ell\times\ell\times\fm}{\es{\Sigma_0}}}
        \le 12\sqrt[12]{\fm\,\peierls{\P}(\xre{\ell\times\ell\times 1}\es{G^\dom_{(0,0)}})}.
    \end{equation}
    Since $\Theta_{(0,0)}\subseteq\xre{\ell\times\ell\times 1}\es{G^\dom_{(0,0)}}$ (see~\eqref{eqn: parity random set}) and $\Phase{\P}=(x_1x_2,0,0)$, we conclude using monotonicity and Proposition~\ref{prop: four parities} that
    \begin{equation}
        \peierls{\P}(\xre{\ell\times\ell\times 1}\es{G^\dom_{(0,0)}})
        \le \peierls{\P}(\Theta_{(0,0)})
    \end{equation}
    can be made arbitrarily small by taking $\beta$ sufficiently large.
    Combining the above estimates completes the proof.
\end{proof}

\begin{lemma}
    \label{lem: escape leads to broken links}
    Suppose that $(\sigma,\sigma')\in(\Sigma_0\times\Sigma_0)\setminus\Sigma_1$.
    Then, there exist an integer $q\ge\min\set{\ell,\fm}/8$ and a $\boxtimes$-walk $v_1,\dots,v_{N},v_{N+1}$ in $\Z^3$ of length $N\le 48q-1$ such that the following statements hold:
    \begin{enumerate}[label=(\roman*)]
        \item \label{itm: escape leads to broken links - starting point} $v_1\in R$;
        \item \label{itm: escape leads to broken links - localization} $v_i\in S$ for all $2\le i\le N$, and $v_{N+1}\in\R^3\setminus S$;
        \item \label{itm: escape leads to broken links - broken links} There exist $q$ distinct vertices $v\in\set{v_1,\dots,v_{N}}$ for which $t_v$ is an endpoint of an $S$-local broken link of $\sigma$ or $\sigma'$.
    \end{enumerate}
\end{lemma}

\begin{proof}
    Since $(\sigma,\sigma')\not\in\Sigma_1$, there exists a $\Box$-path $u_1,\dots,u_{M},u_{M+1}$ connecting $R$ to $\R^3\setminus S$ such that:
    \begin{enumerate}
        \item $u_1\in R$, $u_{M+1}\in\R^3\setminus S$, and $u_i\in S$ for all $2\le i\le M$;
        \item $t_{u_i}\not\in\sigma\cap\sigma'$ for all $1\le i\le M$.
    \end{enumerate}
    Let $1\le i\le M$.
    Choose $\sigma_i\in\set{\sigma,\sigma'}$ and 
    \begin{equation}
        \label{eqn: escape leads to broken links - slice choices}
        L_i=\eta_i L\in\set[\big]{\eta_{(\ell k_1,\ell k_2,k_3)}L\mid -1\le k_1,k_2\le 0,\,-\fm\le k_3\le 2\fm-1}
    \end{equation}
    such that $\sigma_i(t_{u_i})=0$ and $u_i\in L_i$.
    Label each vertex of $\Z^2\times\set{0}$ by $(0,0)$ if it is covered by an $x_1x_2$-oriented tetramer of parity $(0,0)$ in $\eta_i^{-1}\sigma_i$, or by the fixed label ``else'' otherwise.
    In particular, the vertex $\eta_i^{-1}u_i$ is labeled ``else''.
    In the setting of Lemma~\ref{lem: vertex planarity component boundary}, let $n=2\ell$ and $V\subseteq\sV_n$ consist of the vertices labeled ``else''.
    Since $\sigma_i\in\Sigma_0$, $\sG_n\setminus V$ contains a unique $\Box$-component $C_i$ crossing $\sG_n$ in both directions.
    Let $H_i\subseteq V$ be the $\Box$-component of $V$ containing $\eta_i^{-1}u_i$.
    Let $\delta_{C_i}H_i$ be as in the statement of Lemma~\ref{lem: vertex planarity component boundary}, noting that its assumptions are satisfied.

    \smallskip
    
    \textbf{Construction of the $\boxtimes$-walk}\quad
    Consider the sequence of $\Box$-components $(\eta_i H_i)_{1\le i\le M}$.
    Since $u_i\in\eta_i H_i$ for all $1\le i\le M$, we have that $\eta_i H_i$ and $\eta_{i+1} H_{i+1}$ are $\Box$-adjacent for all $1\le i\le M-1$.
    By viewing the sequence as a walk on the vertex set $\set{\eta_i H_i\mid 1\le i\le M}$ and using the fact that any walk between two vertices contains a path between them~\cite[Section~1.3]{diestel2025graphtheory}, we may pass to a subsequence $(\eta_{i_j} H_{i_j})_{1\le j\le M'}$ satisfying the following properties:
    \begin{enumerate}
        \item the subsequence consists of distinct $\Box$-components and has the same first and last terms as the original sequence, and
        \item $\eta_{i_j} H_{i_j}$ and $\eta_{i_{j+1}} H_{i_{j+1}}$ are $\Box$-adjacent for all $1\le j\le M'-1$.
    \end{enumerate}
    For each $1\le j\le M'$, choose $w_{i_j}\in \eta_{i_j} \delta_{C_{i_j}}H_{i_j}$, and note that the set $\eta_{i_j}\delta_{C_{i_j}}H_{i_j}$ is $\boxtimes$-connected by Lemma~\ref{lem: vertex planarity component boundary}.
    Connect $u_1$ to $w_{i_1}$, $w_{i_j}$ to $w_{i_{j+1}}$ for $1\le j\le M'-1$, and $w_{i_{M'}}$ to $u_{M}$, each by a shortest $\boxtimes$-path in $S$.
    The union of these $\boxtimes$-paths and $\eta_{i_j}\delta_{C_{i_j}}H_{i_j}$, $1\le j\le M'$, is $\boxtimes$-connected and contained in $S$.
    Consider a spanning $\boxtimes$-tree of this union.
    Let $v_1,\dots,v_{N}$ be a spanning $\boxtimes$-walk in this tree from $u_1$ to $u_{M}$ that traverses each edge at most twice~\cite[Lemma~3.38]{friedli2017statistical}, and set $v_{N+1}=u_{M+1}$.
    Since $u_M$ and $u_{M+1}$ are $\Box$-adjacent, $v_1,\dots,v_N,v_{N+1}$ is a $\boxtimes$-walk.

    \smallskip

    \textbf{Properties of the $\boxtimes$-walk}\quad
    Properties~\ref{itm: escape leads to broken links - starting point} and~\ref{itm: escape leads to broken links - localization} are clear from the construction. 
    With $q$ set to the cardinality of the set $\bigcup_{j=1}^{M'}\eta_{i_j}\delta_{C_{i_j}}H_{i_j}$, Property~\ref{itm: escape leads to broken links - broken links} also follows by construction.

    \smallskip

    \textbf{Lower bound on $q$}\quad
    Observe that each $v\in\bigcup_{j=1}^{M'}\eta_{i_j}\delta_{C_{i_j}}H_{i_j}$ belongs to $\eta_{i_j}\delta_{C_{i_j}}H_{i_j}$ for at most eight distinct $1\le j\le M'$.
    Indeed, there are at most two choices of the configuration (out of $\sigma$ and $\sigma'$) in which $t_v$ is the endpoint of an $S$-local broken link, and $v$ is contained in at most four possible slices in~\eqref{eqn: escape leads to broken links - slice choices}.
    Hence, we have that
    \begin{equation}
        \label{eqn: escape leads to broken links - q repetition}
        \sum_{j=1}^{M'}\bars[\big]{\delta_{C_{i_j}}H_{i_j}} \le 8q.
    \end{equation}
    We deduce that
    \begin{align}
        \min\set{\ell,\fm}
        {}&\le \norm{u_{M}-u_1}_\infty
        {}&& \begin{subarray}{t}
            \text{since $u_1\in R$ and $u_M$ is $\Box$-adjacent}
            \\
            \text{to $\R^3\setminus S$, using the choices of $R$ and $S$}
        \end{subarray}
        \nonumber
        \\
        {}&\le \sum_{j=1}^{M'}\parentheses[\Big]{\diam_\infty(H_{i_j})+1}-1
        {}&& \begin{subarray}{t}
            \text{since $u_1\in\eta_{i_1}H_{i_1}$, $u_M\in\eta_{i_{M'}}H_{i_{M'}}$, and $\eta_{i_j}H_{i_j}$}
            \\
            \text{and $\eta_{i_{j+1}}H_{i_{j+1}}$ are $\Box$-adjacent for $1\le j\le M'-1$}
        \end{subarray}
        \nonumber
        \\
        {}&\le \sum_{j=1}^{M'}\bars[\big]{\delta_{C_{i_j}}H_{i_j}}-1
        {}&& \text{by Lemma~\ref{lem: vertex planarity component boundary}\ref{itm: vertex planarity component boundary - cardinality}}
        \nonumber
        \\
        {}&\le 8q
        {}&&\text{by~\eqref{eqn: escape leads to broken links - q repetition}},
    \end{align}
    which yields the desired lower bound on $q$.

    \smallskip

    \textbf{Upper bound on $N$}\quad
    Since $S$ is a box, the length of the shortest $\boxtimes$-path between any two vertices $u,v\in S$ is exactly $\norm{u-v}_\infty$.
    Hence, the number of vertices in the $\boxtimes$-walk satisfies
    \begin{equation}
        \label{eqn: escape leads to broken links - upper bound on N - first step}
        N+1\le 
        1
        +2\brackets[\bigg]{
            \dbars[\big]{u_1-w_{i_1}}_\infty
            +\sum_{j=1}^{M'}\parentheses[\Big]{\bars[\big]{\delta_{C_{i_j}}H_{i_j}}-1}
            +\sum_{j=1}^{M'-1}\dbars[\big]{w_{i_j}-w_{i_{j+1}}}_\infty
            +\dbars[\big]{w_{i_{M'}}-u_M}_\infty
        }
        +1,
    \end{equation}
    where the factor of $2$ accounts for the fact that each edge of the spanning $\boxtimes$-tree is traversed at most twice.
    We bound
    \begin{align}
        {}&\dbars[\big]{u_1-w_{i_1}}_\infty
        +\sum_{j=1}^{M'-1}\dbars[\big]{w_{i_j}-w_{i_{j+1}}}_\infty
        +\dbars[\big]{w_{i_{M'}}-u_M}_\infty
        \nonumber
        \\
        \le{}& \diam_\infty(H_{i_1})
        +\sum_{j=1}^{M'-1}\parentheses[\Big]{\diam_\infty(H_{i_j})+1+\diam_\infty(H_{i_{j+1}})}
        +\diam_\infty(H_{i_{M'}})
        \nonumber
        \\
        ={}& 2\,\sum_{j=1}^{M'}\diam_\infty(H_{i_j})+M'-1.
        \label{eqn: escape leads to broken links - upper bound on N - second step}
    \end{align}
    Hence, we obtain
    \begin{align}
        N+1 {}&\le 4\,\sum_{j=1}^{M'}\diam_\infty(H_{i_j})+2\,\sum_{j=1}^{M'}\bars[\big]{\delta_{C_{i_j}}H_{i_j}}
        {}&&\text{by~\eqref{eqn: escape leads to broken links - upper bound on N - first step} and~\eqref{eqn: escape leads to broken links - upper bound on N - second step}}
        \nonumber
        \\
        {}&\le 6\,\sum_{j=1}^{M'}\bars[\big]{\delta_{C_{i_j}}H_{i_j}}-4M'
        {}&&\text{by Lemma~\ref{lem: vertex planarity component boundary}\ref{itm: vertex planarity component boundary - cardinality}}
        \nonumber
        \\
        {}&\le 48q
        {}&&\text{by~\eqref{eqn: escape leads to broken links - q repetition}},
    \end{align}
    as desired.
\end{proof}

\begin{proposition}
    \label{prop: sealed boxes - no escape}
    For every $\eps>0$, one may choose $\fm\ge 2$ and then $\beta_0>0$ such that, for all $\beta\ge\beta_0$ and periodic Gibbs measures $\P,\P'$,
    \begin{equation}
        \peierls{\P\otimes\P'}(\xre{\ell\times\ell\times\fm}{\es{\Sigma_1\cup(\Sigma_0\times\Sigma_0)^c}})
        \le \eps.
    \end{equation}
\end{proposition}

\begin{proof}
    Let $\eps>0$.
    Since the event $\Sigma_1\cup(\Sigma_0\times\Sigma_0)^c$ is $S$-local and invariant under reflections across the axis-aligned planes through the center of $S$, by Lemma~\ref{lem:strongly percolating grid events - chessboard bound for product measures}, it suffices to prove that 
    \begin{equation}
        \norm{(\Sigma_0\times\Sigma_0)\setminus\Sigma_1}_S
        \le \eps^{27}.
    \end{equation}

    We follow a similar strategy to that in the proof of Proposition~\ref{prop: long chains of broken links are rare}.
    Let $n$ be sufficiently large so that $S$ is a block of $\Lambda_n\coloneqq\mathrm{R}_{n!\times n!\times n!}$.
    By definition,
    \begin{equation}
        \norm{(\Sigma_0\times\Sigma_0)\setminus\Sigma_1}_{S\mid\Lambda_n}
        =\parentheses[\big]{\P^\per_{\Lambda_n;\beta}\otimes\P^\per_{\Lambda_n;\beta}}\parentheses[\Bigg]{\bigcap_{\tau\in T^S_{\Lambda_n}}\tau\parentheses[\big]{(\Sigma_0\times\Sigma_0)\setminus\Sigma_1}}^{1/\bars{T^S_{\Lambda_n}}}.
    \end{equation}
    On the disseminated event $\bigcap_{\tau\in T^S_{\Lambda_n}}\tau\parentheses[\big]{(\Sigma_0\times\Sigma_0)\setminus\Sigma_1}$, for each $\tau\in T^S_{\Lambda_n}$, Lemma~\ref{lem: escape leads to broken links} supplies an integer $q_\tau\ge\min\set{\ell,\fm}/8$ and a $\boxtimes$-walk of length at most $48q_\tau$ starting in $\tau R$ and intersecting at least $q_\tau$ distinct vertices $v\in\tau S$ such that $t_v$ is the endpoint of a $\tau S$-local broken link of either $\sigma$ or $\sigma'$.
    It is not difficult to see that there exists a universal constant $C>0$ such that, for each $\tau\in T^S_{\Lambda_n}$, there are at most $\ell^2\fm C^{q_\tau}$ choices for the $\boxtimes$-walk, the $q_\tau$ distinguished vertices $v$, the configuration in which each $t_v$ is the endpoint of a $\tau S$-local broken link, and the broken link itself.

    Fix such a set of choices for every $\tau\in T^S_{\Lambda_n}$.
    Since at most eight vertices are associated with the same broken link in the same configuration, the number of distinct associated broken links is at least $1/8$ the total number of distinguished vertices.
    Since all the broken links in question are between $x_1x_2$-oriented plaquettes of parity $(0,0)$, by the pigeonhole principle applied as in the proof of Proposition~\ref{prop: long chains of broken links are rare}, there exists $z\in\set{0,1}^2\times\set{0}$ such that at least $1/4$ of the associated broken links $e$, across both configurations, are such that $R_e$ has corners in $\eta_{z}[(1/2+2\Z)^2\times(1/2+\Z)]$.
    Choose such a $z$ deterministically, and consider the corresponding subfamily of broken links.
    Applying the finite-volume version of Proposition~\ref{prop: infinite-volume chessboard estimate} to the event that every $\bA$-edge in this subfamily is a broken link in its respective configuration, using the finite-volume version of Corollary~\ref{cor: elementary defect chessboard seminorms}\ref{itm: broken link chessboard seminorm} (whose adaptation to joint configurations is straightforward), and applying a union bound over all original sets of choices, we get that
    \begin{equation}
        \parentheses[\big]{\P^\per_{\Lambda_n;\beta}\otimes\P^\per_{\Lambda_n;\beta}}\parentheses[\Bigg]{\bigcap_{\tau\in T^S_{\Lambda_n}}\tau\parentheses[\big]{(\Sigma_0\times\Sigma_0)\setminus\Sigma_1}}
        \le \brackets[\Bigg]{\sum_{q\ge\min\set{\ell,\fm}/8} \ell^2\fm\,C^q \parentheses[\big]{C_{\ref{cor: elementary defect chessboard seminorms}} e^{-\beta a/2}}^{q/32}}^{\bars{T^S_{\Lambda_n}}}.
    \end{equation}
    Taking the limit superior of the $\bars{T_{\Lambda_n}^S}$-th root of both sides as $n\to\infty$ shows that
    \begin{equation}
        \norm{(\Sigma_0\times\Sigma_0)\setminus\Sigma_1}_S
        \le \ell^2\fm\,\sum_{q\ge\min\set{\ell,\fm}/8} C^q \parentheses[\big]{C_{\ref{cor: elementary defect chessboard seminorms}} e^{-\beta a/2}}^{q/32}.
    \end{equation}

    Now, choose
    \begin{equation}
        \fm\coloneqq\max\set[\bigg]{2,\ceil[\bigg]{\frac{2048(E_0+\kappa/8)}{a}}}.
    \end{equation}
    For all sufficiently large $\beta$, we have that $\ell\ge\fm$ and $C(C_{\ref{cor: elementary defect chessboard seminorms}} e^{-\beta a/2})^{1/32}\le 1/2$, so that
    \begin{equation}
        \sum_{q\ge\min\set{\ell,\fm}/8} C^q \parentheses[\big]{C_{\ref{cor: elementary defect chessboard seminorms}} e^{-\beta a/2}}^{q/32}
        \le 2\brackets[\big]{C(C_{\ref{cor: elementary defect chessboard seminorms}} e^{-\beta a/2})^{1/32}}^{\fm/8}
        =\order{\ell^{-4}}.
    \end{equation}
    Hence, for all sufficiently large $\beta$, we have that $\norm{(\Sigma_0\times\Sigma_0)\setminus\Sigma_1}_S\le\eps^{27}$, as desired.
\end{proof}

\begin{corollary}
    \label{cor: sealed boxes strongly percolate}
    For every $\eps>0$, one may choose $\fm\ge 2$ and then $\beta_0>0$ such that, for all $\beta\ge\beta_0$, if $\P$ and $\P'$ are $\ell\Z^3$-ergodic Gibbs measures with $\Phase{\P}=\Phase{\P'}=(x_1x_2,0,0)$, then
    \begin{equation}
        \peierls{\P\otimes\P'}(\Pi)\le \eps.
    \end{equation}
\end{corollary}

\begin{proof}
    Let $\delta>0$.
    First, by taking $\fm\ge 2$ sufficiently large and choosing $\beta_0>0$ accordingly, we may ensure that the bound given by Proposition~\ref{prop: sealed boxes - no escape} is no greater than $\delta$.
    By increasing $\beta_0$ if necessary, we may also ensure that the bound given by Proposition~\ref{prop: sealed boxes - favorable environment} is no greater than $\delta$.
    Now, it is straightforward to show that
    \begin{equation}
        \label{eqn: sealed boxes strongly percolate - component}
        \peierls{\P\otimes\P'}(\xre{\ell\times\ell\times\fm}{\es{\Sigma_0\times\Omega}})
        \le \peierls{\P}(\xre{\ell\times\ell\times\fm}{\es{\Sigma_0}})
        \quad\text{and}\quad
        \peierls{\P\otimes\P'}(\xre{\ell\times\ell\times\fm}{\es{\Omega\times\Sigma_0}})
        \le \peierls{\P'}(\xre{\ell\times\ell\times\fm}{\es{\Sigma_0}})
        .
    \end{equation}
    Since $\Sigma_1\supseteq \parentheses{\Sigma_0\times\Sigma_0}\cap\brackets{\Sigma_1\cup\parentheses{\Sigma_0\times\Sigma_0}^c}$, we have that
    \begin{align}
        \peierls{\P\otimes\P'}(\Pi)
        \le{}&\peierls{\P\otimes\P'}(\xre{\ell\times\ell\times\fm}{\es{\parentheses{\Sigma_0\times\Sigma_0}\cap\brackets{\Sigma_1\cup\parentheses{\Sigma_0\times\Sigma_0}^c}}})
        {}&&\text{by monotonicity}
        \nonumber
        \\
        \le{}& 3\sqrt[3]{\delta}
        {}&&\begin{subarray}{t}
            \text{by Lemma~\ref{lem:intersection of strongly percolating sets}, \eqref{eqn: sealed boxes strongly percolate - component}, and}
            \\
            \text{Propositions~\ref{prop: sealed boxes - favorable environment} and \ref{prop: sealed boxes - no escape}}.
        \end{subarray}
    \end{align}
    The proof is complete by taking $\delta$ sufficiently small.
\end{proof}

\subsubsection{Proof of Proposition~\ref{prop:disagreement percolation decay}}

With $\eps=\eps_3/4$, choose $\fm\ge 2$ and $\beta_0>0$ so that the conclusion of Corollary~\ref{cor: sealed boxes strongly percolate} holds.
Let $\alpha_{x_1x_2}(s,t)$ be defined as in~\eqref{eqn: Heilmann-Lieb decay}, where $C\ge 1$ and $c>0$ will be chosen at the end of the proof.

\begin{claim}
    \label{clm: paths of disagreement tetramers}
    Let $(\sigma,\sigma')\in\Omega^2$ and $\gamma\subseteq\Delta_{\sigma,\sigma'}$ be $\ddag$-connected.
    Then, the set of vertices of $\Z^3$ covered by $\gamma$ is $\Box$-connected and disjoint from $\sigma\cap\sigma'$.
\end{claim}

\begin{proof}
    The first assertion follows from the fact that $\ddag$-adjacent plaquettes either intersect or contain $\Box$-adjacent vertices.
    For the second assertion, suppose by contradiction that there exist $v\in\Z^3$, $s\in\gamma$, and $t\in\sigma\cap\sigma'$ such that $v$ is covered by both $s$ and $t$.
    By the hard-core constraint, we must have $s=t$, which contradicts the assumption that $s\in\gamma\subseteq\Delta_{\sigma,\sigma'}$.
\end{proof}

Define $f:\F\to\Z^3$ by $f(t):=(\floor{t_1/\ell},\floor{t_2/\ell},\floor{t_3/\fm})$.
Let $(\sigma,\sigma')$ be sampled from $\P\otimes\P'$.
We will rely on the following claim that long $\ddag$-paths in $\Delta_{\sigma,\sigma'}$ induce long sequences of non-sealed boxes.

\begin{claim}
    \label{clm: disagreement paths induce boxtimes paths}
    Let $s,t\in\F$.
    Suppose that $s-t\not\in[-3\ell,3\ell]^2\times[-3\fm,3\fm]$ and that $s,t$ are connected by a $\ddag$-path in $\Delta_{\sigma,\sigma'}$.
    Then, $f(s),f(t)$ are connected by a $\boxtimes$-path in $\Z^3\setminus\Pi$.
\end{claim}

\begin{proof}
    Let $P$ be a $\ddag$-path in $\Delta_{\sigma,\sigma'}$ connecting $s$ and $t$.
    Since $\ell,\fm\ge 2$, the map $f$ sends $\ddag$-adjacent plaquettes to either the same vertex or $\boxtimes$-adjacent vertices of $\Z^3$.
    Thus, $f(P)$ contains a $\boxtimes$-walk $Q$ in $\Z^3$ connecting $f(s)$ to $f(t)$.
    It suffices to show that $Q\subseteq\Z^3\setminus\Pi$.

    First, note that it follows from the definition of $f$ that for each $v\in Q$, there exists a plaquette $r\in P$ such that $r\in R_v:=\eta_{(\ell v_1,\ell v_2,\fm v_3)}\mathrm{R}_{\ell\times\ell\times\fm}$.

    By contradiction, suppose that $Q\cap\Pi\ne\emptyset$.
    Let $v\in Q\cap\Pi$, and choose $r\in P$ as in the previous paragraph.
    Let $V_r$ be the set of vertices covered by the $\ddag$-component of $r$ in $\Delta_{\sigma,\sigma'}$.
    Then, $R_v$ is sealed.
    By Claim~\ref{clm: paths of disagreement tetramers}, $V_r$ is $\Box$-connected and disjoint from $\sigma\cap\sigma'$.
    By the definition of $\Sigma_1$, $V_r$ is contained in $\eta_{(\ell (v_1-1),\ell (v_2-1),\fm (v_3-1))}\mathrm{R}_{3\ell\times3\ell\times3\fm}$.
    However, since $s$ and $t$ both belong to the $\ddag$-component of $r$ in $\Delta_{\sigma,\sigma'}$, this contradicts the assumption that $s-t\not\in[-3\ell,3\ell]^2\times[-3\fm,3\fm]$.
\end{proof}

Let $s\in A$.
By a union bound, it suffices to show that
\begin{equation}
    (\P\otimes\P')(\set{s}\text{ and }B\text{ are connected by a $\ddag$-path in }\Delta_{\sigma,\sigma'})\le \sup_{t\in B}\alpha_{x_1x_2}(s,t).
\end{equation}
Suppose that $s$ is connected by a $\ddag$-path in $\Delta_{\sigma,\sigma'}$ to some $t\in B$.
If there exists such a $t$ with $s-t\in[-3\ell,3\ell]^2\times[-3\fm,3\fm]$, then
\begin{equation}
    \alpha_{x_1x_2}(s,t)\ge C\exp{-(6+3\fm)c}.
\end{equation}
Otherwise, by Claim~\ref{clm: disagreement paths induce boxtimes paths}, $f(s)$ is connected to $f(t)$ by a $\boxtimes$-path in $\Z^3\setminus\Pi$.
By Corollary~\ref{cor: sealed boxes strongly percolate} and Lemma~\ref{lem:quantitative Peierls argument}, the probability of this event is bounded above by
\begin{equation}
    \begin{multlined}
        (\eps/\eps_3)^{1+\inf_{t\in B}\norm{f(s)-f(t)}_\infty}
        \le (\eps/\eps_3)^{\frac{1}{3}\inf_{t\in B}\left(\ell^{-1}\sum_{i=1}^2\abs{s_i-t_i}+\fm^{-1}\abs{s_3-t_3}\right)}
        \\
        =\sup_{t\in B}(\eps/\eps_3)^{\frac{1}{3}\left(\ell^{-1}\sum_{i=1}^2\abs{s_i-t_i}+\fm^{-1}\abs{s_3-t_3}\right)}.
    \end{multlined}
\end{equation}
Choosing $C\ge 1,\,c>0$ so that 
\begin{equation}
    C\exp{-(6+3\fm)c}\ge 1,
    \quad
    (\eps/\eps_3)^{\frac{1}{3}}\le e^{-c},
    \quad\text{and}\quad
    (\eps/\eps_3)^{\frac{1}{3\fm}}\le e^{-c}
\end{equation}
completes the proof.

\subsection{Microscopic characterizations of discotic columnar order}

Finally, we derive microscopic characterizations of discotic columnar order in terms of the probability of observing a tetramer of the dominant orientation and parity versus one with a different parity or a different orientation, thus establishing Theorem~\ref{thm:main}\ref{itm:main - tetramer probabilities}.
The symmetry relations of Proposition~\ref{prop:twelve phases}\ref{itm:twelve phases - transformation} allow us to restrict to the $(x_1x_2,0,0)$-phase, where it suffices to show that each reference tetramer is missing with probability $\order{e^{-\beta a/2}}$.
We note that this is not immediate from the results obtained earlier, since the events defining the phase label only guarantee the presence of a fixed fraction of reference tetramers in a typical slice.
Instead, we exploit the $2\Z\times2\Z\times\Z$-invariance of $\P_{(x_1x_2,0,0)}$ to represent this probability in terms of the expected fraction of vertices in a slice not covered by a reference tetramer, and then use linearity of expectation to reduce it to the probabilistic cost of a single broken link.

\begin{proposition}
    \label{prop: good tetramers should not be missing}
    Suppose that $\beta\ge\beta_0$.
    If $t\in\F$ is $x_1x_2$-oriented of parity $(0,0)$, then
    \begin{equation}
        \P_{x_1x_2,(0,0)}(\sigma(t)=0)=\order{e^{-\beta a/2}}.
    \end{equation}
\end{proposition}

\begin{proof}
    To lighten the notation, write $G=G^\dom_{(0,0)}\setminus B^{\mathrm{long}}_{(0,0),r}$, $\P=\P_{x_1x_2,(0,0)}$, and $\E$ for the corresponding expectation.
    Recall the slice $L$ from Section~\ref{sec: mesoscopic slices have a dominant parity}.
    Define the random variable $X:\Omega\to[0,1]$ as the fraction of vertices of $L$ \textit{not} covered by $x_1x_2$-oriented tetramers of parity $(0,0)$.
    By Proposition~\ref{prop:twelve phases}\ref{itm:twelve phases - invariance}, $\P$ is $2\Z\times2\Z\times\Z$-invariant, so 
    \begin{equation}
        \P(\sigma(t)=0)
        =\frac{1}{\abs{\set{s\in\F\mid s\subseteq L,\,x_1x_2\text{-oriented of parity }(0,0)}}}\sum_{\substack{s\in\F:\,s\subseteq L\\x_1x_2\text{-oriented}\\\text{parity }(0,0)}}\P(\sigma(s)=0)
        =\E[X].
    \end{equation}
    We split
    \begin{equation}
        \E[X]=\E[X\indicator{G}]+\E[X\indicator{G^c}]
        \le \E[X\indicator{G}]+\P(G^c)
    \end{equation}
    and bound each term on the right-hand side separately.

    We start with the first term.
    Recall the edge set $\cE_{\sigma,p}$ introduced in Section~\ref{sec: long chains of broken links}.
    In the setting of Lemma~\ref{lem: edge planarity}, let $n=2\ell$ and $E=\cE_{\sigma,(0,0)}\cap\sE_n$.
    On the event $G$, the exclusion $\parentheses[\big]{B^{\mathrm{long}}_{(0,0),r}}^c$ ensures that every $F\subseteq E$ with connected dual satisfies $\diam_\infty^\ast(F^\ast)\le r$ (see~\eqref{eqn: exclusion of long chains of broken links}), while parity dominance $G^\dom_{(0,0)}$ ensures that the unique component of $\sG_n\setminus E$ crossing $\sG_n$ in both directions consists of vertices labeled $(0,0)$.
    Hence, by Lemma~\ref{lem: edge planarity}\ref{itm: edge planarity - cardinality}, on the event $G$, the number of vertices of $L$ not covered by $x_1x_2$-oriented tetramers of parity $(0,0)$ is at most $\frac{r}{2}\abs{\cE_{\sigma,(0,0)}\cap\sE_{2\ell}}$.
    Thus,
    \begin{equation}
        \E[X\indicator{G}]
        \le \frac{r}{2\abs{\sV_{2\ell}}} \E\brackets[\big]{\abs{\cE_{\sigma,(0,0)}\cap\sE_{2\ell}}\indicator{G}}
        \le \frac{r}{2\abs{\sV_{2\ell}}} \sum_{e\in\sE_{2\ell}}\P(e\in\cE_{\sigma,(0,0)})
        \le \frac{r\abs{\sE_{2\ell}}}{2\abs{\sV_{2\ell}}} C_{\ref{cor: elementary defect chessboard seminorms}}e^{-\frac{1}{2}\beta a},
    \end{equation}
    having used Proposition~\ref{prop: infinite-volume chessboard estimate} and Corollary~\ref{cor: elementary defect chessboard seminorms}\ref{itm: broken link chessboard seminorm} in the last step.

    For the second term, by a union bound, we have that
    \begin{equation}
        \P(G^c)\le \P\parentheses[\big]{(G^\dom_{(0,0)})^c}+\P\parentheses[\big]{B^{\mathrm{long}}_{(0,0),r}}.
    \end{equation}
    On the one hand, by Lemma~\ref{lem:quantitative Peierls argument}, Proposition~\ref{prop: four parities}, and~\eqref{eqn: four parities - cubic bound}, we have that
    \begin{equation}
        \label{eqn: good tetramers should not be missing - parity non-dominance}
        \P\parentheses[\big]{(G^\dom_{(0,0)})^c}
        \le \frac{\peierls{\P}(\Theta_{(0,0)})}{\eps_3}
        \le \frac{3\sqrt[3]{\max\set*{\eps_\reg(\beta,r,\ell),\eps_\dom(\beta,r,\ell),\eps_\syn(\beta,r,\ell)}}}{\eps_3}.
    \end{equation}
    On the other hand, by Propositions~\ref{prop: infinite-volume chessboard estimate} and~\ref{prop: long chains of broken links are rare},
    \begin{equation}
        \label{eqn: good tetramers should not be missing - long chains of broken links}
        \P\parentheses[\big]{B^{\mathrm{long}}_{(0,0),r}}
        \le \dbars[\big]{B^{\mathrm{long}}_{(0,0),r}}_L
        \le 4\ell(2\ell-1) 6^{2r} \parentheses[\big]{C_{\ref{cor: elementary defect chessboard seminorms}}e^{-\beta a/2}}^{r/8}.
    \end{equation}

    The proof is complete by combining the above estimates, noting that the right-hand sides of both~\eqref{eqn: good tetramers should not be missing - parity non-dominance} and~\eqref{eqn: good tetramers should not be missing - long chains of broken links} are $\order{\ell^{-1/3}}=o(e^{-\beta a/2})$ by the choices of $\ell$ and $r$ (see Section~\ref{sec: parity synchronization between adjacent mesoscopic slices}).
\end{proof}

We now deduce Theorem~\ref{thm:main}\ref{itm:main - tetramer probabilities}.

\begin{corollary}
    Suppose that $\beta\ge\beta_0$.
    If $t\in\F$ is $x_1x_2$-oriented of parity $(0,0)$, then
    \begin{equation}
        \P_{x_1x_2,(0,0)}(\sigma(t)=1)=1-\order{e^{-\beta a/2}}.
    \end{equation}
\end{corollary}

\begin{proof}
    This is immediate from Proposition~\ref{prop: good tetramers should not be missing}, since $\Omega=\set{\sigma(t)=0}\sqcup\set{\sigma(t)=1}$.
\end{proof}

\begin{corollary}
    \label{cor: out-of-place tetramers are rare}
    Suppose that $\beta\ge\beta_0$.
    If $t\in\F$ is not $x_1x_2$-oriented or not of parity $(0,0)$, then
    \begin{equation}
        \P_{x_1x_2,(0,0)}(\sigma(t)=1)=\order{e^{-\beta a/2}}.
    \end{equation}
\end{corollary}

\begin{proof}
    Let $s\in\F$ be an $x_1x_2$-oriented tetramer of parity $(0,0)$ that intersects $t$.
    Then, $\set{\sigma(t)=1}\subseteq\set{\sigma(s)=0}$.
    The corollary now follows from Proposition~\ref{prop: good tetramers should not be missing}. 
\end{proof}

\section*{Acknowledgments}

We thank Deepak Dhar for helpful discussions on lattice hard-core models in three dimensions, Daniel Hadas for answering a question about the adaptability of the framework of strongly percolating sets to higher dimensions, and Matan Harel for related discussions.

\bibliographystyle{plain}
\bibliography{bibliography}

\begin{thebibliography}{10}

\bibitem{abraham1980interacting}
Douglas~B. Abraham and Ole~J. Heilmann.
\newblock Interacting dimers on the simple cubic lattice as a model for liquid crystals.
\newblock {\em Journal of Physics A: Mathematical and General}, 13(3):1051--1062, 1980.

\bibitem{alberici2016cluster}
Diego Alberici.
\newblock {A cluster expansion approach to the Heilmann--Lieb liquid crystal model}.
\newblock {\em Journal of Statistical Physics}, 162(3):761--791, 2016.

\bibitem{alencar2026slab}
George~L.D. Alencar, Nicolau~C. Saldanha, and Arthur~M.M. Vieira.
\newblock Slab tilings, flips and the triple twist.
\newblock {\em Discrete \& Computational Geometry}, 75(4):1266--1287, 2026.

\bibitem{angelescu1995long}
N.~Angelescu, Silvano Romano, and V.A. Zagrebnov.
\newblock On long-range order in low-dimensional lattice-gas models of nematic liquid crystals.
\newblock {\em Physics Letters A}, 200(6):433--437, 1995.

\bibitem{angelescu1982lattice}
N.~Angelescu and V.A. Zagrebnov.
\newblock A lattice model of liquid crystals with matrix order parameter.
\newblock {\em Journal of Physics A: Mathematical and General}, 15(11):L639--L643, 1982.

\bibitem{barlow2004random}
Martin~T. Barlow.
\newblock {Random walks on supercritical percolation clusters}.
\newblock {\em The Annals of Probability}, 32(4):3024--3084, 2004.

\bibitem{boden1995mechanism}
N.~Boden, R.J. Bushby, J.~Clements, B.~Movaghar, K.J. Donovan, and T.~Kreouzis.
\newblock Mechanism of charge transport in discotic liquid crystals.
\newblock {\em Physical Review B}, 52(18):13274--13280, 1995.

\bibitem{bollobas2006short}
B{\'e}la Bollob{\'a}s and Oliver Riordan.
\newblock {A short proof of the Harris--Kesten theorem}.
\newblock {\em Bulletin of the London Mathematical Society}, 38(3):470--484, 2006.

\bibitem{bollobas2006percolation}
B{\'e}la Bollob{\'a}s and Oliver Riordan.
\newblock {\em Percolation}.
\newblock Cambridge University Press, 2006.

\bibitem{bricmont1989phase}
J.~Bricmont and J.~Slawny.
\newblock Phase transitions in systems with a finite number of dominant ground states.
\newblock {\em Journal of Statistical Physics}, 54(1):89--161, 1989.

\bibitem{chandrasekhar1977liquid}
S.~Chandrasekhar, B.K. Sadashiva, and K.A. Suresh.
\newblock Liquid crystals of disc-like molecules.
\newblock {\em Pramana}, 9(5):471--480, 1977.

\bibitem{deuschel1996surface}
Jean-Dominique Deuschel and Agoston Pisztora.
\newblock Surface order large deviations for high-density percolation.
\newblock {\em Probability Theory and Related Fields}, 104(4):467--482, 1996.

\bibitem{diestel2025graphtheory}
Reinhard Diestel.
\newblock {\em Graph Theory}, volume 173 of {\em Graduate Texts in Mathematics}.
\newblock Springer, 6th edition, 2025.

\bibitem{disertori2013nematic}
Margherita Disertori and Alessandro Giuliani.
\newblock The nematic phase of a system of long hard rods.
\newblock {\em Communications in Mathematical Physics}, 323(1):143--175, 2013.

\bibitem{disertori2020plate}
Margherita Disertori, Alessandro Giuliani, and Ian Jauslin.
\newblock Plate-nematic phase in three dimensions.
\newblock {\em Communications in Mathematical Physics}, 373(1):327--356, 2020.

\bibitem{dobrushin1968problem}
Roland~L'vovich Dobrushin.
\newblock {The problem of uniqueness of a Gibbsian random field and the problem of phase transitions}.
\newblock {\em Functional Analysis and Its Applications}, 2(4):302--312, 1968.

\bibitem{duminil2017exponential}
Hugo Duminil-Copin, Ron Peled, Wojciech Samotij, and Yinon Spinka.
\newblock {Exponential decay of loop lengths in the loop $O(n)$ model with large $n$}.
\newblock {\em Communications in Mathematical Physics}, 349(3):777--817, 2017.

\bibitem{friedli2017statistical}
Sacha Friedli and Yvan Velenik.
\newblock {\em Statistical mechanics of lattice systems: a concrete mathematical introduction}.
\newblock Cambridge University Press, 2017.

\bibitem{galvin2004phase}
David Galvin and Jeff Kahn.
\newblock On phase transition in the hard-core model on $\mathbb{Z}^d$.
\newblock {\em Combinatorics, Probability and Computing}, 13(2):137--164, 2004.

\bibitem{degennes1993physics}
P.~G.~de Gennes and J.~Prost.
\newblock {\em The Physics of Liquid Crystals}.
\newblock Oxford University Press, 1993.

\bibitem{georgii2011gibbs}
Hans-Otto Georgii.
\newblock {\em Gibbs measures and phase transitions}, volume~9.
\newblock Walter de Gruyter, 2011.

\bibitem{giuliani2015columnar}
Alessandro Giuliani and Elliott~H. Lieb.
\newblock Columnar phase in quantum dimer models.
\newblock {\em Journal of Physics A: Mathematical and Theoretical}, 48(23):235203, 2015.

\bibitem{hadas2026columnar}
Daniel Hadas and Ron Peled.
\newblock Columnar order in random packings of $2\times2$ squares on the square lattice.
\newblock {\em Journal of the European Mathematical Society}, 28(11):5023--5106, 2026.

\bibitem{he2025extended}
Qidong He.
\newblock {Extended regime of nematic order in an interacting monomer-dimer model of Heilmann and Lieb}.
\newblock {\em arXiv preprint arXiv:2511.23437}, 2025.

\bibitem{he2026nematic}
Qidong He.
\newblock {Nematic order in monomer-dimer models of Heilmann and Lieb with lateral attraction in dimensions $d\ge 2$}.
\newblock {\em arXiv preprint arXiv:2610.08600}, 2026.

\bibitem{heilmann1979lattice}
Ole~J. Heilmann and Elliott~H. Lieb.
\newblock Lattice models for liquid crystals.
\newblock {\em Journal of Statistical Physics}, 20(6):679--693, 1979.

\bibitem{jauslin2018nematic}
Ian Jauslin and Elliott~H. Lieb.
\newblock Nematic liquid crystal phase in a system of interacting dimers and monomers.
\newblock {\em Communications in Mathematical Physics}, 363(3):955--1002, 2018.

\bibitem{klivans2026domino}
Caroline~J. Klivans and Nicolau Saldanha.
\newblock {Domino tilings beyond 2D}.
\newblock In {\em Surveys in Combinatorics 2026}, London Mathematical Society Lecture Note Series, pages 267--292. Cambridge University Press, 2026.

\bibitem{kotecky1993spin}
R.~Koteck{\'y}, L.~Laanait, A.~Messager, and S.~Miracle-Sol{\'e}.
\newblock A spin-1 lattice model of microemulsions at low temperatures.
\newblock {\em Journal of Physics A: Mathematical and General}, 26(20):5285--5293, 1993.

\bibitem{letawe1994module}
I.~Letawe.
\newblock {Le mod\`ele de cristaux liquides de Heilmann et Lieb}.
\newblock {\em M{\'e}moire de Licenci{\'e}e en Sciences, Universit{\'e} Catholique de Louvain, Louvain-la-Neuve}, 1994.

\bibitem{mandal2023phases}
Dipanjan Mandal, Geet Rakala, Kedar Damle, Deepak Dhar, and R.~Rajesh.
\newblock Phases of the hard-plate lattice gas on a three-dimensional cubic lattice.
\newblock {\em Physical Review E}, 107(6):064136, 2023.

\bibitem{pak2013complexity}
Igor Pak and Jed Yang.
\newblock The complexity of generalized domino tilings.
\newblock {\em The Electronic Journal of Combinatorics}, 20(4):\#P12, Oct. 2013.

\bibitem{peled2014odd}
Ron Peled and Wojciech Samotij.
\newblock Odd cutsets and the hard-core model on $\mathbb{Z}^d$.
\newblock {\em Annales de l'IHP Probabilit{\'e}s et Statistiques}, 50(3):975--998, 2014.

\bibitem{peled2019lectures}
Ron Peled and Yinon Spinka.
\newblock {Lectures on the spin and loop $O(n)$ models}.
\newblock In {\em Sojourns in Probability Theory and Statistical Physics-I: Spin Glasses and Statistical Mechanics, A Festschrift for Charles M. Newman}, pages 246--320. Springer, 2019.

\bibitem{pirogov1975phase}
Sergey~Anatol'evich Pirogov and Yakov~Grigor'evich Sinai.
\newblock Phase diagrams of classical lattice systems.
\newblock {\em Teoreticheskaya i Matematicheskaya Fizika}, 25(3):358--369, 1975.
\newblock (Russian, English translation in Theoret. and Math. Phys. 25 (1975), no. 3, 1185--1192).

\bibitem{pirogov1976phase}
Sergey~Anatol'evich Pirogov and Yakov~Grigor'evich Sinai.
\newblock Phase diagrams of classical lattice systems (continuation).
\newblock {\em Teoreticheskaya i Matematicheskaya Fizika}, 26(1):61--76, 1976.
\newblock (Russian, English translation in Theoret. and Math. Phys. 26 (1976), no. 1, 39--49).

\bibitem{schmidt2001self}
Lukas Schmidt-Mende, A.~Fechtenk{\"o}tter, K.~M{\"u}llen, Ellen Moons, Richard~H. Friend, and J.~Devin MacKenzie.
\newblock Self-organized discotic liquid crystals for high-efficiency organic photovoltaics.
\newblock {\em Science}, 293(5532):1119--1122, 2001.

\bibitem{sentker2019self}
Kathrin Sentker, Arda Yildirim, Milena Lippmann, Arne~W. Zantop, Florian Bertram, Tommy Hofmann, Oliver~H. Seeck, Andriy~V. Kityk, Marco~G. Mazza, Andreas Sch{\"o}nhals, et~al.
\newblock Self-assembly of liquid crystals in nanoporous solids for adaptive photonic metamaterials.
\newblock {\em Nanoscale}, 11(48):23304--23317, 2019.

\bibitem{sheffield2005random}
Scott Sheffield.
\newblock Random surfaces.
\newblock In {\em Ast{\'e}risque}, volume 304. Soci{\'e}t{\'e} math{\'e}matique de France Paris, 2005.

\bibitem{timar2013boundary}
{\'A}d{\'a}m Tim{\'a}r.
\newblock Boundary-connectivity via graph theory.
\newblock {\em Proceedings of the American Mathematical Society}, 141(2):475--480, 2013.

\bibitem{van1993uniqueness}
Jacob van~den Berg.
\newblock {A uniqueness condition for Gibbs measures, with application to the 2-dimensional Ising antiferromagnet}.
\newblock {\em Communications in Mathematical Physics}, 152(1):161--166, 1993.

\bibitem{van1994disagreement}
Jacob van~den Berg and Christian Maes.
\newblock {Disagreement percolation in the study of Markov fields}.
\newblock {\em The Annals of Probability}, 22(2):749--763, 1994.

\bibitem{vigneshwar2019phase}
N.~Vigneshwar, Dipanjan Mandal, Kedar Damle, Deepak Dhar, and R.~Rajesh.
\newblock Phase diagram of a system of hard cubes on the cubic lattice.
\newblock {\em Physical Review E}, 99(5):052129, 2019.

\bibitem{zagrebnov1996long}
V.A. Zagrebnov.
\newblock Long-range order in a lattice-gas model of nematic liquid crystals.
\newblock {\em Physica A: Statistical Mechanics and its Applications}, 232(3-4):737--746, 1996.

\bibitem{zahradnik1984alternate}
Milo{\v s} Zahradn{\'\i}k.
\newblock {An alternate version of Pirogov--Sinai theory}.
\newblock {\em Communications in Mathematical Physics}, 93(4):559--581, 1984.

\bibitem{zbiri2010density}
Mohamed Zbiri, Mark~R. Johnson, Gordon~J. Kearley, and Fokko~M. Mulder.
\newblock {Density functional calculations of potential energy surface and charge transfer integrals in molecular triphenylene derivative HAT${}_6$}.
\newblock {\em Theoretical Chemistry Accounts}, 125(3):445--451, 2010.

\end{thebibliography}

\appendix

\section{Properties of strongly percolating sets}
\label{appx: properties of strongly percolating sets}

In this appendix, we collect some properties of strongly percolating sets (defined in Section~\ref{sec: strongly percolating sets}) for the reader's convenience, all of which are imported from previous works.
While these results were originally formulated by Hadas--Peled~\cite[Section~6]{hadas2026columnar} in two dimensions, they extend straightforwardly to three (and higher) dimensions, so we will not belabor their proofs here.

\begin{lemma}[{\cite[Lemma 6.2]{hadas2026columnar}}]
    \label{lem:Peierls argument}
    If $\eps<\eps_3$ and $B$ is $\eps$-rare with respect to a probability measure $\P$, then $\Z^3\setminus B$ $\P$-almost surely contains a unique infinite $\Box$-component $I$.
    Moreover, each $\boxtimes$-component of $\Z^3\setminus I$ is finite.
\end{lemma}

\begin{lemma}[{\cite[Corollary 6.3]{hadas2026columnar}}]
    \label{lem:strong percolation of complement of rare set}
    If $\eps\ge 0$ and $B$ is $\eps$-rare with respect to a probability measure $\P$, then $\Z^3\setminus B$ is $\eps$-strongly percolating, and in particular $\peierls{\P}(\Z^3\setminus B)\le\eps$.
\end{lemma}

\begin{lemma}[{\cite[Lemma 6.4]{hadas2026columnar}}]
    \label{lem:quantitative Peierls argument}
    Let $u\in\Z^3$, $d\ge 0$, and $\Psi\subseteq\Z^3$ be a random set on a probability space $(\Omega,\sF,\P)$.
    Let $E$ be the event that there exists a $\boxtimes$-path in $\Z^3\setminus\Psi$ starting at $u$ and ending at some point in $\set{v\in\Z^3\mid\norm{v-u}_\infty\ge d}$.
    Then, 
    \begin{equation}
        \P(E)\le\left(\frac{\peierls{\P}(\Psi)}{\eps_3}\right)^{d+1}.
    \end{equation}
\end{lemma}

\begin{lemma}[{\cite[Lemma 6.6]{hadas2026columnar}}]
    \label{lem:intersection of strongly percolating sets}
    For all $k\ge 1$ and random sets $\Psi_1,\dots,\Psi_k\subseteq\Z^3$ on a common probability space $(\Omega,\sF,\P)$,
    \begin{equation}
        \peierls{\P}\parentheses[\Bigg]{\bigcap_{i=1}^k\Psi_i}\le k\sqrt[k]{\max_{1\le i\le k}\peierls{\P}(\Psi_i)}.
    \end{equation}
\end{lemma}

\begin{lemma}[{\cite[Lemma 6.7]{hadas2026columnar}}]
    \label{lem:strongly percolating grid events}
    Let $E\subseteq\Omega$ be an event.
    Let $l_1,l_2,l_3,L_1,L_2,L_3\in\Z_{>0}$ be such that $l_i\mid L_i$ for all $i\in\set{1,2,3}$.
    Denote $H\coloneqq\set{\eta_{z}\mid z\in(l_1\Z\times l_2\Z\times l_3\Z)\cap([0,L_1)\times[0,L_2)\times[0,L_3))}$ and $r\coloneqq\abs{H}=\frac{L_1L_2L_3}{l_1l_2l_3}$.
    Denote $\Psi\coloneqq\xre{l_1\times l_2\times l_3}{\es{E}}$, $\Psi'_\eta\coloneqq\xre{L_1\times L_2\times L_3}{\es{\eta E}}$ for each $\eta\in H$, and $\Psi'\coloneqq\bigcap_{\eta\in H}\Psi'_\eta$.
    Then,
    \begin{enumerate}[label=(\roman*)]
        \item \label{itm:strongly percolating grid events - subgrid over grid} $\peierls{\P}(\Psi')\le r\peierls{\P}(\Psi)$.
        \item $\peierls{\P}(\Psi)\le \sqrt[r]{\peierls{\P}(\Psi')}$.
        \item \label{itm:strongly percolating grid events - grid over subgrid}  $\peierls{\P}(\Psi)\le \sqrt[r]{r\sqrt[r]{\max_{\eta\in H}\peierls{\P}(\Psi'_\eta)}}$.
        \item \label{itm:strongly percolating grid events - chessboard bound} If $E\subseteq\Omega$ is $R$-local for $R=\rect{L_1\times L_2\times L_3}{z}$ and is invariant under reflections across all three axis-aligned planes passing through the center of $R$, then, for all periodic Gibbs measures $\P$,
        \begin{equation}
            \peierls{\P}(\Psi)\le\sqrt[r]{\norm{\Omega\setminus E}_R}.
        \end{equation}
    \end{enumerate}
\end{lemma}

\begin{lemma}[{\cite[Lemma 2.18]{he2025extended}}]
    \label{lem:strongly percolating grid events - chessboard bound for product measures}
    In the setting of Lemma~\ref{lem:strongly percolating grid events}, if $E\subseteq\Omega^k$ is $R$-local for $R=\rect{L_1\times L_2\times L_3}{z}$ and is invariant under reflections across the axis-aligned planes passing through the center of $R$, then, given periodic Gibbs measures $(\P_i)_{i=1}^k$,
        \begin{equation}
            \peierls{\otimes_{i=1}^k\P_i}(\Psi)\le \sqrt[r]{\norm{\Omega^k\setminus E}_R}.
        \end{equation}
\end{lemma}

\begin{lemma}[{\cite[Proposition 6.8]{hadas2026columnar}}]
    \label{lem:splitting strongly percolating sets}
    Let $\P$ be an $\cL$-ergodic probability measure on $\Omega$ for some lattice $\cL\subseteq L_1\Z\times L_2\Z\times L_3\Z$.
    Let $E,F\subseteq\Omega$ be events.
    If
    \begin{equation}
        \label{eqn: splitting strongly percolating sets - condition}
        E\cap F=E\cap\eta_{L_i\vec{e}_i}F=\eta_{L_i\vec{e}_i}E\cap F=\emptyset
        \quad\text{for all }i\in\set{1,2,3},
    \end{equation}
    then 
    \begin{equation}
        \min\set{\peierls{\P}(\xre{L_1\times L_2\times L_3}{\es{E}}),\peierls{\P}(\xre{L_1\times L_2\times L_3}{\es{F}})}\le\peierls{\P}(\xre{L_1\times L_2\times L_3}{\es{E\cup F}}).
    \end{equation}
\end{lemma}

\end{document}